\documentclass[reprint,superscriptaddress,amsmath,amssymb,aps,nofootinbib]{revtex4-2}
\usepackage{dsfont}
\usepackage{graphicx}
\usepackage[caption=false]{subfig}
\usepackage{enumitem}
\usepackage{dcolumn}
\usepackage{bm}
\usepackage{xcolor}
\usepackage{epsfig}
\usepackage{soul}
\usepackage[colorlinks=true,urlcolor=brown,linkcolor=blue,citecolor=magenta]{hyperref}
\usepackage{float}
\usepackage{footmisc}
\usepackage{siunitx}
\usepackage{booktabs}
\usepackage{orcidlink}
\usepackage[capitalise]{cleveref}
\usepackage{cancel}
\usepackage{bbold}
\usepackage{bbm}
\usepackage{caption}
\usepackage{subcaption}
\usepackage{mathtools}
\usepackage{slashed}
\usepackage{appendix}
\usepackage{amsmath,amssymb}
\usepackage{todonotes}
\usepackage{multirow}
\usepackage[T1]{fontenc}
\usepackage{colortbl}
\usepackage{xspace}
\usepackage[normalem]{ulem}

\newcommand{\GeV}{~\mathrm{GeV}}

\allowdisplaybreaks

\begin{document}
  \title{
  Simulation-based inference for gravitational-wave cosmology:\\[0.2em]
  parameter reconstruction of a first-order electroweak phase transition
}

\author{Thomas Biekötter}
\email{thomas.biekotter@csic.es}
\affiliation{
  Instituto de F\'isica Te\'orica UAM/CSIC,
  Calle Nicolás Cabrera 13-15,
  Cantoblanco, 28049, Madrid, Spain
}

\author{Andrii Dashko}
\email{andrii.dashko@ugr.es}
\affiliation{
  Departamento de Física Teórica y del Cosmos,
  Universidad de Granada, E-18071 Granada, Spain
}

\author{Kateryna Radchenko}
\email{kateryna.radchenko@desy.de}
\affiliation{
  Instituto de F\'isica Te\'orica UAM/CSIC,
  Calle Nicolás Cabrera 13-15,
  Cantoblanco, 28049, Madrid, Spain
}

{\hfill\footnotesize IFT-UAM/CSIC-26-169}
\vspace*{1em}

\begin{abstract}
The detection of a stochastic gravitational-wave background (SGWB) from a cosmological
first-order phase transition would be unambiguous evidence for physics
beyond the Standard Model, with a strong electroweak phase transition~(EWPT)
at LISA as a prime example. Reconstructing the underlying model from
such a signal is challenging, since the map from the model parameters
to the spectrum is degenerate and strong transitions occupy only small
and sparse regions of parameter space. We show that simulation-based inference
(SBI), a likelihood-free approach applicable to generic SGWB sources,
provides a reliable parameter reconstruction despite these
difficulties. We first apply SBI to the thermodynamic parameters
characterizing the transition, namely its strength, inverse duration,
temperature and bubble wall velocity. As a concrete application, we then
consider a strong
EWPT in a singlet extension of the Standard Model
and reconstruct the model parameters
from the resulting SGWB. We find that
the SBI posteriors agree with those of a likelihood-based Markov chain Monte Carlo analysis, at
negligible cost per data set once the network has
been trained. 
Using the network posterior as the proposal for importance sampling with the exact likelihood further corrects its remaining deviations, at a sampling efficiency several orders of magnitude higher than sampling from the prior. SBI thus offers a fast and flexible framework for the
cosmological science case of LISA. 

\end{abstract}

\maketitle

  \section{Introduction}
\label{sec:intro}

A cosmological first-order phase transition proceeds through the
nucleation, expansion and collision of bubbles of the broken phase,
and thereby sources a stochastic gravitational-wave background
(SGWB)~\cite{Witten:1984rs,Hogan:1986qda,Kosowsky:1992rz,Kamionkowski:1993fg}.
The peak frequency of the signal today is set by the temperature at which the
transition takes place,
such that gravitational-wave (GW) observatories operating in different
frequency bands probe phase transitions at very different energy
scales~\cite{Grojean:2006bp}. Pulsar timing arrays, which have
recently found evidence for a SGWB in the nanohertz
band~\cite{NANOGrav:2023gor,EPTA:2023fyk}, are sensitive to
transitions at the MeV to GeV scale.
The Laser Interferometer
Space Antenna (LISA)~\cite{LISA:2017pwj,LISA:2024hlh}, as well as
the proposed space missions Taiji~\cite{Hu:2017mde} and
TianQin~\cite{TianQin:2015yph}, will probe the millihertz band,
which coincides with the characteristic frequencies of transitions
taking place at temperatures around the electroweak
scale~\cite{Caprini:2015zlo,Caprini:2019egz}. Deci-hertz
observatories such as DECIGO~\cite{Seto:2001qf,Kawamura:2020pcg} and
BBO~\cite{Crowder:2005nr}, and third-generation ground-based
detectors such as the Einstein Telescope~\cite{Punturo:2010zz,ET:2019dnz}
and Cosmic Explorer~\cite{Reitze:2019iox}, will extend the reach to
transitions at the TeV scale and above.

Within the Standard Model
(SM), however, lattice simulations have shown that the electroweak
phase transition is of first order only for Higgs-boson masses below
about $70\GeV$. For the measured mass of $125\GeV$, electroweak
symmetry breaking proceeds through a smooth
crossover~\cite{Kajantie:1996mn,Csikor:1998eu,DOnofrio:2015gop}. The
onset of confinement in QCD is a crossover as
well~\cite{Aoki:2006we}.
Hence, the observation of a SGWB from a cosmological phase
transition, in whichever frequency band, would constitute unambiguous
evidence for physics beyond the SM~(BSM).

In this work we focus on a strong first-order electroweak phase
transition (EWPT), which is particularly well motivated for two
reasons. First, it supplies the departure from thermal equilibrium required by the
Sakharov conditions~\cite{Sakharov:1967dj} and is therefore a
prerequisite for electroweak
baryogenesis~\cite{Kuzmin:1985mm,Morrissey:2012db,Cline:2006ts}, one
of the most compelling explanations of the observed baryon asymmetry
of the Universe. Second, it connects GW observations with collider
physics. A first-order EWPT requires new fields coupled to the Higgs
doublet, most commonly
additional scalars~\cite{Profumo:2007wc,Espinosa:2011ax}. The LHC
probes such extended scalar sectors at zero temperature, through
precision measurements of the Higgs boson~\cite{ATLAS:2022vkf,CMS:2022dwd}
and searches for additional scalars. A SGWB, in contrast, probes the
same sector at finite temperature, since it gives direct access to the
dynamics of the transition in the early Universe. Determining the
thermal history of electroweak symmetry breaking is therefore a prime
science target of both LISA and the LHC.

Interpreting a putative signal at any of these observatories is, however, far from straightforward.
The SGWB produced by a cosmological first-order phase transition
is a broad and smooth
spectrum that can be characterized by only a handful of
geometric parameters (e.g.~peak frequency, peak amplitude and
spectral slopes) or thermodynamic
quantities (e.g.~the strength of the transition $\alpha$,
its inverse duration in units of the Hubble rate, $\beta/H_*$,
the transition temperature $T_n$ and the bubble
wall velocity $v_w$)~\cite{Caprini:2024hue}.
The spectral density
exhibits few features in which the underlying microphysics could
imprint itself. In contrast,
BSM theories capable of predicting a first-order phase transition
and a SGWB that is sufficiently strong
to be detectable typically come with a
large number of free parameters. The map from the fundamental
parameters to the observable spectrum is therefore
plagued by degeneracies.
The reconstruction of model parameters from a measured SGWB
is further complicated by the fact that the
relevant regions of parameter space are small.
This is a consequence of the fact that
a phase transition is a critical phenomenon, and
the parameter points of interest lie in the vicinity of the
first-order line of the phase diagram of the theory. Sampling such
sparse regions of parameter space properly and robustly is a
challenging task for any inference method.

In addition to these conceptual difficulties, the reconstruction is
computationally expensive, since already the prediction of the SGWB
for a single parameter point of a BSM theory requires a chain of
demanding steps. One first constructs the
finite-temperature effective potential of the model and identifies the
region of parameter space in which a first-order phase transition is realized. For
each viable parameter point, the thermodynamic parameters characterizing the
transition
have to be computed. This involves the evaluation
of the bounce solution~\cite{Coleman:1977py} at
various temperatures in order to determine the onset
of the transition.
The thermodynamic parameters are then
converted into a prediction for the SGWB spectrum by means of spectral
templates calibrated on numerical simulations, such as those adopted by
the LISA Cosmology Working Group~\cite{Caprini:2024hue}, and the resulting signal is
finally injected into simulated detector data together with a model of the
instrumental noise. Parameter reconstruction amounts to inverting this
chain, i.e.~to inferring the fundamental parameters of the BSM theory
from a simulated GW measurement~\cite{Caprini:2019pxz,Caprini:2024hue}.

We perform such a reconstruction for a strong first-order EWPT
observed with LISA and compare two conceptually different approaches. The first is the traditional
likelihood-based route, in which Bayesian parameter inference is
carried out with a Markov chain Monte Carlo (MCMC) method.
This approach is
well established and has been used to reconstruct the parameters of
first-order phase transitions from simulated LISA
data~\cite{Gowling:2022pzb,Caprini:2024hue}. It is, however,
computationally expensive. Each likelihood evaluation requires the full theory
prediction described above, and the whole analysis has to be repeated
from scratch for every new data set. The second approach that we
employ is simulation-based
inference (SBI), a likelihood-free technique in which a neural density
estimator is trained on simulated data. Its cost is concentrated
entirely in the training stage. Once trained, the network provides an
amortized posterior, i.e.~it yields the reconstructed parameters
essentially instantaneously for any given measurement.
Neither approach is specific to the EWPT or to LISA. The
underlying model enters only through the prediction of the SGWB
spectrum, and the detector only through its noise model and
response, such that the framework carries over directly to the SGWB
from other cosmological phase transitions and to other GW
observatories, such as DECIGO or the Einstein Telescope.

Since every step of an MCMC chain involves
the evaluation of the likelihood function,
we have developed a fast and numerically stable
Fortran implementation of the theoretical predictions for SGWBs.
The sampling itself has to be
preconditioned as well, because outside the narrow region in which the
transition is of first order no signal is predicted at all.
As a result, an MCMC
chain started from a uniform prior would in most parts
explore an essentially flat
likelihood and would be prone to missing the phase-transition
region altogether. We
therefore first scan the parameter space with a genetic algorithm (GA)
using \texttt{evortran}~\cite{Biekotter:2025gkp,evortran}, which locates the
relevant regions efficiently and robustly. The parameter points collected in
this way are not distributed in a statistically meaningful manner and
serve only to delimit the domain in which we subsequently perform the
actual MCMC sampling, based on the Affine Invariant Ensemble Sampler
algorithm~\cite{Goodman:2010dyf} with the parallel stretch
move~\cite{Foreman-Mackey:2012any}, as implemented in
\texttt{copa}~\cite{Biekotter:2025gkp,copa}.

The second approach, SBI, dispenses with an explicit likelihood
altogether. The term denotes the general strategy of carrying out the
inference with the help of a simulator instead, which has developed
rapidly in recent years~\cite{Cranmer:2019eaq}. It encompasses a range of concrete realizations, from approximate
Bayesian computation to the methods based on neural networks that we employ here. In
these methods, a neural network, typically a normalizing
flow~\cite{Papamakarios:2019fms}, is trained on simulated pairs of
parameter values and data to represent a conditional probability
density, in our case the posterior of the parameters given the
data. The network is thus not
used as a regression tool that maps a spectrum onto a set of parameter
values, but as a flexible parameterization of a probability
distribution, whose weights are adjusted such that the distribution
reproduces the statistical properties of the simulated pairs. The
combination of the two is what is called a neural density estimator.
The central advantage of this approach is that any
effect that can be simulated (the instrumental noise, the response of
the detector, or nuisance parameters describing systematic
uncertainties) can be accounted for without the simplifying
assumptions that writing down a tractable likelihood would enforce. The amortization moreover renders analyses
feasible that are prohibitively expensive in the likelihood-based
approach, such as the repeated inference on large numbers of simulated
data sets that is required for a statistical validation of the
reconstruction. These features have led to a wide range of
applications in the physical sciences, among them particle
physics~\cite{Brehmer:2020cvb,ATLAS:2024jry,Chatterjee:2025gej,ATLAS:2025clx,Bahl:2026zdd}.
In GW astronomy, SBI has been used for the parameter estimation of
compact binaries observed by LIGO and
Virgo~\cite{Dax:2021tsq,Dax:2022pxd,Bhardwaj:2023xph}. For LISA, it
has been applied to massive black-hole
binaries~\cite{Vilchez:2024qnw,Spadaro:2026evb} and to the population
of Galactic binaries~\cite{Srinivasan:2025etu}, as well as to the
reconstruction of
SGWBs~\cite{Alvey:2023npw,Dimitriou:2023knw,Dimitriou:2026agw}.
SBI has moreover been used to detect anisotropies of a
SGWB~\cite{Lemke:2026gxj}.
To our knowledge, however, SBI has not yet been used to reconstruct
the thermodynamic parameters of a first-order EWPT, or the parameters
of the underlying BSM model, from LISA data.

In this work we employ SBI to determine the credible regions for the
model parameters from a simulated LISA measurement, at a computational
cost that, once the network has been trained, is negligible compared
to that of the MCMC analysis. The training set is built from a large
number of simulated LISA observations. For each of
them we take a point in the
parameter space of the model,
compute the SGWB spectrum that it predicts by means of the chain of
steps outlined above, and superimpose this spectrum on a realization
of the instrumental noise. Every training example is therefore a pair
consisting of values of the model parameters
and the noisy spectral shape that
they give rise to. 
The neural density estimator is then trained on these pairs. Once the
training is completed, the reconstruction amounts to a single
evaluation of the network: a spectrum is passed as input, and the
posterior for the
model parameters is returned, from which the credible regions follow
without any further simulation. We validate the regions obtained in
this way against those of the likelihood-based MCMC analysis of the
same injected signal.

As a concrete example we consider the SM extended by a complex gauge
singlet scalar, the so-called
cxSM~\cite{Barger:2008jx,Gonderinger:2012rd}.\footnote{The prediction of the SGWB spectrum for
a given parameter point is subject to sizable theory uncertainties,
see Ref.~\cite{Biekotter:2025npc} for a discussion in the cxSM.
These uncertainties arise, for example, from the choice of
perturbative treatment used to calculate the finite-temperature effective
potential and the bubble nucleation rate, and can have a significant impact on the
inferred model parameters~\cite{Lewicki:2024xan}. Within
the SBI framework, such uncertainties can be accounted for in a
well-defined manner, by
including them among the quantities that are varied in the simulations
from which the network is trained. This is an important
additional motivation for
exploring the usefulness of SBI in the present context. In the
analysis presented here the theory uncertainties are not yet taken
into account, which we leave for future work.}
Adding a gauge singlet $S$ is the simplest way of obtaining a strong
first-order EWPT~\cite{Profumo:2007wc,Espinosa:2011ax}, and imposing a
global $\text{U(1)}$ symmetry under which $S$ transforms leaves only three free parameters: the
mass $m_S$ of the additional scalar, the portal coupling
$\lambda_{HS}$ to the Higgs doublet $H$, and the singlet self-coupling
$\lambda_S$. This choice is particularly interesting. Since the
$\text{U(1)}$ symmetry is unbroken at zero temperature, the singlet
does not mix with the Higgs boson and is
notoriously difficult to probe at colliders~\cite{Ashoorioon:2009nf},
a situation often
referred to as the \textit{nightmare scenario}~\cite{Curtin:2014jma}. It is
precisely in this case that the information LISA could provide is most
valuable. Despite its simplicity, the cxSM exhibits prominently the
two main obstacles for the parameter reconstruction discussed above,
namely the degeneracies among the parameters and the smallness of the
region in which an EWPT occurs.

The remainder of this paper is organized as follows.
In \cref{sec:theory} we present the theoretical framework and the
inference methods. The cxSM is introduced in \cref{sec:cxsm}, and
\cref{sec:simulation} summarizes the calculation of the SGWB spectrum
and the simulation of the LISA observations. We discuss the likelihood
function in \cref{sec:likelihood}, describe the MCMC analysis and the
SBI approach in \cref{sec:mcmc,sec:sbi}, respectively, and specify
the priors in \cref{sec:priors}. Our results are presented in
\cref{sec:results}. We first reconstruct the thermodynamic parameters
and the cxSM parameters for a fixed bubble wall velocity in
\cref{sec:results_thermal,sec:results_cxsm}, respectively, and then
include the bubble wall velocity as an additional free parameter in
\cref{sec:vw}. 
We further study the inference of the LISA noise parameters in
\cref{sec:noise}.
The validity and robustness of the methods are
discussed in \cref{sec:results_validation}, and their computational cost is compared in \cref{sec:cost}.
We conclude in
\cref{sec:conclusions}. Technical details on the implementation of the
MCMC and SBI analyses are collected in \cref{sec:app_mcmc,sec:app_sbi}.

  \section{Theoretical framework and methods}
\label{sec:theory}

\subsection{The cxSM}
\label{sec:cxsm}

As a benchmark, we consider an
extension of the SM by a complex
gauge-singlet scalar~(cxSM), which is able to accommodate a strong first-order EWPT.
In addition to the Higgs doublet $\Phi$, the scalar sector
contains the complex singlet $S$.
We impose a global U(1) symmetry acting on $S$,
such that the scalar Lagrangian reads
\begin{equation}
\begin{aligned}
	&\mathcal{L}_{\Phi S} ={} (D_\mu \Phi)^\dagger D^\mu \Phi
    - \mu_H^2 \Phi^\dagger \Phi - \lambda_H (\Phi^\dagger \Phi)^2 \\
	&+ |\partial_\mu S|^2 - \frac{1}{2} \mu_S^2 |S|^2
    - \frac{1}{4}\lambda_S |S|^4
	- \frac{1}{2} \lambda_{HS} |S|^2 \Phi^\dagger \Phi \, ,
\end{aligned}
\end{equation}
where
$D_\mu = \partial_\mu - \frac{i}{2} g \tau^a W_\mu^a + i\frac{Y}{2} g' B_\mu$
is the covariant derivative, with $g$ and $g'$ denoting the gauge couplings 
of the $\text{SU(2)}_L$ and $\text{U}(1)_Y$ symmetries, respectively.
Since a strong first-order EWPT in this model requires a positive portal 
coupling~\cite{Chiang:2017nmu}, we restrict ourselves to
$\lambda_{HS}>0$, and the scalar potential to be bounded from
below requires $\lambda_S > 0$.
We further focus on
a thermal history of the form
$(0,0) \to (0,v_S(T)) \to (v_H(T),0)$, in which the EW symmetry is broken 
only in the second step.

At $T = 0$, the Higgs boson does not mix with the singlet.
The scalar spectrum consists of two physical states,
where we write $h$ for the SM-like Higgs boson whose mass
we set to $125\GeV$, and $s$ for the gauge singlet state.
The singlet is
stable due to the $\text{U(1)}$ symmetry.
Beyond the SM parameters, the model is thus specified by three
independent parameters,
which we take to be
the singlet mass $m_S$, the singlet self-coupling $\lambda_S$ 
and the portal coupling $\lambda_{HS}$.

The 
singlet field in principle provides a stable
dark-matter candidate,
but direct-detection limits combined with searches for 
invisible Higgs-boson decays leave only the resonance region, 
$m_S \approx 62.5\GeV$~\cite{LZ:2022lsv,Biekotter:2022ckj}, and 
the heavy singlet regime,
$m_S \gtrsim 2\,\text{TeV}$~\cite{Arcadi:2024wwg,EscuderoAbenza:2025cfj}, viable.
As the dark-matter interpretation is not the subject of this work,
we do not take into account such constraints here, and
instead assume that $S$
decays on cosmological timescales, either via a tiny mixing
with the Higgs boson that does not have an impact on the EWPT,
or via decays into other particles that could be present in the dark sector.
For $m_S > 62.5\GeV$, no new decay channels of the Higgs boson open up, 
and the model is hardly distinguishable from the SM at colliders, 
realizing the \textit{nightmare scenario}~\cite{Curtin:2014jma} mentioned in \cref{sec:intro}.

The Lagrangian parameters $\{\mu_H^2,\allowbreak \lambda_H,\allowbreak \mu_S^2,\allowbreak
g,\allowbreak g',\allowbreak y_t,\allowbreak \lambda_S,\allowbreak \lambda_{HS}\}$ can be 
mapped onto the set of physical parameters $\{G_F,\allowbreak m_H,\allowbreak m_S,\allowbreak
m_Z,\allowbreak m_W,\allowbreak m_t, \lambda_S,\allowbreak \lambda_{HS}\}$,\allowbreak
where a lowercase $m_p$ stands for the physical mass of the particle~$p$; 
the corresponding relations are listed in Ref.~\cite{Biekotter:2025npc}.
We adopt the tree-level form of these relations at the input scale $\mu = m_Z$.
Including their one-loop corrections would only slightly
shift the mapping between
the input and the Lagrangian parameters, which does not
affect the discussion below.

To describe the dynamics of the EWPT, we employ the high-temperature effective field theory (EFT) framework 
of Ref.~\cite{Biekotter:2025npc}, which is based on the $3D$ effective theory 
obtained through dimensional reduction~\cite{Farakos:1994kx,Kajantie:1995dw}.
Starting from $m_Z$, we run the Lagrangian parameters with their
one-loop $\beta$-functions up to the hard scale $\mu =
\pi T$ and perform the matching onto the $3D$ EFT there, using the matching
relations for the cxSM given in Ref.~\cite{Schicho:2022wty}.
It was shown that this framework yields GW predictions with a 
considerably weaker residual dependence on the renormalization scale and on 
the choice of gauge than the conventional $4D$ 
treatment~\cite{Croon:2020cgk,Gould:2021oba,Schicho:2022wty,Biekotter:2025npc}.
A remaining source of uncertainty is the truncation of the EFT at a finite operator dimension.
Its impact grows towards low transition temperatures, i.e.\ precisely 
in the region of the strongest GW signals~\cite{Biekotter:2025npc}.
We leave it to future work to incorporate these theoretical
uncertainties in the parameter reconstruction in a systematic way.

The first step in the computation of the transition parameters is the
determination of the vacuum structure as a function of the temperature.
We locate the local minima of the $3D$ potential
that correspond to the singlet phase $(0,v_S(T))$ and to the EW phase
$(v_H(T),0)$ by numerical minimization, using the minimum found at one
temperature as the starting point at the next.
A first-order transition between the two phases
$(0,v_S)$ and $(v_H,0)$ requires that both minima
coexist within a common temperature interval.
Inside this interval, the critical temperature $T_c$ is the temperature at
which the two minima are degenerate.
Below $T_c$, the EW minimum is the deeper one, and the singlet phase becomes
metastable.

The onset of the transition
is set by the nucleation of bubbles of the EW phase
at sufficient rate,
\begin{equation}
  \Gamma(T) \approx A_B \, e^{-S_B^{3D}} \, ,
  \label{eq:decay_rate}
\end{equation}
where
$A_B \approx T^4(S_B^{3D}/2\pi )^{3/2}$~\cite{Linde:1980tt} and
$S_B^{3D}$ is the $3D$ bounce action~\cite{Croon:2020cgk},
\begin{equation}
  S^{3D}_B = 4\pi\int_0^\infty d\rho\rho^2 \left[
    \frac{1}{2}\left(\frac{d\phi^B_i}{d\rho}\right)^2 +
    V^{3D}_0(\phi^B_i)\right] .
\end{equation}
Here, $\phi^B_i(\rho)$ is the $O(3)$-symmetric bounce solution
obtained from the tree-level $3D$ potential $V_0^{3D}$.

The exponential form of $\Gamma(T)$ in \cref{eq:decay_rate} results from a semiclassical evaluation of
the path integral, i.e.\ a saddle-point approximation around the bounce
solution, with the fluctuations around the bounce contributing to the
prefactor~$A_B$~\cite{Coleman:1977py,Callan:1977pt,Linde:1980tt}.
We identify the onset of the transition with the nucleation 
temperature $T_n$,
at which on average one 
bubble per Hubble volume has formed,
\begin{equation}
\int_{T_n}^{T_c} \frac{dT}{T} \frac{\Gamma(T)}{H(T)^4} \approx 1 \, ,
\end{equation}
with $T_c$ being the temperature at which both minima are degenerate.
In practice, we approximate this condition by $\Gamma(T_n) \approx
H(T_n)^4$.
Inserting \cref{eq:decay_rate} with $A_B \approx T^4$ and using
$H \propto T^2/M_{\rm Pl}$, with the Planck mass $M_{\rm Pl}$, this
translates into the condition $S_B^{3D}(T_n) \approx 140$ for transition
temperatures around the EW scale~\cite{Quiros:1999jp,Athron:2023xlk}.
The approximation assumes that $S_B^{3D}$ decreases
relatively rapidly below $T_c$,
and it becomes unreliable for strongly supercooled
transitions.
We determine $T_n$ by bisection in the temperature interval between the onset
of the coexistence of the two minima and $T_c$, computing the bounce action
at each step.
Besides $T_n$, the GW spectrum depends on the inverse duration
of the transition in units of the Hubble rate
$\beta/H_*$, where $H_*$ denotes the Hubble rate at $T_n$,
and the strength $\alpha$ 
of the transition~\cite{Caprini:2019egz,LISACosmologyWorkingGroup:2022jok},
\begin{equation}
\begin{aligned}
    \frac{\beta}{H_*} &= T \frac{d S^{3D}_B}{dT} \bigg|_{T=T_n} , \\
    \alpha &= \frac{1}{\rho_{\rm rad}}
    \left( \frac{3}{4} T \Delta V^{3D} - \frac{1}{4} T^2
    \frac{d\Delta V^{3D}}{dT} \right) \bigg|_{T=T_n} ,
\end{aligned}
\end{equation}
in which $\Delta V^{3D}$ denotes the difference of the $3D$ potential between 
the high- and low-temperature phases, and $\rho_{\rm rad} = g_* \pi^2 T^4/30$ is 
the energy density of the radiation bath.

Finally, the GW spectrum depends on the bubble wall velocity $v_w$, the
terminal velocity of the expanding bubble walls relative to the plasma far in
front of them.
Its determination is considerably more involved than that of $T_n$, $\alpha$
and $\beta/H_*$, which follow from the equilibrium thermodynamics of the plasma as described by the $3D$ EFT.
The wall velocity is instead set by the balance between the pressure difference
between the two phases, which drives the expansion, and the friction exerted by
the plasma on the wall.
Computing it requires solving the Boltzmann equations for their distribution
functions together with the hydrodynamic equations of the plasma around the
wall, and the result is still subject to sizable theoretical
uncertainties
(see, e.g.~Refs.~\cite{DeCurtis:2024hvh,Branchina:2025jou,
vandeVis:2025plm,Tang:2026dhj,Dorsch:2026lmo}
for recent discussion).
We therefore do not compute $v_w$ in our simulator.

For the reconstruction of the thermodynamic and the cxSM parameters in
\cref{sec:results_thermal,sec:results_cxsm}, we assume that a reliable
prediction of $v_w$ will be available by the time an SGWB is measured, and we
set $v_w$ to the hypothetical value $v_w = 0.8$.
Using the same value for all parameter points in the scan over the cxSM
parameters implies the assumption that $v_w$ varies only mildly within the
region of parameter space that is compatible with the measured signal.
We choose a supersonic wall velocity well above the speed of sound
$c_s = 1/\sqrt{3}$, where the GW template does not reliably describe
the spectrum (see \cref{sec:vw}), and below the ultrarelativistic regime
$v_w \to 1$.
In \cref{sec:vw}, we relax the assumption of a known $v_w$ and infer it as an
additional free parameter.
This tests whether $v_w$ could be measured from the SGWB spectrum itself
instead of being predicted from either the thermodynamic
parameters or the model parameters.
Inferring $v_w$ also quantifies how much the precision
of the reconstructed thermodynamic
and cxSM parameters degrades if a
sufficiently precise prediction of $v_w$ in terms of either
set of parameters is not available.

The chain of computations described above, from the vacuum structure over
$T_c$ and $T_n$ to $\alpha$ and $\beta/H_*$, has to be evaluated for every point in the space of the model
parameters $\{m_S, \lambda_S, \lambda_{HS}\}$, and it constitutes the
bottleneck in runtime of both inference approaches.
In the MCMC analysis, it enters every evaluation of the likelihood.
In the SBI, it is required to generate the training set, which maps
$\{m_S, \lambda_S, \lambda_{HS}\}$ onto the predicted SGWB spectra (see
\cref{sec:sbi,sec:cost}).
The most expensive step is the computation of the bounce action, which has to
be repeated at many temperatures to locate $T_n$ and to determine $\beta/H_*$.
We have therefore implemented the full chain in an in-house Fortran code
optimized for performance.
It includes a Fortran port of \texttt{CosmoTransitions}~\cite{Wainwright:2011kj,
fortransition},
which computes the bounce solution with the path-deformation method and which
we developed with the help of \texttt{Claude Fable v.5.1}.

\subsection{Simulated LISA data}
\label{sec:simulation}

A strong EWPT
would give rise to a SGWB in the LISA frequency band.
The main sources of GWs are the collision
of expanding bubbles and the associated plasma motion in the form of 
sound waves~\cite{Hindmarsh:2013xza,Hindmarsh:2015qta,Hindmarsh:2016lnk,Hindmarsh:2017gnf}, 
together with the turbulence generated after the 
collisions~\cite{Caprini:2006jb,Gogoberidze:2007an,Caprini:2009yp,RoperPol:2019wvy,Auclair:2022jod} 
(see Ref.~\cite{Caprini:2019egz} for a review). 
We model the SGWB using the doubly broken power-law (DBPL) templates of
Ref.~\cite{Caprini:2024hue}.
In the regime relevant for the transitions considered here, 
the dominant contribution is expected to arise from the long-lasting 
acoustic motion of the plasma and the subsequent magnetohydrodynamic (MHD) turbulence. 
We
fix the fraction of the total kinetic energy transferred to MHD turbulence to
$\epsilon_{\rm turb}=0.5$ throughout our analysis.
The value of $\epsilon_{\rm turb}$ is still under
investigation~\cite{RoperPol:2019wvy,Caprini:2019egz,Auclair:2022jod,Dahl:2021wyk,Dahl:2024eup},
and Ref.~\cite{Caprini:2024hue} used $\epsilon_{\rm turb}=1$ in its
parameter inference. Even for this value, the turbulence contribution
remains subdominant compared with the sound-wave contribution in the
frequency range around the peak of the
spectrum~\cite{Hindmarsh:2015qta,Caprini:2019egz}, which dominates the
sensitivity of LISA to the signal. The precise value of
$\epsilon_{\rm turb}$ is therefore of minor importance for the
reconstruction of the parameters in this work.

The DBPL template
for the sound-wave contribution
provides a compact description of the resulting spectrum
in terms of a small number of parameters~\cite{Caprini:2024hue}.
Its spectral shape is characterized by two break frequencies,
$f_1 < f_2$, and by the amplitude $\Omega_2$ at $f_2$,
close to which the spectrum peaks.
The remaining spectral slopes are fixed by the sound-shell dynamics
encoded in the template~\cite{Hindmarsh:2016lnk}.
The two characteristic frequencies have a direct
interpretation in the sound-shell picture.
As bubbles nucleate and expand,
they accelerate the surrounding plasma and generate fluid shells whose
characteristic scales are set by the mean bubble separation, $R_*$,
and the sound-shell thickness, $\Delta_w R_*$, with $\Delta_w = |v_w - c_s|/\max(v_w, c_s)$. 
These length scales translate into the two characteristic frequencies of the GW spectrum, where
the mean bubble separation determines the lower break $f_1$,
whereas the shell thickness determines $f_2$ and thus the position of the peak.
These scales are, in turn, controlled by the macroscopic
properties of the phase transition. The mean bubble separation depends on the
inverse duration of the transition and the bubble wall velocity,
$H_* R_* = (8\pi)^{1/3}\max(v_w, c_s)/(\beta / H_*)$,
while $\Delta_w$, and thus the shell thickness relative to $R_*$, depends on the bubble wall velocity alone. 
The positions and relative separation of the spectral
features consequently encode information about the underlying transition dynamics.
The MHD turbulence contribution is described by a template of similar form.
Its peak is set by $R_*$,
and its lower break by the kinetic energy of the turbulence, i.e.~by the fraction $\epsilon_{\rm turb}$ of the kinetic energy of the fluid~\cite{Caprini:2024hue}.

\begin{figure*}[t]
  \centering
  \includegraphics[width=0.32\textwidth]{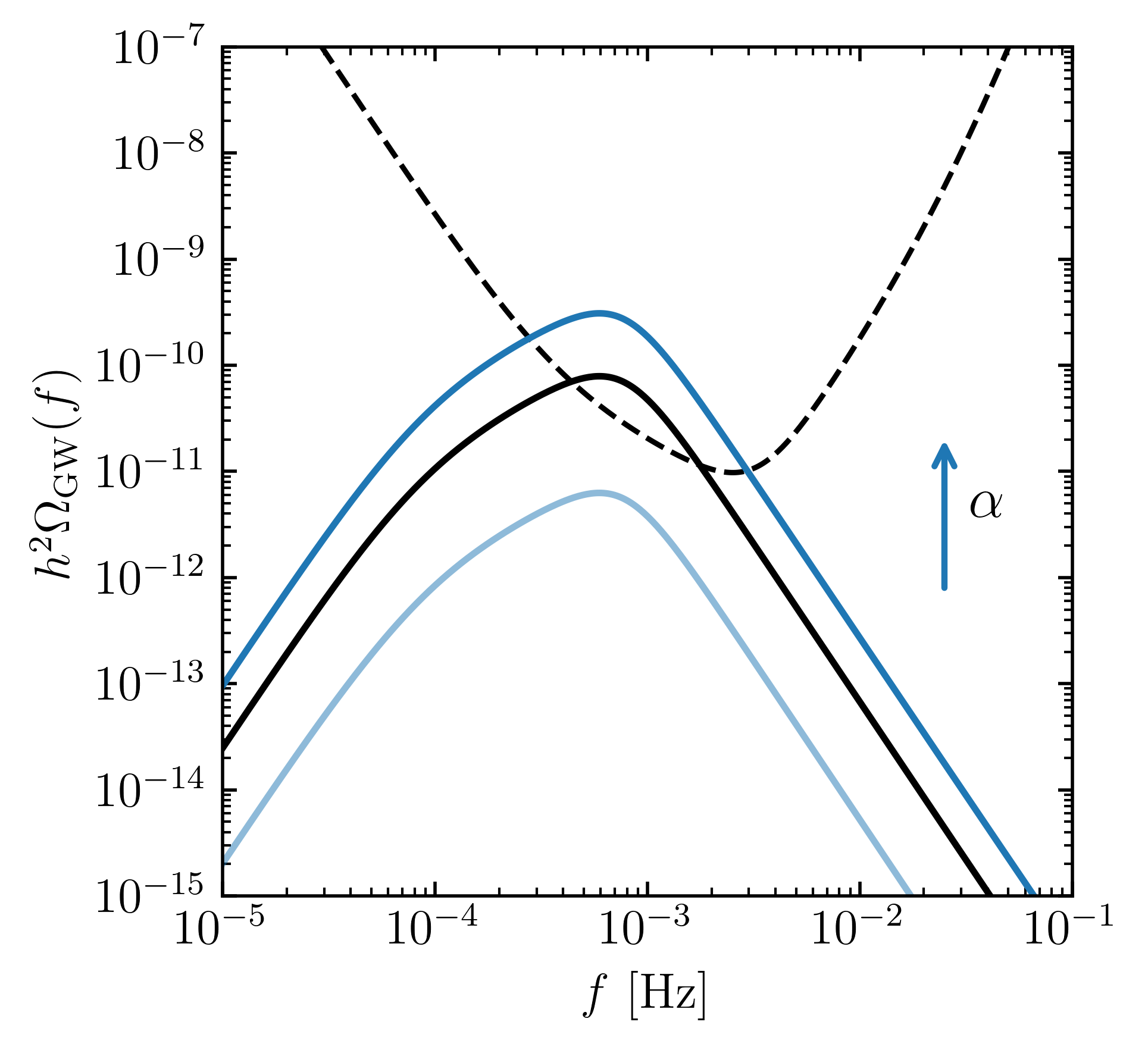}
  \includegraphics[width=0.32\textwidth]{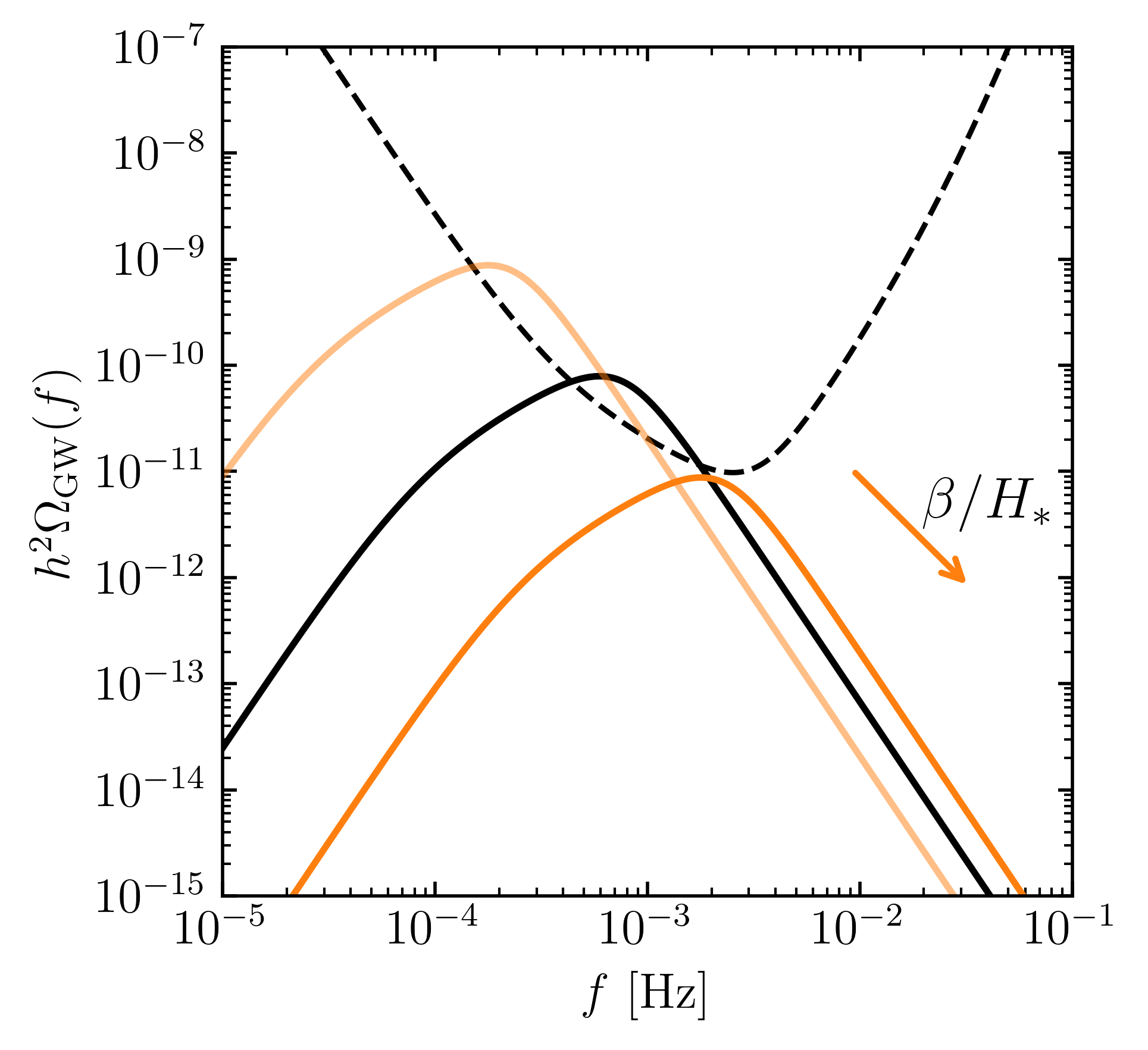}
  \includegraphics[width=0.32\textwidth]{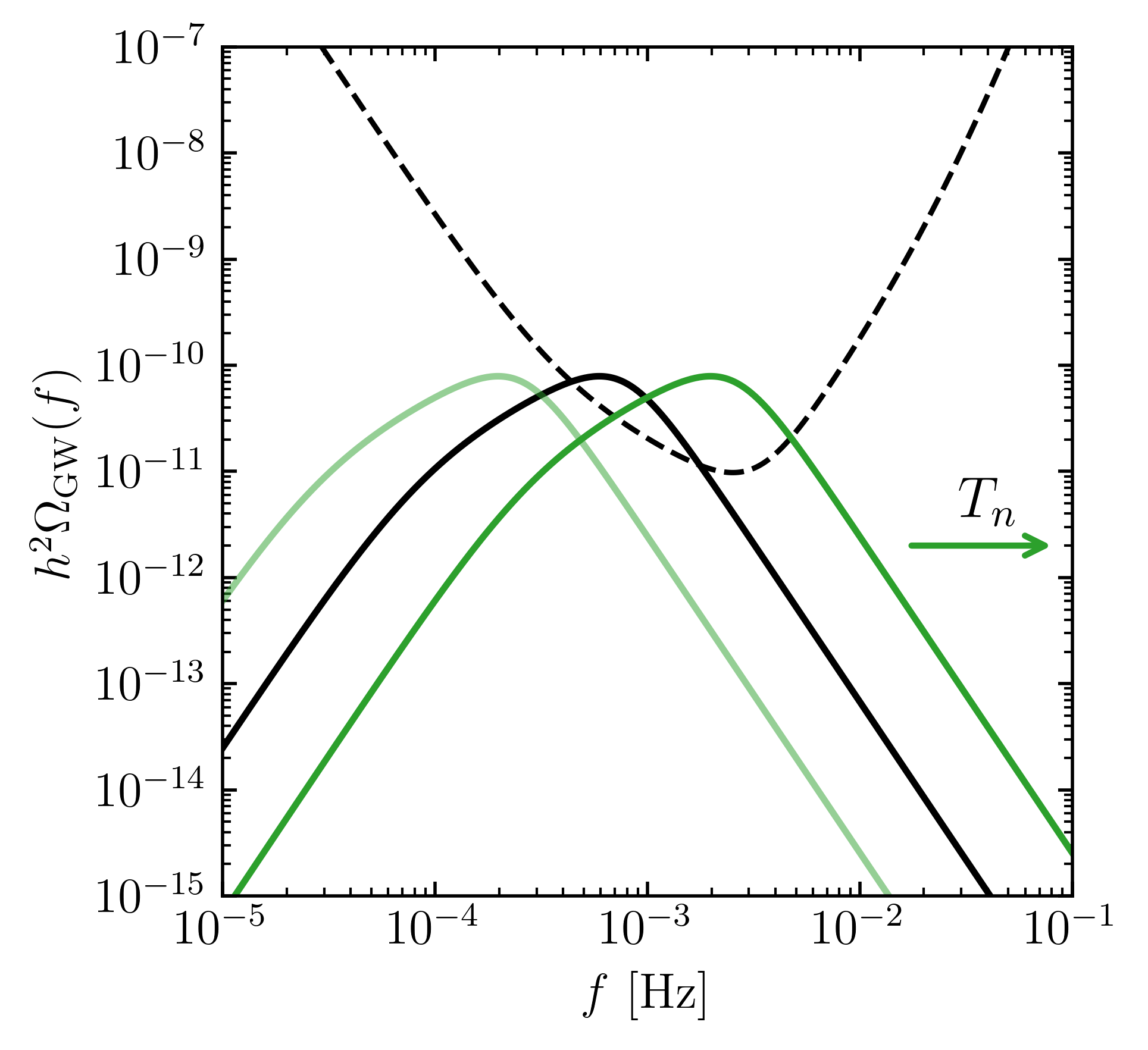}

  \includegraphics[width=0.32\textwidth]{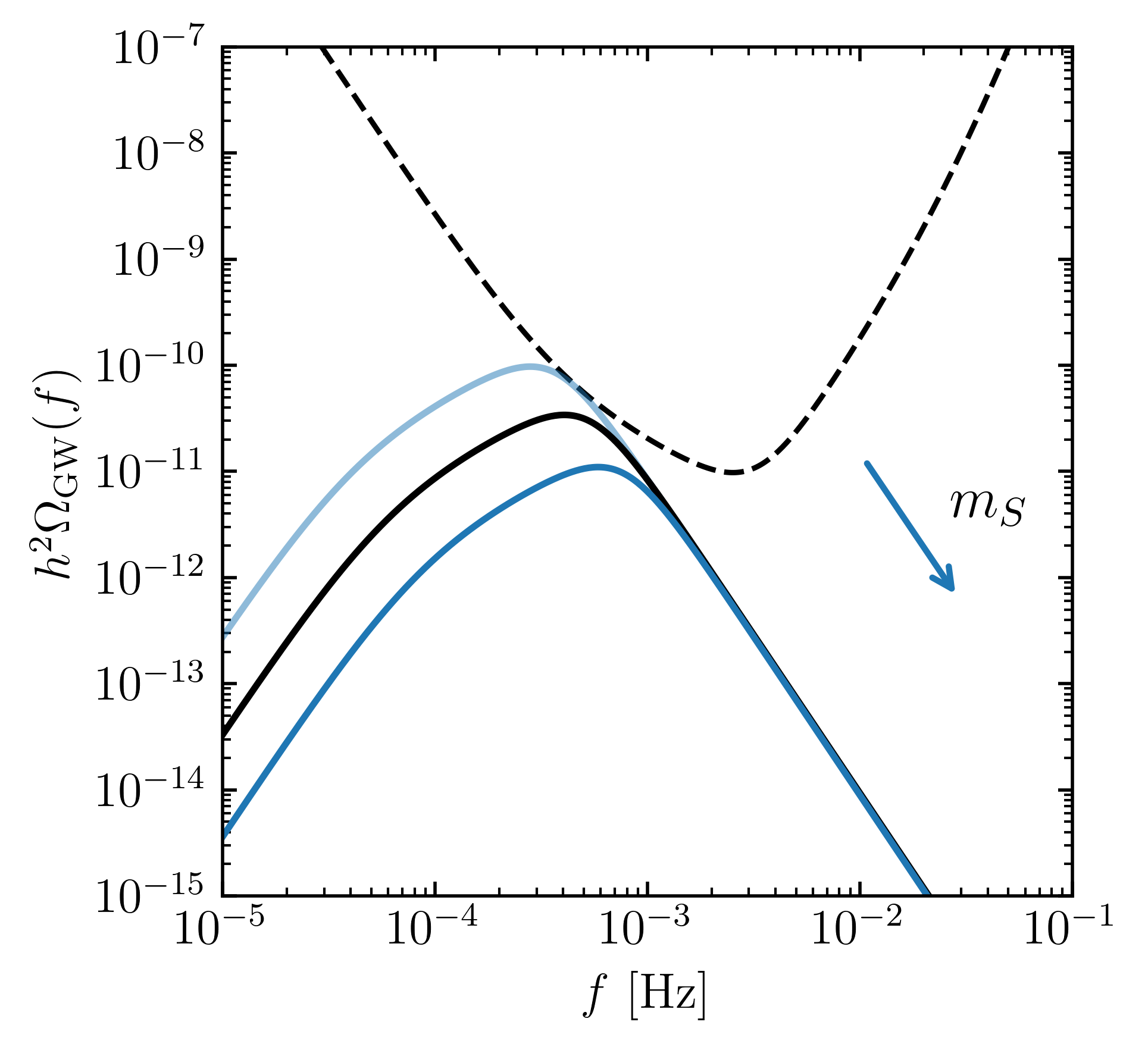}
  \includegraphics[width=0.32\textwidth]{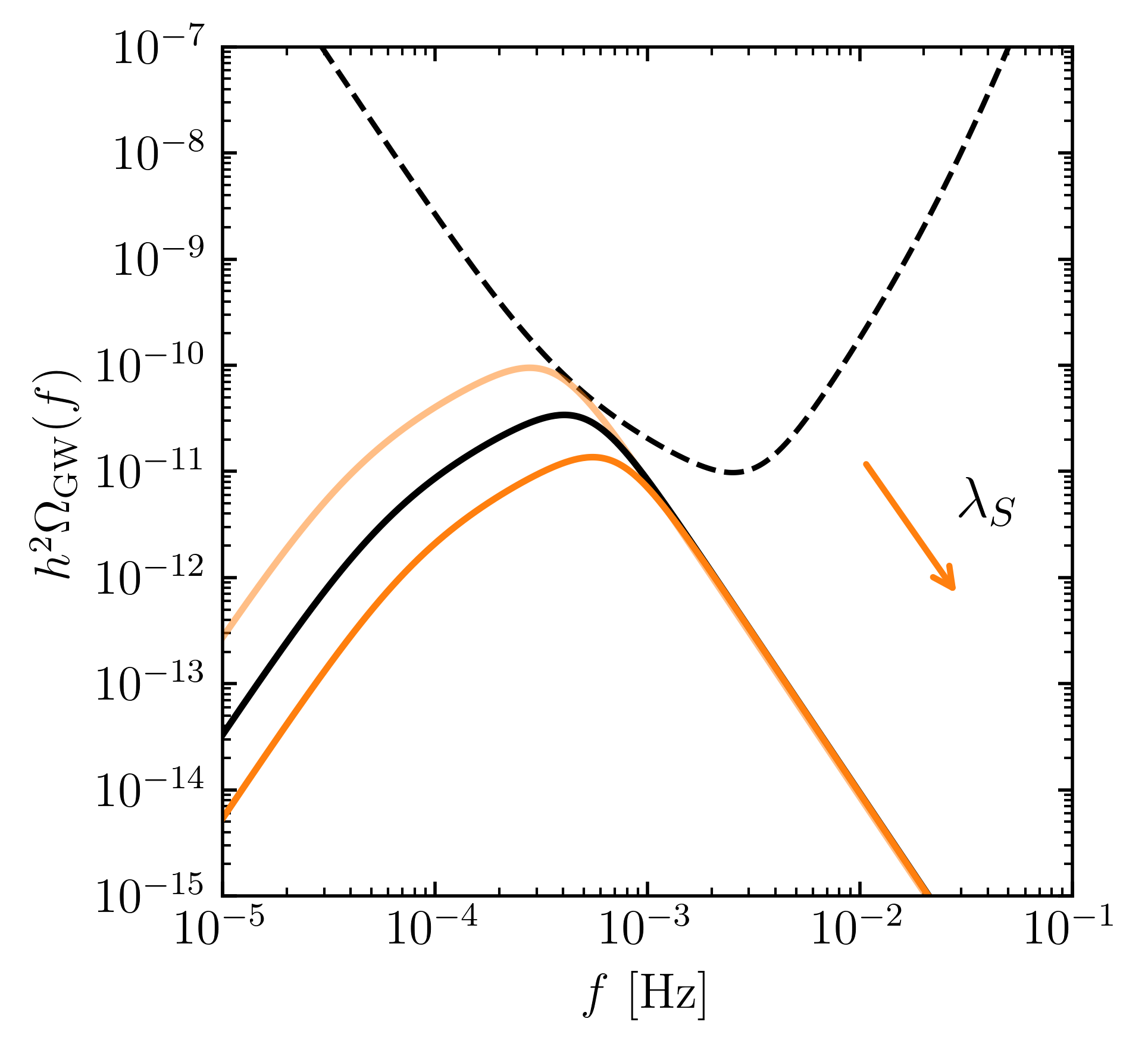}
  \includegraphics[width=0.32\textwidth]{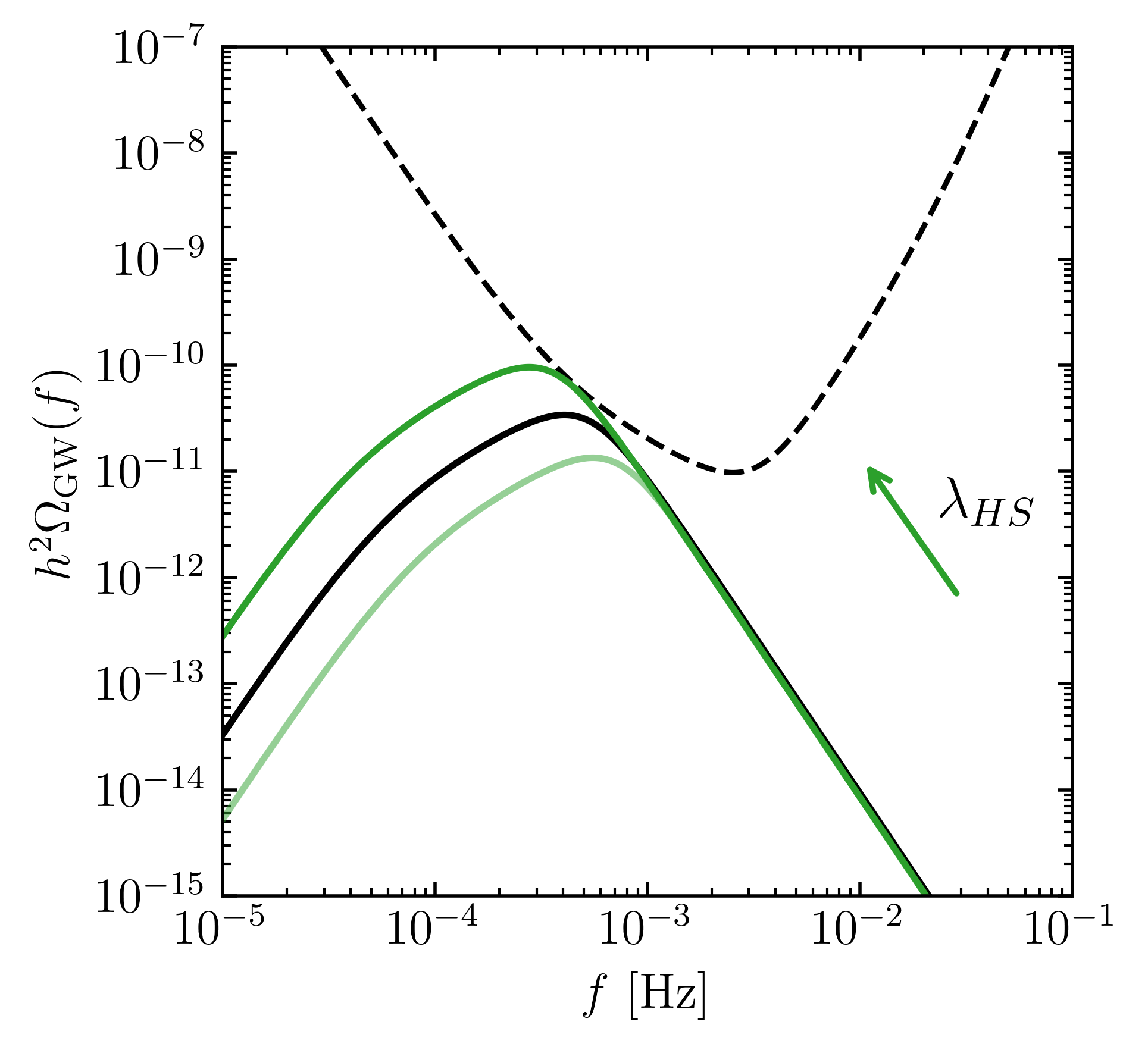}

  \caption{Dependence of the GW spectrum
    on the thermodynamic parameters (upper row) and
    the cxSM parameters (lower row).
    In the top row, the black curves depict the signal
    $\alpha = 1$, $\beta / H_* = 100$ and $T_n = 60 \GeV$,
    while in the bottom row they correspond to
    $m_S = 68 \GeV $, $\lambda_S = 0.46$ and $\lambda_{HS} = 0.66$.
    The colored lines depict the signals after variation
    of the parameter shown in each plot, with the other parameters
    kept fixed. The arrows indicate the directions of the
    changes of the signals for increasing values of the parameters. The black dashed line shows the LISA noise spectrum $h^2\Omega_{\rm n}$.}
  \label{fig:analytic_GW_dep}
\end{figure*}

The DBPL templates defined
in Ref.~\cite{Caprini:2024hue} to describe 
the sound-wave and MHD contributions depend on the thermodynamic parameters 
$\alpha$, $\beta/H_*$, $T_{n}$, and $v_{w}$ in a correlated manner. 
In \cref{fig:analytic_GW_dep}, we illustrate how variations of the 
macroscopic parameters $\alpha$, $\beta/H_*$, and $T_n$ (top row)
as well as of the microscopic 
cxSM parameters $m_S$, $\lambda_S$, and $\lambda_{HS}$
(bottom row), modify the resulting GW spectrum. 
In each plot, we vary one of the parameters,
while the others are kept fixed.
The bubble wall velocity is set to
$v_w = 0.8$. 
The black dashed line shows the LISA noise spectrum,
see the discussion below.
The figure shows that a change in one parameter
can be largely compensated by a change in another,
such that different sets of parameters predict very similar
spectra, which become indistinguishable once the
instrumental noise of LISA is taken into account.
As a consequence, the posterior distributions of the thermodynamic parameters, and of the cxSM parameters, extend along these degenerate directions in parameter space, as we find in \cref{sec:results}.

Our goal is to reconstruct the parameters that characterize
the EWPT from a measured SGWB at LISA.
In order to make our analysis as realistic as possible, we include instrumental noise 
in the injected signals.
Specifically, we account for the two dominant contributions 
to the LISA instrumental noise: residual acceleration noise associated with deviations 
of the test masses from ideal free fall, and noise from the optical metrology system 
used to measure their relative positions.
Following Ref.~\cite{Caprini:2024hue},
these contributions
are expressed in terms of the LISA noise
spectrum $h^2\Omega_{\rm n}(f; A, P)$, which is
parameterized by the 
amplitudes $A$ and $P$ for the noise from
the test masses and the optical measurement system (OMS), respectively.
We adopt the fiducial values $A=3~\mathrm{fm} \, \mathrm{s^{-2}}/\sqrt{\mathrm{Hz}}$
and $P=15~\mathrm{pm}/\sqrt{\mathrm{Hz}}$,
and the standard LISA noise model.
These parameters are assumed to be independently
constrained very precisely by the low- and high-frequency
regions of the LISA spectrum, where the contribution from an
EWPT
signal is negligible, such that we do not take into
account an uncertainty on $A$ and $P$ in our analysis.
The validity of this assumption is discussed in
\cref{sec:noise}.

Following Ref.~\cite{Caprini:2019pxz}, we simulate the data
at the frequencies $f_j$ between $f_{\rm min} = 3 \times 10^{-5}\,\mathrm{Hz}$
and $f_{\rm max} = 0.5\,\mathrm{Hz}$, with a resolution of
$\Delta f = 10^{-6}\,\mathrm{Hz}$ set by the duration
$1/\Delta f = 10^{6}\,\mathrm{s}$ of a single data chunk.
This amounts to about $5 \times 10^{5}$ frequencies~$f_j$.
A four-year LISA mission with an observing efficiency of about $75\%$
yields $N_c = 94$ such chunks, i.e.~$N_c$ statistically independent
measurements of the spectrum at each frequency~$f_j$~\cite{Caprini:2019pxz}.\footnote{The
analysis of Ref.~\cite{Caprini:2024hue} instead uses 126 data chunks
of about 11.4 days each, corresponding to four effective years of data.
This choice only rescales the statistical uncertainty of the simulated
spectrum. It does not affect our conclusions, since the SBI and the
MCMC analysis are applied to the same simulated data.}
We simulate each of these measurements by drawing the stochastic
signal and noise contributions from Gaussian distributions.
From these chunks, we compute the mean spectrum $\overline{D}(f)$ and its 
variance $\sigma^2(f)$. 

If the spectrum value at each frequency~$f_j$ were used as an input,
the SBI would require a very large neural network. We therefore reduce the number of frequencies
by binning the data (see \cref{sec:sbi} for details on the network).
In the MCMC analysis, the number of frequencies enters the
likelihood as a sum over all frequencies. Here the binning speeds up
the evaluation of the likelihood, but it is not required, and the
MCMC analysis can also be performed on the unbinned
data~\cite{Biekotter:2025gkp}.
We have checked that the posterior distributions of the parameters
obtained with the MCMC analysis are effectively unchanged when using
the binned instead of the unbinned data, such that the binning preserves the information
on these parameters contained in the data. We therefore use the binned data in both
the MCMC analysis and the~SBI.

For the binning, we follow the coarse-graining prescription described
in Ref.~\cite{Caprini:2019pxz}.
We keep the frequencies below the cutoff $f_{\rm cut}=10^{-4}\,\mathrm{Hz}$ unchanged, while the 
frequency range above $f_{\rm cut}$ is divided into 500 bins whose edges are
equally spaced on a logarithmic frequency scale.
Within each high-frequency bin, the spectrum is combined using inverse-variance weighting, 
with weights $w_j=1/\sigma^2_j$, where $\sigma^2_j$ is the variance estimated from the noisy chunks 
at frequency $f_j$.
The representative frequency and spectrum value of each bin are 
therefore given by

\begin{equation}
f^{\rm bin}_k=\frac{\sum_j f_j/\sigma^2_j}{\sum_j 1/\sigma^2_j}, \quad
D^{\rm bin}_k=\frac{\sum_j \overline{D}_j/\sigma^2_j}{\sum_j 1/\sigma^2_j} .
\label{eq:binning_def}
\end{equation}
This coarse-graining reduces the original $\mathcal{O}(10^5)$ frequency points to a
set of 571 bins, which constitute the input features for
both the MCMC and the SBI analysis.\footnote{
While we follow the coarse-graining prescription of Ref.~\cite{Caprini:2019pxz}, 
that work uses $f_{\rm cut}=10^{-3}\,\mathrm{Hz}$ and 1000 logarithmically spaced bins. 
This would result in 1971 input features and substantially increase the computational 
cost of the inference. We adopt a coarser graining, at the possible cost of some loss of spectral  information. 
}
The
effect of the binning
on the parameter inference
is discussed in detail in \cref{sec:likelihood}.

\subsection{Likelihood function}
\label{sec:likelihood}

For a given set of parameters $\theta$, the simulator
(see \cref{sec:simulation}) outputs the distribution of spectra,
described by the probability density $p(D \mid \theta)$ of obtaining
a data set (spectrum) $D$.
Depending on the inference analysis, in this 
work $\theta$ denotes either the thermodynamic parameters $\{T_n, \alpha, \beta/H_*\}$ of 
the transition, the parameters $\{m_S, \lambda_S, \lambda_{HS}\}$ of the cxSM, or each additionally extended by the bubble wall velocity $v_w$.
The likelihood $\mathcal{L}(\theta)$ is the same density with the $D$ fixed to the measured one,
which is then a function of only the model parameters $\theta$.

An explicit expression for the likelihood therefore requires not only the
predicted spectrum, but also the probability distribution of the measured
spectrum around this prediction.
The MCMC analysis (\cref{sec:mcmc}), as well as the importance sampling
and the reweighting with the likelihood (\cref{sec:sbi}), evaluate
$\mathcal{L}(\theta)$ explicitly for each sampled parameter point.
For these methods, a precise definition of $\mathcal{L}(\theta)$ is
therefore essential, since any inaccuracy in $\mathcal{L}(\theta)$
directly biases the inferred posterior.

The SBI approach, on the other hand, performs the inference without an explicit expression for the likelihood.
Here, the statistical properties of the data enter through the simulator that generates them, which is another key advantage of this approach.
The
SBI may even be used to infer the likelihood function itself~\cite{Papamakarios:2018zoy,Alsing:2019xrx}.

As discussed in \cref{sec:simulation}, we simulate a SGWB as
a stationary Gaussian random field.
Then, their Fourier amplitudes at a frequency $f_j$ are complex Gaussian variables whose variance is set by the total spectral density
\begin{equation}
    \mathcal{M}_j(\theta) = h^2\Omega_{\rm GW}(f_j; \theta) + h^2\Omega_{\rm n}(f_j; A, P) \, ,
    \label{eq:total_level}
\end{equation}
with $h^2\Omega_{\rm n}$ the instrumental noise introduced above,
and $\theta$ are the parameters
that the spectrum $h^2\Omega_{\rm GW}$ is predicted from.
The measurement is then the variance of this field, and
the power of a single chunk 
is exponentially distributed with expectation value $\mathcal{M}_j$. 
The mean $\overline{D}_j$ of the $N_c$ chunks follows a Gamma distribution of shape $N_c$,
\begin{equation}
    \overline{D}_j \sim \Gamma\left(N_c, \frac{\mathcal{M}_j}{N_c}\right) ,
    \label{eq:gamma_data}
\end{equation}
with two degrees of freedom per chunk, corresponding to the real and imaginary parts of the amplitude.
In particular, the expectation value of $\overline{D}_j$ over noise realizations
is $\langle \overline{D}_j \rangle = \mathcal{M}_j$, such
that the mean spectrum is an unbiased estimator of the predicted level, with a standard deviation $\mathcal{M}_j/\sqrt{N_c}$.
The fluctuations are multiplicative,
\begin{equation}
    \overline{D}_j = \mathcal{M}_j X_j \, , \qquad X_j \sim \Gamma\left(N_c, \frac{1}{N_c}\right) ,
    \label{eq:multiplicative}
\end{equation}
where the random factor $X_j$, with $\langle X_j \rangle = 1$, does not depend on $\mathcal{M}_j$.
The relative uncertainty is the same at all frequencies, irrespective of whether a frequency is dominated by the signal or by the noise.
This implies that the size of the uncertainty is itself a prediction of the model, in contrast to the case of a deterministic signal observed in instrumental noise of known spectral density, for which the error bars $\sigma_j$ are external to the model and the likelihood reduces to the usual Gaussian form,
$-2\ln\mathcal{L} = \chi^2 = \sum_j (D_j - \mathcal{M}_j)^2/\sigma_j^2$ up to a constant.

For an observation that is long compared to the inverse frequency resolution, the Fourier amplitudes at different frequencies of a stationary Gaussian process are statistically independent to a good approximation~\cite{Whittle:1953}.
In this case, the log-likelihood is
the sum of the individual contributions at each frequency
$f_j$ shown in \cref{eq:gamma_data},
\begin{equation}
    \ln \mathcal{L}(\theta) = - N_c \sum_j \left[ \ln \mathcal{M}_j(\theta) + \frac{\overline{D}_j}{\mathcal{M}_j(\theta)} \right] + \text{const} \, .
    \label{eq:whittle}
\end{equation}
This
is the \textit{Whittle likelihood} of a stationary Gaussian process,
with $N_c$ averaged chunks~\cite{Whittle:1953}.
The assumption of independent frequencies is violated by windowing, data gaps or non-stationarities of the instrument~\cite{Franciolini:2025leq}, none of which are included in the simulated data sets used here.

For $N_c \gg 1$ the distribution in \cref{eq:gamma_data} approaches a Gaussian and it is common to replace \cref{eq:whittle} by a $\chi^2$ constructed either with the variance $\sigma^2_j$ estimated from the chunks~\cite{Caprini:2019pxz,Flauger:2020qyi} or with the variance predicted by the model.
Both replacements introduce a bias of order $1/N_c$ in the inferred spectral density~\cite{Franciolini:2025leq}.
These biases are enhanced for the signals whose power is compatible with the instrumental noise power, since a relative shift of $\mathcal{M}_j$ translates into a relative shift of $h^2\Omega_{\rm GW}$ enhanced by $\langle \mathcal{M}_j / h^2\Omega_{\rm GW}(f_j)\rangle$.
As the biases do not decrease with the number of frequencies $N_f$, while the statistical uncertainty does, they exceed the statistical precision of the measurement for $N_f \gg N_c$~\cite{Franciolini:2025leq}.
We therefore use the exact Whittle likelihood of \cref{eq:whittle} throughout this work.

The data entering our inference analysis are, however,
not the individual frequencies $f_j$ and the spectral
density $\overline{D}_j$, but the
binned spectrum of \cref{eq:binning_def}.
To compare it with the model, we replace the bin frequency of \cref{eq:binning_def}, which differs between noise realizations, by a fixed frequency computed with the noise-only weights, which are the same for all spectra, as used for the coarse-graining in Refs.~\cite{Flauger:2020qyi,Caprini:2024hue}
\begin{equation}
f^{\rm bin}_k = \frac{\sum_j f_j\,\Omega_{\rm n}^{-2}(f_j)}{\sum_j \Omega_{\rm n}^{-2}(f_j)} \, ,
\label{eq:fbin}
\end{equation}
and define the binned model spectrum $\mathcal{M}_k \equiv \mathcal{M}(f^{\rm bin}_k)$, with $\mathcal{M}$ of \cref{eq:total_level}.
This choice does not affect the SBI, as the networks only see the spectrum values $D^{\rm bin}_k$, and the bin frequency enters only the evaluation of the likelihood.

However, $\mathcal{M}_k$ is not the expectation value of $D^{\rm bin}_k$, for two reasons.
First,
the weights $w_j = 1/\sigma^2_j$ modify the statistical description in a
non-trivial way, which has to be taken into account to
avoid a bias in the parameter inference.
Since the chunks are exponentially distributed, the variance estimated from them scales as $\sigma^2_j \propto \overline{D}_j^{2}$ (see \cref{eq:multiplicative}),
such that each
weight $w_j$ is
computed from the same noisy value $\overline{D}_j$ that
it multiplies in \cref{eq:binning_def}.
Downward fluctuations tend to receive larger weights, biasing the binned spectrum downward, and the variance of $D^{\rm bin}_k$ is larger than that of a plain average of the same number of chunks.
Both effects of the data reduction are of the same origin as the $\mathcal{O}(1/N_c)$ bias discussed above regarding the choice of the likelihood function.
Second, $\mathcal{M}(f)$ is curved across a bin, so that its value at $f^{\rm bin}_k$ differs from its average over the bin.
To our knowledge, neither effect has been taken into account in previous analyses of coarse-grained LISA spectra, which evaluate the model at the bin frequency~\cite{Caprini:2019pxz,Flauger:2020qyi,Caprini:2024hue}.

We account for these biases
by retaining the Gamma form of \cref{eq:gamma_data} for each bin, but with the modified expectation value $b_k \mathcal{M}_k$ and an effective number of degrees of freedom $N_k$,
\begin{align}
    D^{\rm bin}_k &\sim \Gamma\left(N_k, \frac{b_k \mathcal{M}_k}{N_k}\right) ,
    \label{eq:binned_gamma}
    \\
    \ln \mathcal{L}(\theta) &= - \sum_k N_k \left[ \ln \big(b_k \mathcal{M}_k(\theta)\big) + \frac{D^{\rm bin}_k}{b_k \mathcal{M}_k(\theta)} \right] + \text{const} .
    \label{eq:binned_likelihood}
\end{align}

We determine $b_k$ and $N_k$ from $10^5$ noise-only realizations of the full simulator of \cref{sec:simulation}, with the noise fixed to the nominal LISA sensitivity, binned in the same way as the data,
\begin{equation}
b_k = \frac{\langle D^{\rm bin}_k \rangle}{h^2 \Omega_{\rm n}(f^{\rm bin}_k)} \, , \qquad
N_k = \frac{\langle D^{\rm bin}_k \rangle^2}{\langle (D^{\rm bin}_k)^2 \rangle - \langle D^{\rm bin}_k \rangle^2} \, ,
\label{eq:bk_Nk}
\end{equation}
where $\langle \cdot \rangle$ denotes the average over the realizations
(see also Ref.~\cite{Satterthwaite1946AnAD}).
In this way $b_k$ accounts for both the bias from the weights and the curvature of the noise spectrum within each bin.
One finds $b_k = 1$ and $N_k = N_c$ for the frequencies below $f_{\rm cut}$, and $b_k \to 1 - 2/N_c \simeq 0.979$ for the widest bins at high frequencies.
The values of $N_k$ range between $N_c$ and $\simeq 7\times 10^{5}$, about $12\%$ below the plain average of the $n_k$ frequencies in the bin, $N_c n_k$.

We have checked that calibrating with the nominal LISA noise instead of the actual spectrum of a given data set is a good approximation.
For the GW signals considered here, the mean of the binned spectrum agrees with $b_k \mathcal{M}_k$ to better than $10^{-5}$.
Changing the noise amplitudes by $\pm 10\%$ changes $b_k$ by less than $5\times 10^{-5}$, which shifts the inferred noise amplitudes by less than $0.01\sigma$.

Calibrating with a constant spectral density captures the effect of the weights but not the curvature.
The resulting $b_k \mathcal{M}(f^{\rm bin}_k)$ differs from the exact expectation value by only about $2\times10^{-4}$ above $10\,$mHz, which is negligible for the
inference of the thermodynamic or cxSM parameters, which is the focus
of this work.
However, we stress that a calibration of the likelihood without the curvature effect
shifts the inferred OMS noise amplitude $P$ by $1.4\sigma$, since $P$ is determined to
relative precision at the level of $10^{-4}$,
as is discussed in detail in
\cref{sec:noise}.

The effective number of degrees of freedom $N_k$ also quantifies the information lost by the binning. Using the unbinned spectra would decrease the statistical uncertainty by about $1/\sqrt{0.88}-1 \simeq 6.6\%$, while increasing the data volume by almost three orders of magnitude.
We have confirmed this with fits of the noise amplitudes to $10^4$ simulated realizations, once to the unbinned spectra with \cref{eq:whittle} and once to the binned spectra with \cref{eq:binned_likelihood}. 
The binned fits are unbiased, and their spread is larger by about $6\%$, in agreement with the estimate from $N_k$.

The steps outlined above establish a chain of
mappings that connect the parameters of the cxSM,
via the macroscopic thermodynamic quantities 
characterizing the transition, to the SGWB spectrum as it would be observed by LISA.
Assuming the bubble wall velocity $v_w$ to be known and constant,
this mapping is given by
\begin{equation}
  \{m_S, \lambda_S, \lambda_{HS}\}
  \longleftrightarrow
  \{T_n, \alpha, \beta/H_*\}
  \longleftrightarrow
  \{D^{\rm bin}_k \}
  \, .
  \label{eq:full_direction}
\end{equation}
Assuming that $v_w$ is an additional unknown parameter that has
to be inferred from the data, the mappings are
\begin{align}
  \{m_S, \lambda_S, \lambda_{HS}, v_w\}
  \longleftrightarrow
  \{D^{\rm bin}_k \} \, , \\
  \{T_n, \alpha, \beta/H_*, v_w\}
  \longleftrightarrow
  \{D^{\rm bin}_k \} \, .
\end{align}
In the forward direction shown in \cref{eq:full_direction},
each cxSM parameter point provides a unique prediction 
for the spectrum, but the information about the underlying model can be partially lost, since
distinct points in the space of $\{m_S, \lambda_S, \lambda_{HS}\}$ may lead to similar
values of $\{T_n, \alpha, \beta/H_*\}$, and distinct sets of the latter may produce
spectra that cannot be distinguished within the experimental sensitivity of LISA.
In the inverse direction, which is the central question of 
\cref{sec:results}, we determine to what extent the cxSM 
and thermodynamic parameters can be reconstructed from a simulated LISA measurement, 
and how the degeneracies of the forward map and the parameter region featuring a 
first-order EWPT shape
the resulting predictions.

\subsection{Bayesian inference with MCMC}
\label{sec:mcmc}

In the likelihood-based analysis, we sample the posterior distribution
\begin{equation}
    p(\theta \mid D) = \frac{p(D \mid \theta)\, \pi(\theta)}{p(D)}
    \propto \mathcal{L}(\theta)\, \pi(\theta)
    \label{eq:bayes}
\end{equation}
with a Markov chain
Monte Carlo (MCMC) method, using the binned likelihood of
\cref{eq:binned_likelihood}.
The evidence $p(D)$ does not depend on $\theta$ and is therefore not needed for the sampling. Parameter points for which no first-order phase
transition takes place, such that no GW signal is predicted, are assigned a
vanishing likelihood.

We use affine-invariant ensemble sampling~\cite{Goodman:2010dyf} with the
parallel stretch move~\cite{Foreman-Mackey:2012any}, as implemented in the
Fortran package \texttt{copa}~\cite{Biekotter:2025gkp,copa}.
The sampler evolves an ensemble of walkers whose proposals are constructed from
the current spread of the ensemble, so that it adapts to the
strongly correlated posteriors encountered here, in particular for the cxSM
parameters, whose posteriors contract onto thin strips at sufficiently high
signal-to-noise ratio~(SNR) (see \cref{sec:results_cxsm}).
The details of the sampler and its settings are given in \cref{sec:app_mcmc_sampler}.

In order to verify the convergence of the chains, we have generated several
independent ensembles for each spectrum, starting from different random
initial positions of the walkers, and checked that they yield the same
posterior distributions.

For the thermodynamic parameters $\{T_n, \alpha, \beta/H_*\}$, the
evaluation of the likelihood only requires the GW spectrum predicted for the
given parameters. These scans are therefore much faster than the ones in the
cxSM parameters, for which each likelihood evaluation additionally requires
the costly computation of the thermodynamic parameters from the cxSM
parameters described in \cref{sec:cxsm}.
In the thermodynamic scans, the walkers sample directly the box prior of
\cref{eq:box_prior}, which is uniform in the logarithms of $T_n$, $\alpha$
and $\beta/H_*$, and are initialized according to this prior.

On the other hand, for the cxSM scans, the MCMC sampling has to be
prepared more carefully.
The likelihood vanishes outside the narrow parameter region in which the transition is of first order.
Moreover, even within this region, most parameter points predict a GW signal
far below the LISA sensitivity. 
For the remaining points with a first-order
transition, the costly computation of the thermodynamic parameters has to be
carried out, but the predicted spectrum is indistinguishable from the noise.
Walkers started uniformly within the prior box of \cref{eq:cxsm_box}
would therefore spend most of
the computing time exploring a flat likelihood.
We therefore restrict the MCMC sampling to the region of high likelihood determined beforehand with a genetic algorithm~(GA), implemented
in \texttt{evortran}~\cite{Biekotter:2025gkp,evortran}, see  \cref{sec:app_mcmc_ga} for further discussion.

In the scans of \cref{sec:vw}, in which the bubble wall velocity $v_w$ is
an additional free parameter,
different values of $v_w$ can be
compensated by different combinations of $\lambda_S$ and $\lambda_{HS}$.
As a result,
the correlation between $\lambda_S$ and $\lambda_{HS}$
that the GA identifies for fixed $v_w$, and which we
use to restrict the prior region of the MCMC sampling, is largely lost
(see \cref{sec:app_mcmc_ga} for details).
Thus, we do not use the preparation of the sampling region
with the GA if $v_w$ is jointly inferred.
Instead, the walkers sample directly a prior that is uniform in $m_S$,
$\lambda_S$, $\lambda_{HS}$ and $v_w$, within the box of \cref{eq:cxsm_box}
and $v_w \in [0.1, 1]$ as for the SBI (see \cref{sec:vw}).
Since the sampling region cannot be restricted in advance, and the parameter
space is four-dimensional, these scans require a considerably larger number
of samples to obtain a converged posterior.

\subsection{Simulation-based inference}
\label{sec:sbi}

The inference
using the MCMC method discussed in \cref{sec:mcmc}
is based on the explicit likelihood of \cref{eq:whittle}.
Every evaluation of the likelihood
requires the full chain of calculations from the cxSM parameters to the predicted spectrum.
SBI replaces this by
applying the simulator described in \cref{sec:simulation}
to a large set of parameter points drawn from the prior.
The resulting pairs $\{\theta^{(a)}, D^{(a)}\}$ of parameters and binned spectra, where the index $a$ labels the simulated parameter points, are used to train a neural density estimator~\cite{Cranmer:2019eaq}.
In contrast to a regression network, which would map a spectrum onto a single set of parameter values, the network parameterizes a full probability density, such that the uncertainties and degeneracies of the reconstruction are retained. We realize it as a normalizing flow~\cite{Papamakarios:2019fms}.
The weights $\varphi$ of the flow are adjusted such that the density reproduces the statistical properties of the simulated points.

In order to keep the network size and training cost manageable, we reduce the input dimension by almost three orders of magnitude by binning the spectrum in frequency (see \cref{sec:simulation}).
The binning retains the relevant
the statistical information,
with an estimated increase in statistical uncertainties by only about $7\%$
(see the discussion in \cref{sec:likelihood} for details).
The input data consist of the $571$ bins of \cref{eq:binning_def}, but we condition the flow on a low-dimensional summary of $64$ features produced by an embedding network that is trained simultaneously with the flow.
We use a one-dimensional convolutional network (1D CNN) acting along the frequency direction, which we find to describe the posterior better than a fully connected network with the same number of parameters.
The details of the embedding network and flow architectures and the training settings are provided in \cref{sec:app_sbi}.

There are several variants of the SBI, differing in which density $q_\varphi$
is learned by the flow.
In this work, we use
\emph{neural posterior estimation}~(NPE)~\cite{Greenberg:2019apt}, where
the target is the posterior of \cref{eq:bayes} itself,
\begin{equation}
    q_\varphi(\theta \mid D)  \simeq  p(\theta \mid D) ,
    \label{eq:npe}
\end{equation}
obtained by minimizing $-\sum_a \ln q_\varphi(\theta^{(a)} \mid D^{(a)})$ with respect to the weights $\varphi$, i.e.~by maximizing the probability that the flow assigns to the true parameters $\theta^{(a)}$ of each simulated spectrum $D^{(a)}$.
In NPE, the prior enters through the distribution from which the training points are drawn, and the trained network returns samples of the posterior for any data set in a fraction of a second, without a further call to the simulator.
This amortization is what makes the repeated inference required for a statistical validation of the reconstruction feasible.
The flow is a neural spline flow~\cite{Durkan:2019nsf}, specified together with its training settings in \cref{sec:app_sbi}.

The posterior distribution learned by NPE is a normalized density in $\theta$, i.e.~it integrates to one over the parameter space, and
it can be both sampled from and evaluated directly in $\theta$. It can therefore be used as a proposal in an importance sampling~\cite{Brooks:2016,Le:2016,Dax:2022pxd} in which each sample $\theta_i$ drawn from $q_\varphi(\theta \mid D)$ is assigned the weight $w_i = w(\theta_i)$, with
\begin{equation}
    w(\theta)  \propto  \frac{\mathcal{L}(\theta)  \pi(\theta)}{q_\varphi(\theta \mid D)} ,
    \label{eq:is_weights}
\end{equation}
with $\pi$ the prior and $\mathcal{L}$ the exact likelihood of \cref{eq:binned_likelihood}.
Expectation values of a quantity $f(\theta)$, such as a parameter or the indicator function of a credible region, are estimated from the weighted samples as $\sum_i w_i f(\theta_i) / \sum_i w_i$. They converge to those of the exact posterior of \cref{eq:bayes} as the number of samples increases. This holds irrespective of the quality of the network, provided that the network assigns a nonvanishing probability to all regions of parameter space in which the exact posterior does. The quality of the network only determines how many samples are needed, i.e.~the efficiency of the sampling, measured by the effective sample size (ESS),
\begin{equation}
    \mathrm{ESS} = \frac{\left(\sum_i w_i\right)^2}{\sum_i w_i^2} .
    \label{eq:ess}
\end{equation}
Normalizing the ESS to the number $N$ of proposed samples, we define the efficiency $\epsilon = \mathrm{ESS}/N$,
which equals one for a perfect network
and takes its minimal value $1/N$ if a single sample carries all the weight.
The neural importance sampling (NIS) allows the speed of the amortized network to be combined with the
exact likelihood in the final result. 
In contrast to the NPE alone, it is therefore not likelihood-free. It requires the evaluation of $\mathcal{L}(\theta_i)$ for every sample, and its result relies on the same assumptions as the likelihood. In our analysis, the likelihood describes the simulated data exactly, and we use NIS both to validate the NPE posteriors and to correct their remaining deviations from the exact posterior.

\subsection{Choice of priors}
\label{sec:priors}
The prior $\pi(\theta)$ enters our
analysis
both in the MCMC and the SBI inference.
In the former, it enters in the posterior distribution
by multiplying the likelihood as shown in \cref{eq:bayes}.
In the latter,
it is the distribution from which the training points of
\cref{eq:npe}
for the SBI are drawn,
and it
enters the reweighting in the NIS shown in
\cref{eq:is_weights}.
Because the simulator returns a detectable GW signal only
in a
narrow region of parameter space in which the
EWPT is sufficiently strong, the priors have to be chosen in
such a way as to avoid that almost all samples are placed in a
region in which only instrumental noise is predicted.
We state our prior choices for both the thermodynamic parameters
$\theta = \{ T_n, \alpha, \beta / H_* \}$
and the cxSM parameters $\theta = \{ m_S, \lambda_S, \lambda_{HS} \}$ below.

For the thermodynamic parameters we use a box prior, uniform in the logarithms of $T_n$, $\alpha$ and $\beta/H_*$
over the ranges
\begin{equation}
    T_n \in [15, 75]\GeV , \;
    \alpha \in [0.05, 10] , \;
    \beta/H_* \in [2.5, 2500] \, .
    \label{eq:box_prior}
\end{equation}
These ranges
were chosen to enclose the values that can be predicted in the nightmare
scenario of the cxSM.
This facilitates a direct comparison of
the inference in the thermodynamic parameters and the
inference in the cxSM parameters (see below), since both
are carried out over the same domain.
We impose a lower limit on the nucleation temperature
of $T_n > 15\GeV$,
because the dimensionally reduced three-dimensional effective theory
that is used to compute the properties of the EWPTs in the cxSM
relies on a high-temperature expansion whose accuracy degrades
once $T_n$ is no longer large compared with the mass scales
contained in
the broken phase~\cite{Biekotter:2025npc}
(see also the discussions in Refs.~\cite{Chala:2024xll,Gould:2024jjt,
Bernardo:2025vkz,Chala:2025aiz,Chala:2025oul}).

For the cxSM, the shape of
the parameter space region in which a detectable SGWB is predicted
is not known in closed form.
We determined it by random sampling,
where we considered the parameter ranges
\begin{equation}
    m_S \in [60, 120]\GeV , \;
    \lambda_S \in [0.1, 5] , \;
    \lambda_{HS} \in [0.25, 2.5] \, .
    \label{eq:cxsm_box}
\end{equation}
For the parameter points predicting a first-order EWPT, we
computed the thermodynamic parameters
and retained the
parameter points whose predicted SGWB
 is detectable at LISA, i.e.
\begin{equation}
    \mathrm{SNR} = \left[ T_{\rm obs} \int df \left( \frac{h^2\Omega_{\rm GW}(f;\theta)}{h^2\Omega_{\rm n}(f)} \right)^{2} \right]^{1/2} > 1 ,
    \label{eq:snr}
\end{equation}
with $T_{\rm obs} = N_c/\Delta f = 9.4\times10^{7}\,\mathrm{s} \simeq 3$ years the effective observation time of the simulated data, $N_c = 94$ chunks of duration $1/\Delta f = 10^{6}\,\mathrm{s}$, corresponding to a four-year mission with a $75\%$ duty cycle~\cite{Caprini:2024hue}.
We generated a total number of about
$9.3 \times 10^{6}$ parameter points.
Among those, around $5.5 \times 10^{4}$ ($0.6\%$) 
feature a first-order
EWPT and in addition satisfy \cref{eq:snr}
(shown in
\cref{fig:training} of \cref{sec:app_sbi}), which we use for training, validation and testing of the cxSM NPE.
The prior is thus uniform in the microscopic parameters within the box, restricted to the region of detectable first-order transitions, which is the natural choice in the absence of any preferred scale among them.
In the MCMC analysis, we apply the same bounds on the cxSM parameters,
given in \cref{eq:cxsm_box}. As a result of the preconditioning with the GA,
the MCMC samples $\lambda_S$ and $\lambda_{HS}$ with a prior that is uniform
in their logarithms. We reweight the MCMC samples by $\lambda_S \lambda_{HS}$,
such that the resulting posteriors correspond to the same uniform prior as in
the SBI analysis (see \cref{sec:mcmc} for details).

The bounds in \cref{eq:cxsm_box} are in part choices made for
simplicity. They nevertheless contain all features of the parameter
space that are relevant for our discussion, in particular the
degeneracies among the parameters and the narrowness of the region
of detectable SGWBs. Moreover, since the MCMC and the SBI analyses
use the same bounds, their precise values do not affect the
comparison of the two methods, which is the main focus of this work.
For $m_S$ and $\lambda_S$ within the given ranges, the region of
detectable SGWBs lies entirely within
$0.38 \lesssim \lambda_{HS} \lesssim 2.2$, such that the bounds on
$\lambda_{HS}$ do not restrict it. In $m_S$ and $\lambda_S$, by
contrast, the region extends beyond the bounds. The upper bound
$\lambda_S \leq 5$ keeps the analysis well within the perturbative
regime. The lower bound on $m_S$ lies close to
$m_H/2 \simeq 62.5\GeV$, since for lighter singlets the decay
$h \to ss$ is kinematically open and $\lambda_{HS}$ is tightly
constrained by the signal-rate measurements of the Higgs boson at
$125\GeV$ and by searches for exotic Higgs-boson decays
(see \cref{sec:cxsm}). The upper bound $m_S \leq 120\GeV$ focuses the
analysis on the region with the largest SNRs, although detectable SGWBs
are also predicted for substantially larger singlet masses.

\section{Reconstruction of the parameters}
\label{sec:results}

In this section, we investigate the inference of the
macroscopic/thermodynamic and microscopic/cxSM parameters from
simulated LISA measurements.
We start by assuming a known and constant bubble wall velocity,
fixing it to
$v_w = 0.8$, and reconstruct the thermodynamic parameters in
\cref{sec:results_thermal} and the cxSM parameters in
\cref{sec:results_cxsm}.
Afterwards, in \cref{sec:vw}, we carry out additional parameter
reconstructions in which we
treat the bubble wall velocity as an additional free parameter
that has to be inferred from the data.
The calibration of the networks and the accuracy
of the SBI approach are discussed in \cref{sec:results_validation}.
Finally, in \cref{sec:cost} we discuss
the computational cost of
both the MCMC and the SBI methods.

All parameter reconstructions in this section follow the same
procedure. For a given set of injected parameters, we compute the
predicted GW spectrum and simulate a LISA measurement of this signal,
including the instrumental noise, as described in \cref{sec:simulation}.
The resulting binned spectrum of \cref{eq:binning_def} is the mock data set
from which we reconstruct the parameters. We choose the injected parameters
such that the signals cover a range of SNRs (see \cref{eq:snr}), and in each
case show examples with low, medium and high SNR. For each mock data set,
we determine the posterior distribution of the parameters with three
methods, namely the amortized NPE, the NIS, which uses the NPE posterior as
proposal and reweights it with the exact likelihood (see \cref{sec:sbi}),
and the MCMC analysis of \cref{sec:mcmc}. All three methods are applied to
the same mock data set. Since the NIS and the MCMC analysis are both based on
the likelihood of \cref{eq:binned_likelihood}, they should agree within the
statistical precision of the samples, and their comparison with the NPE
tests the accuracy of the amortized posterior.
We display the posteriors as one-dimensional marginal distributions of each
parameter and as two-dimensional marginal distributions of each pair of
parameters, the latter showing the $68\%$ and $95\%$ credible regions.
The injected parameters are marked by a star, such that one can judge
whether they are recovered within these regions.

\subsection{Thermodynamic parameters}
\label{sec:results_thermal}

\begin{figure*}[t]
  \centering
  \includegraphics[width=0.32\textwidth]{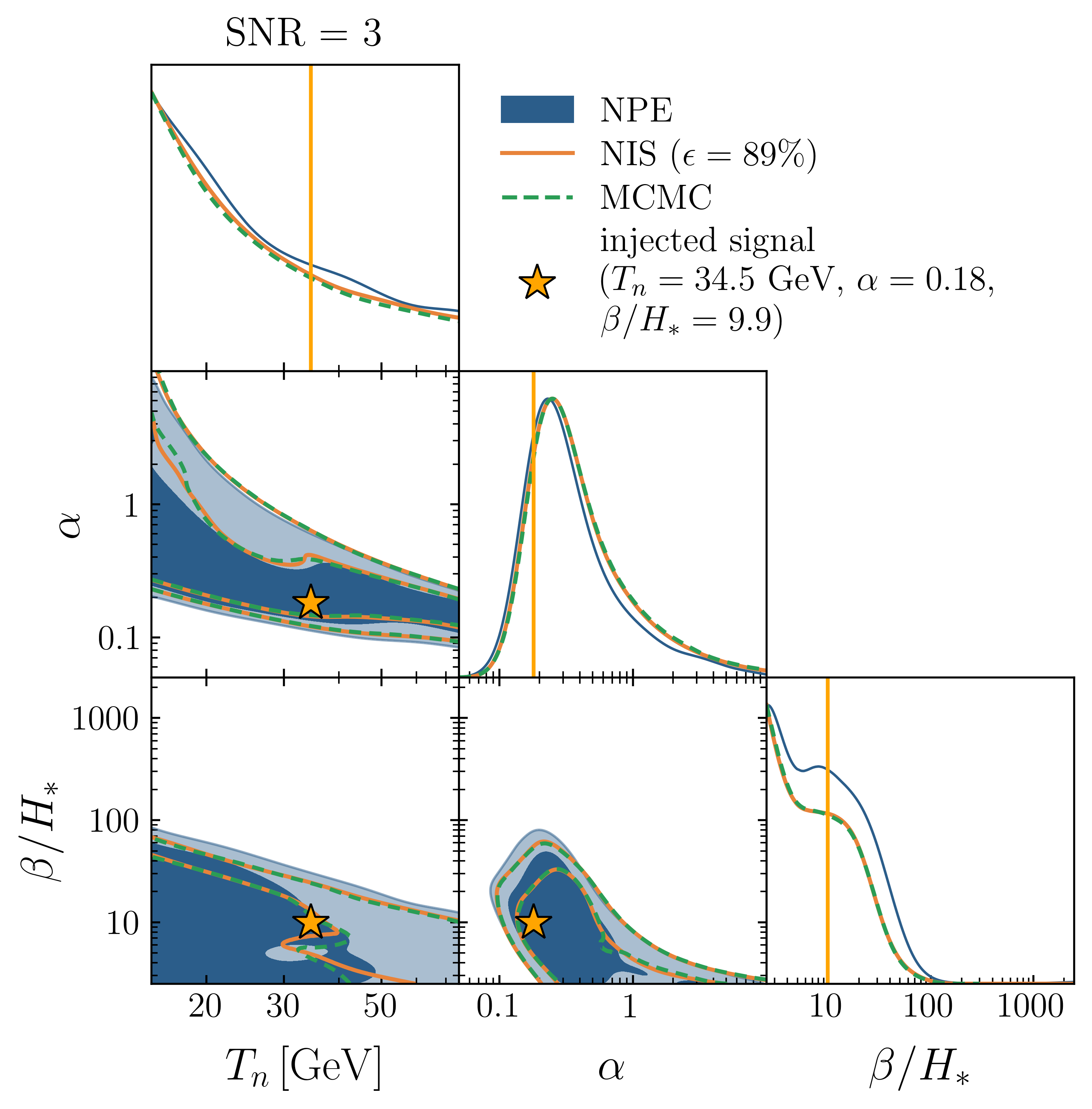}
  \includegraphics[width=0.32\textwidth]{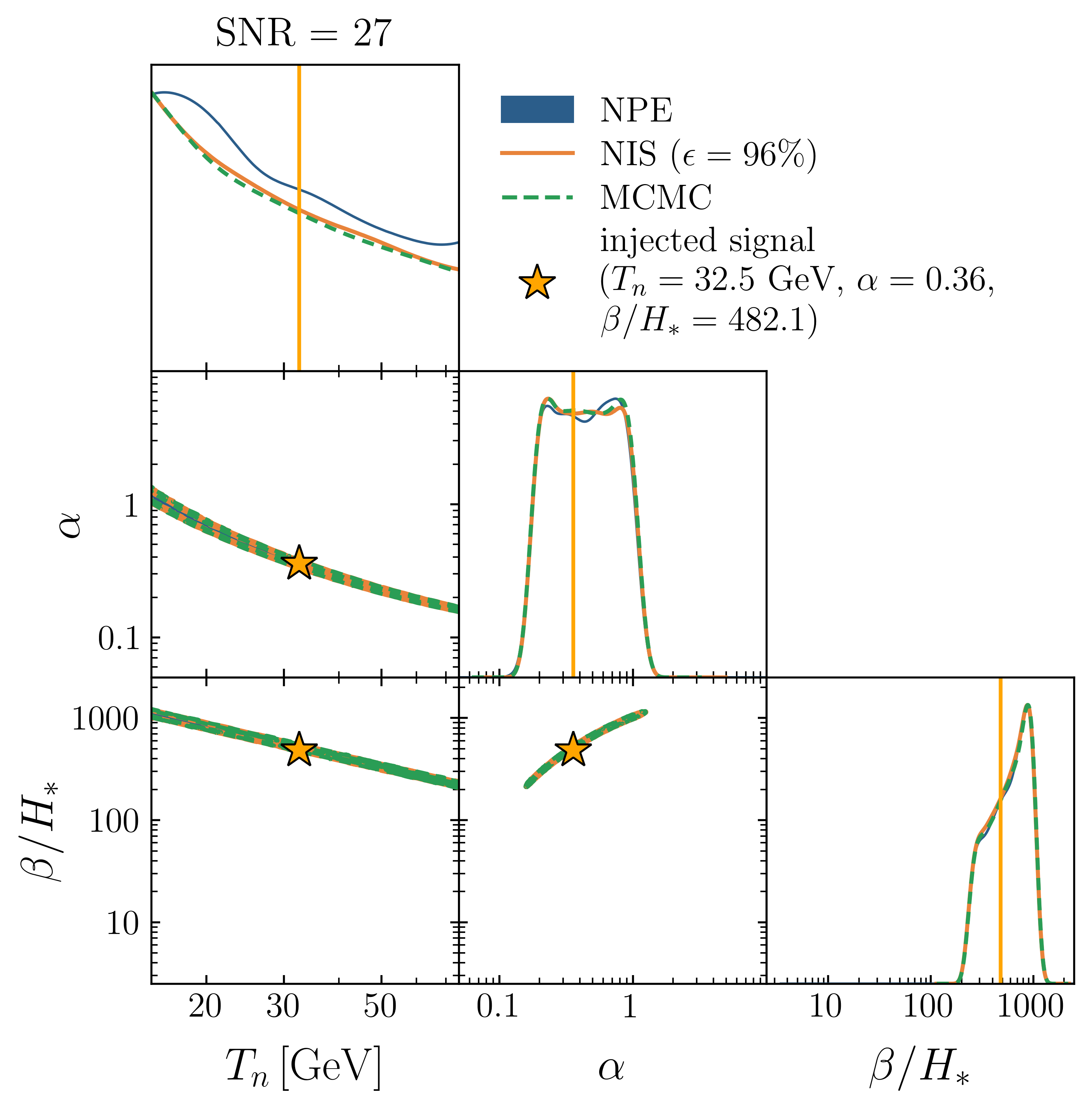}
  \includegraphics[width=0.32\textwidth]{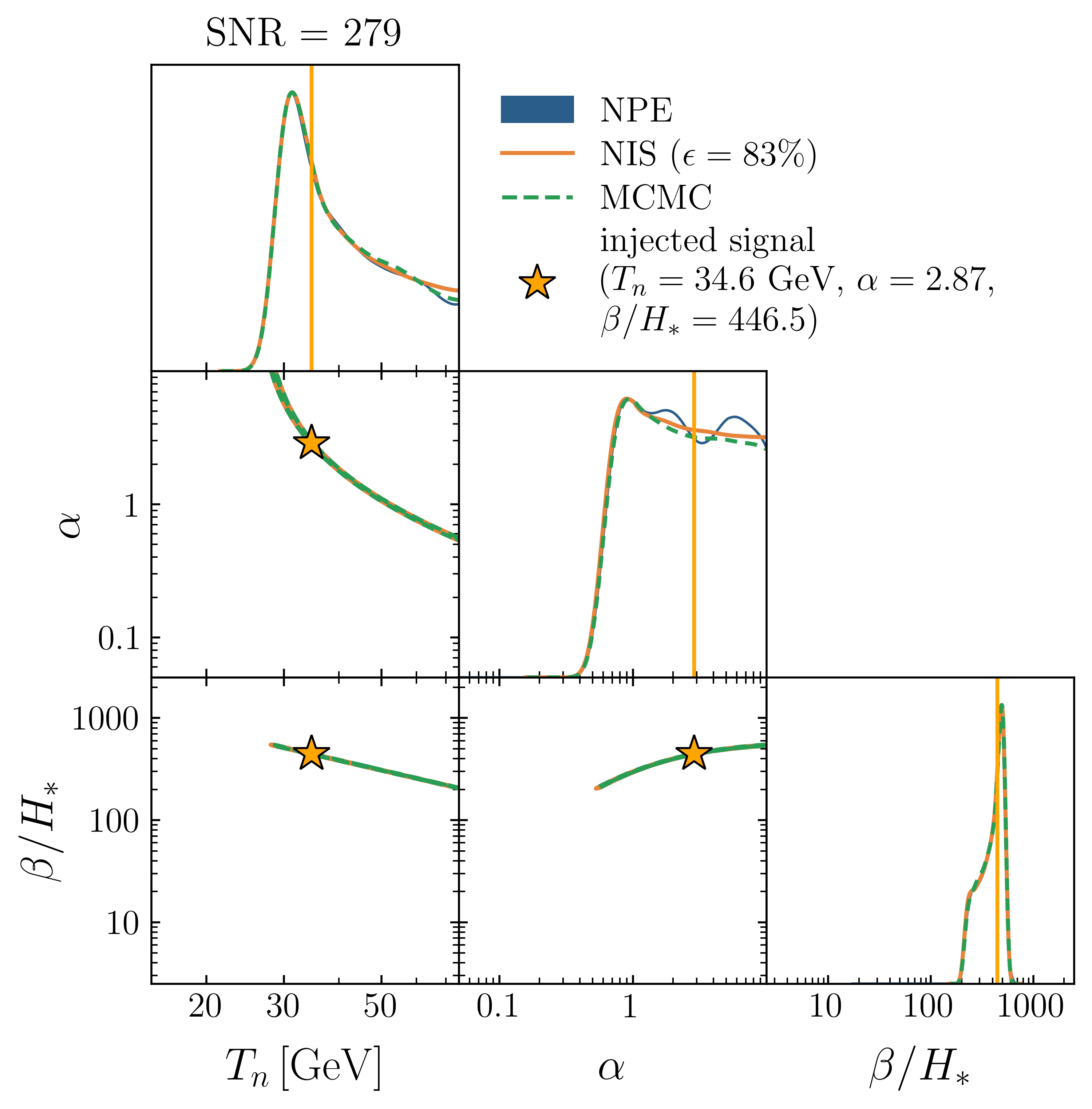}
  \caption{One- and two-dimensional marginal
  distributions of the thermodynamic parameters $\{T_n, \alpha, \beta/H_*\}$
  for three simulated LISA measurements with an injected signal of
  low (left), medium (center) and high (right) SNR,
  obtained with the amortized NPE (blue), NIS based on the NPE proposal
  (orange) and MCMC (green). The orange stars in the
  two-dimensional distributions and the orange vertical lines in
  the one-dimensional distributions mark
  the
  parameters of the injected SGWBs.
  The efficiencies $\epsilon$ of the importance sampling,
  defined in \cref{sec:sbi},
  are given in the legends.}
  \label{fig:th1}
\end{figure*}

We start with the inference of the macroscopic
thermodynamic parameters  $\{T_n, \alpha, \beta/H_*\}$.
In \cref{fig:th1} we show corner plots with
the posterior distributions of these parameters
for three injected signals
with low (left), medium (center) and high SNR (right).

As expected, the posteriors become more informative with increasing SNR
of the injected signal. For the low-SNR spectrum ($\mathrm{SNR} = 3$), the
$68\%$ and $95\%$ credible regions extend over a large part of the prior
range of \cref{eq:box_prior}, and the data mainly disfavor large values
$\beta/H_* \gtrsim 100$. For the medium- and high-SNR spectra
($\mathrm{SNR} = 27$ and $279$), the credible region contracts to
a thin strip in the three-dimensional parameter space, which manifests itself
as the thin strips
in each of the two-dimensional planes.
In all three cases, the injected
parameters lie within the $68\%$ credible regions, such that the
reconstruction is consistent with the injected signal.

The strips follow the degenerate directions of the spectrum discussed in
\cref{sec:simulation},
along which a change of
one thermodynamic parameter can be compensated by a change of the other two.
Along these directions, $\alpha$ and $\beta/H_*$ decrease with increasing
$T_n$, while $\beta/H_*$ increases with $\alpha$. Since the strips extend over
a large part of the prior range, none of the three parameters can be
determined individually from the data, even for the high-SNR spectrum.
For the medium-SNR spectrum, the one-dimensional posterior of $T_n$ covers
the full prior range, and those of $\alpha$ and $\beta/H_*$ extend over
almost an order of magnitude. For the high-SNR spectrum, the posterior of
$T_n$ is cut off below $T_n \approx 28\GeV$, where the strip reaches the
upper prior boundary $\alpha = 10$, but $\alpha$ and $\beta/H_*$ remain
similarly poorly determined. At the same time, the narrow width of the strips
reflects a strong correlation among the parameters. For the medium- and
high-SNR spectra, if one of the three parameters were known, for instance
from a calculation of the transition within a specific model, the other two
could be extracted from the data with good precision.

We now compare the posteriors obtained with the SBI to those of the
MCMC analysis. In the two-dimensional distributions of \cref{fig:th1}, the
NPE posteriors are shown as blue shaded regions, with the darker and lighter
shading indicating the $68\%$ and $95\%$ credible regions, respectively.
The orange and green lines show the boundaries of the same credible regions
obtained with the NIS and the MCMC analysis, respectively. The
one-dimensional distributions use the same colors, where the MCMC results
are shown as dashed lines, since they lie almost on top of the NIS results.
For all three spectra, the NPE posteriors agree very well with the MCMC
posteriors, both in the location and in the extent of the credible regions.
After the importance sampling, the agreement is even better, and the NIS and
the MCMC analysis yield practically identical posteriors.

With the NPE posterior as the proposal, the NIS reaches an
efficiency of $\epsilon = 89\%$, $96\%$ and $83\%$
for the
low-, medium- and high-SNR spectra, respectively.

For the medium- and high-SNR spectra, the medians of the one-dimensional NPE and NIS posteriors of each parameter differ by at most $0.02$ times the half-width of the $68\%$ credible interval, and the widths of these intervals agree to within $2\%$.
For the low-SNR spectrum, whose posterior extends to the boundaries of the prior, the deviations are larger, up to $0.16$ half-widths in the medians of $\alpha$ and $\beta/H_*$ and $10\%$ in the widths.

This shows that the NPE reproduces the degeneracy correctly both in its shape and in its width.
We checked that the same holds for a larger set of $200$ test spectra,
finding that the median efficiency is about $\epsilon = 90\%$ up to an SNR of
$100$. For larger SNRs, the efficiency
decreases to about $\epsilon = 70$--$80\%$,
since the posteriors of
such strong signals become thin strips in the parameter space,
for which small deviations of the NPE from the exact posterior
become increasingly important.

In addition, one can compare the efficiency of drawing parameter points from the NPE posterior with that of drawing them uniformly from the prior and weighting each sample by its likelihood of \cref{eq:binned_likelihood}.
The latter is highly
inefficient since the posterior occupies only a
small fraction of the prior
volume, which decreases with the SNR.
For the three
reconstructions shown in \cref{fig:th1}, the efficiency
would be of the order of
$10^{-2}$, $10^{-4}$ and $10^{-5}$, respectively,
which can be compared with an efficiency of
$80$--$95\%$
when using the NPE as the proposal.
The NPE thus reduces the number of
likelihood evaluations needed to obtain a given number of effective samples
by two to five orders of magnitude compared to sampling from the prior,
while providing an equally accurate parameter reconstruction.

The three injected signals of \cref{fig:th1} were chosen as
representative examples of signals with low, medium and high SNR.
We have, however, carried out the same analysis for many other injected
signals and noise realizations, as well as for several independently
trained networks, and found similar results in all cases. This concerns the
agreement between the likelihood-free SBI and the likelihood-based MCMC
analysis, the improvement achieved with the importance sampling, and the
impact of the degeneracies on the extent to which the thermodynamic
parameters can be reconstructed at a given SNR.

\subsection{cxSM parameters}
\label{sec:results_cxsm}

\begin{figure*}[t]
  \centering
  \includegraphics[width=0.32\textwidth]{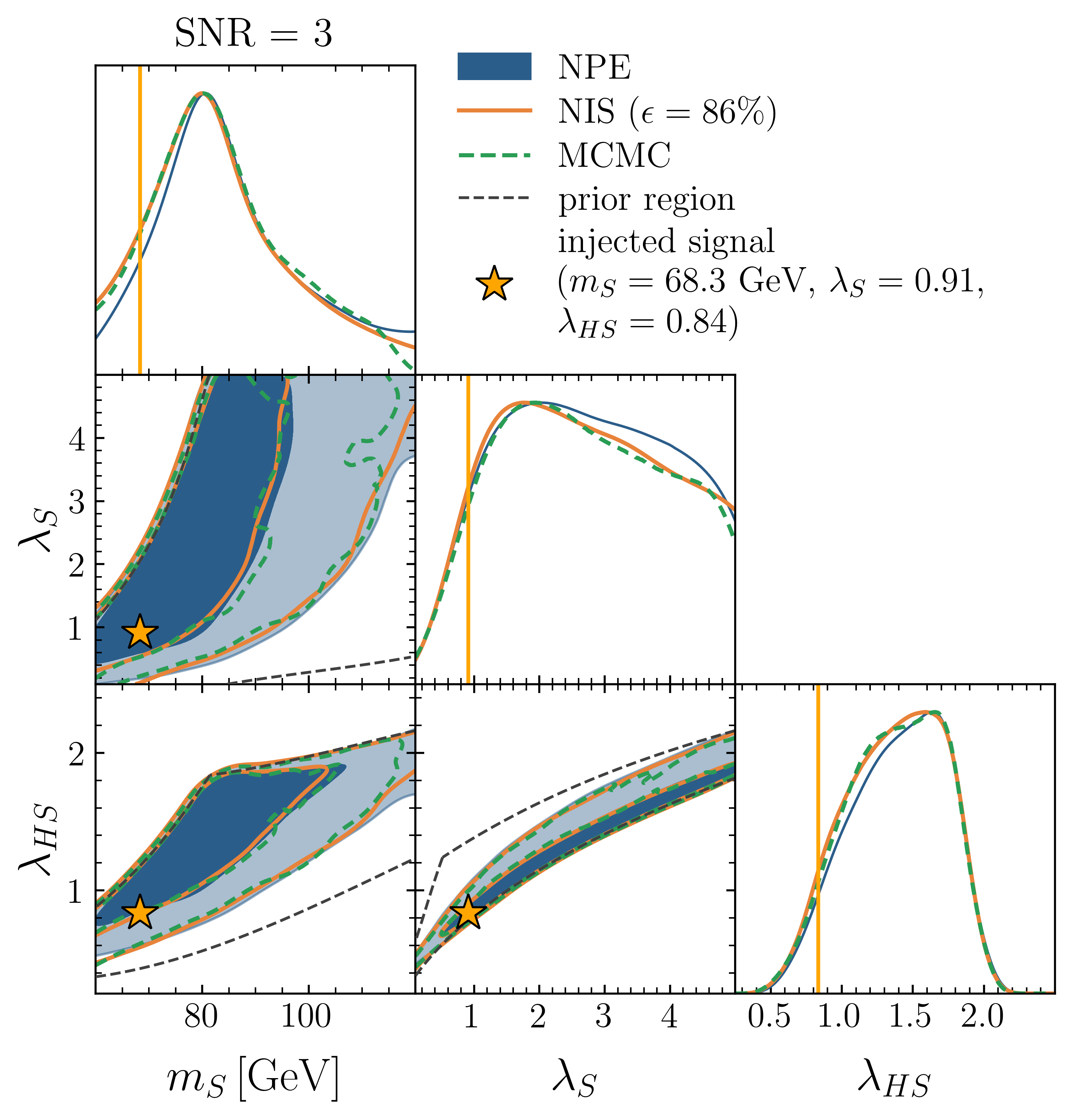}
  \includegraphics[width=0.32\textwidth]{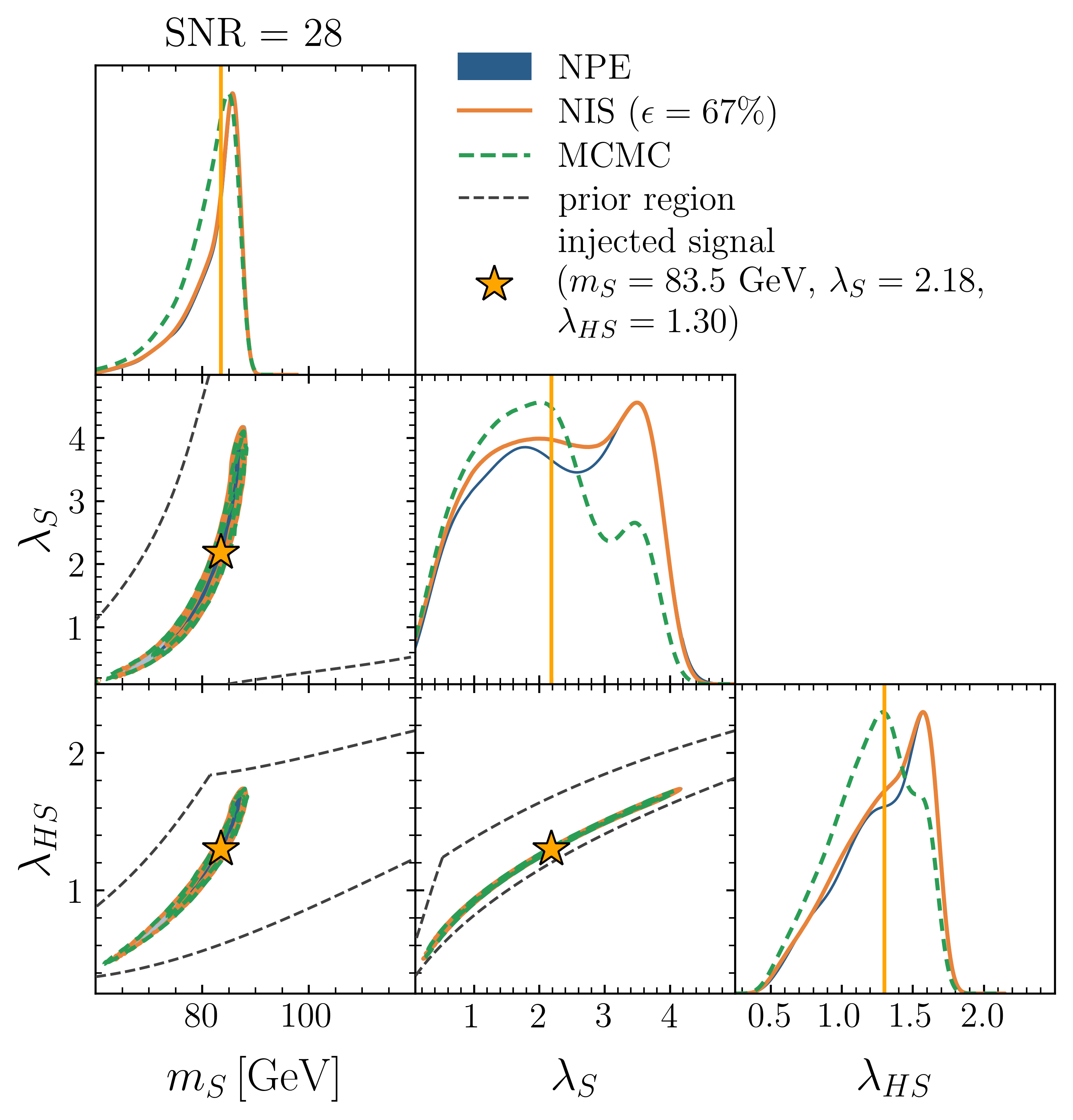}
  \includegraphics[width=0.32\textwidth]{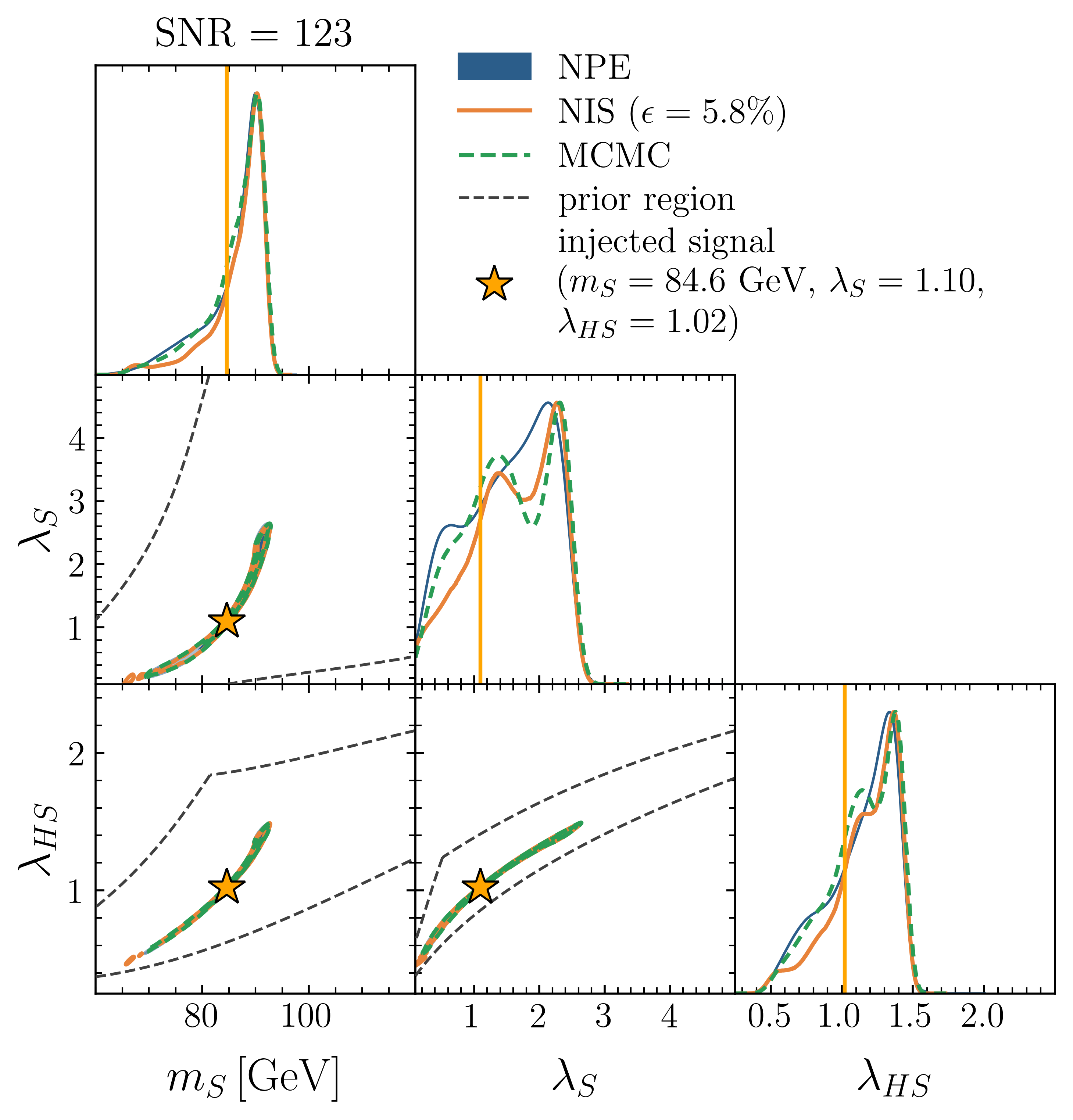}

  \includegraphics[width=0.32\textwidth]{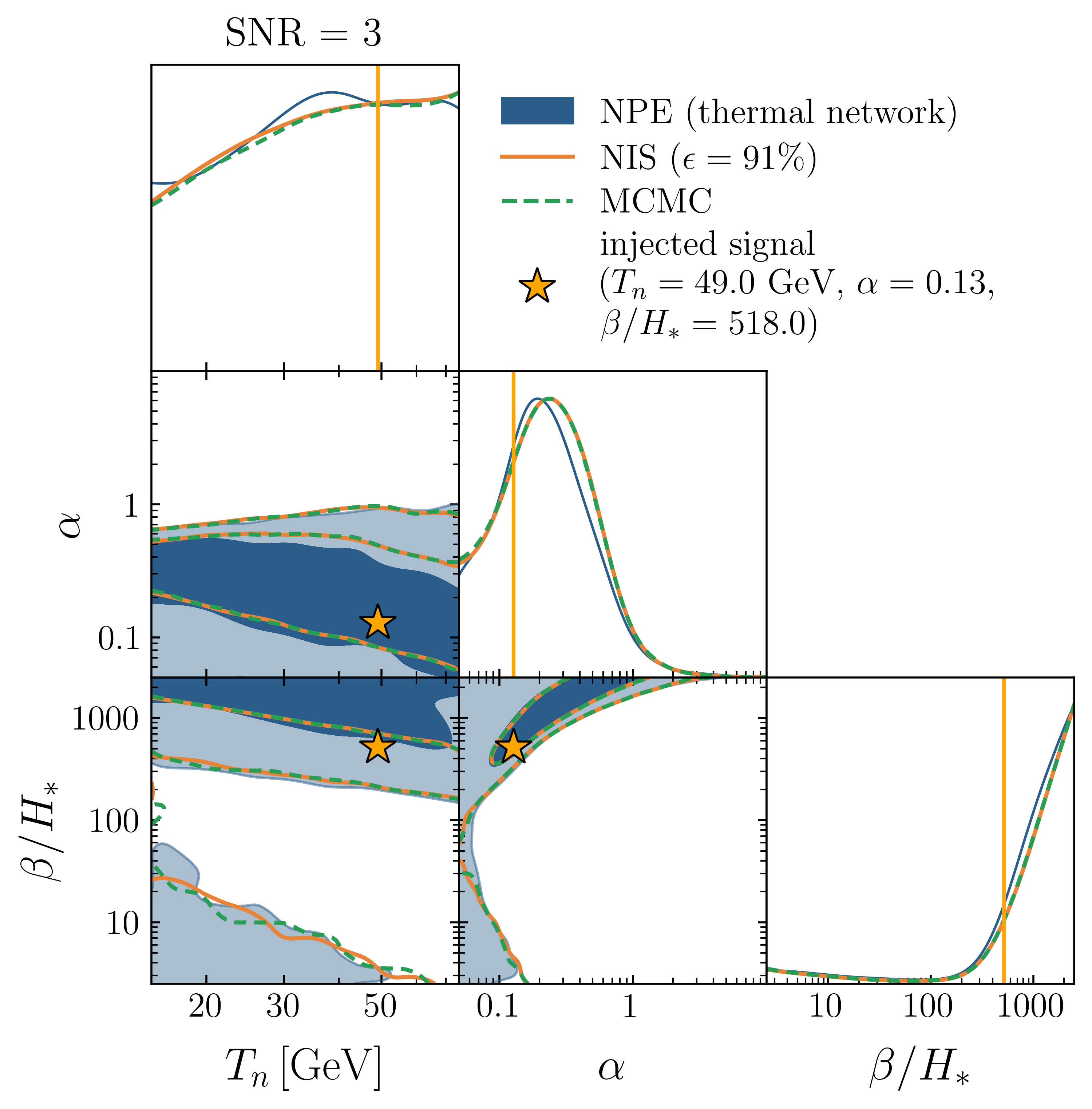}
  \includegraphics[width=0.32\textwidth]{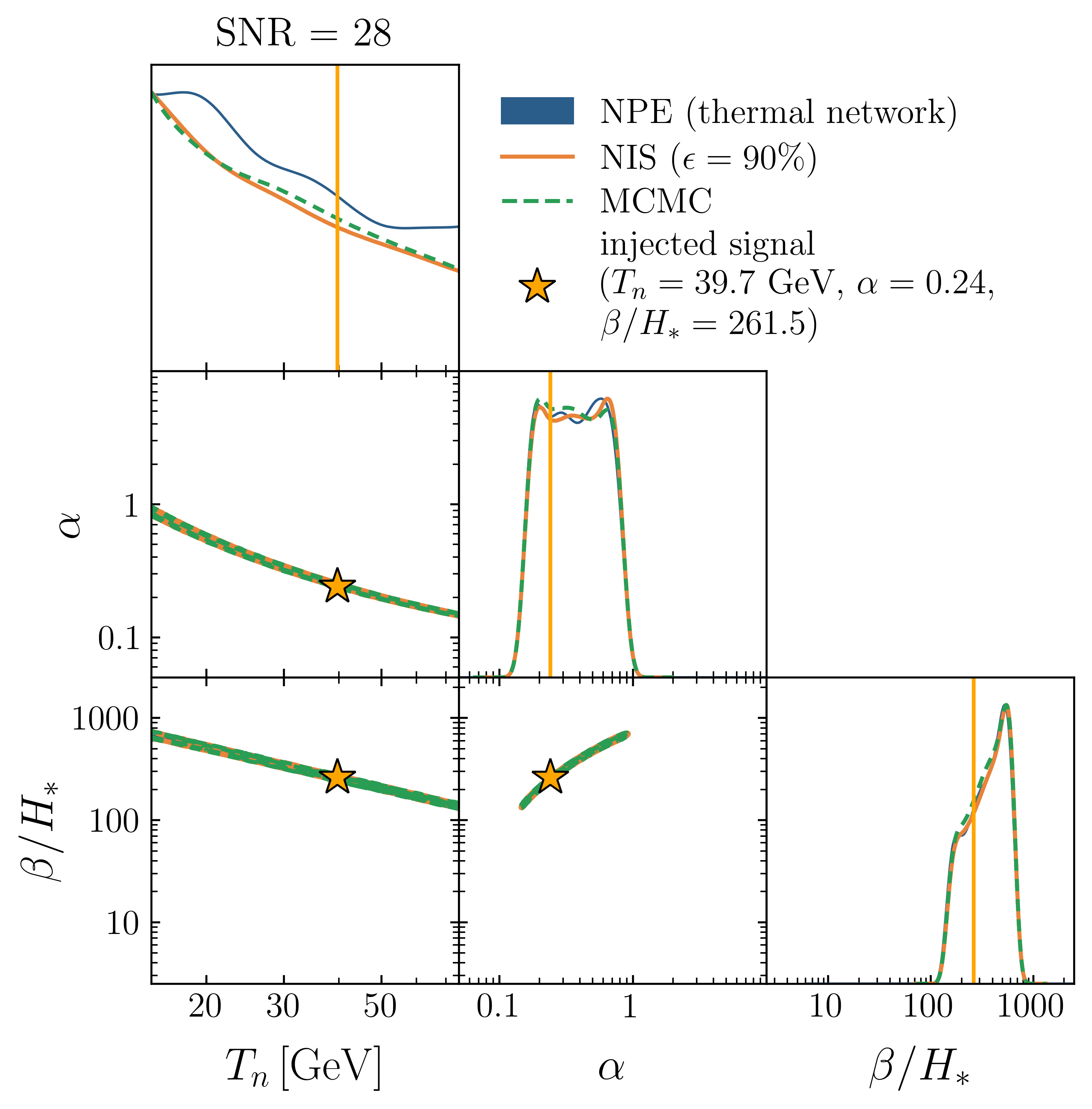}
  \includegraphics[width=0.32\textwidth]{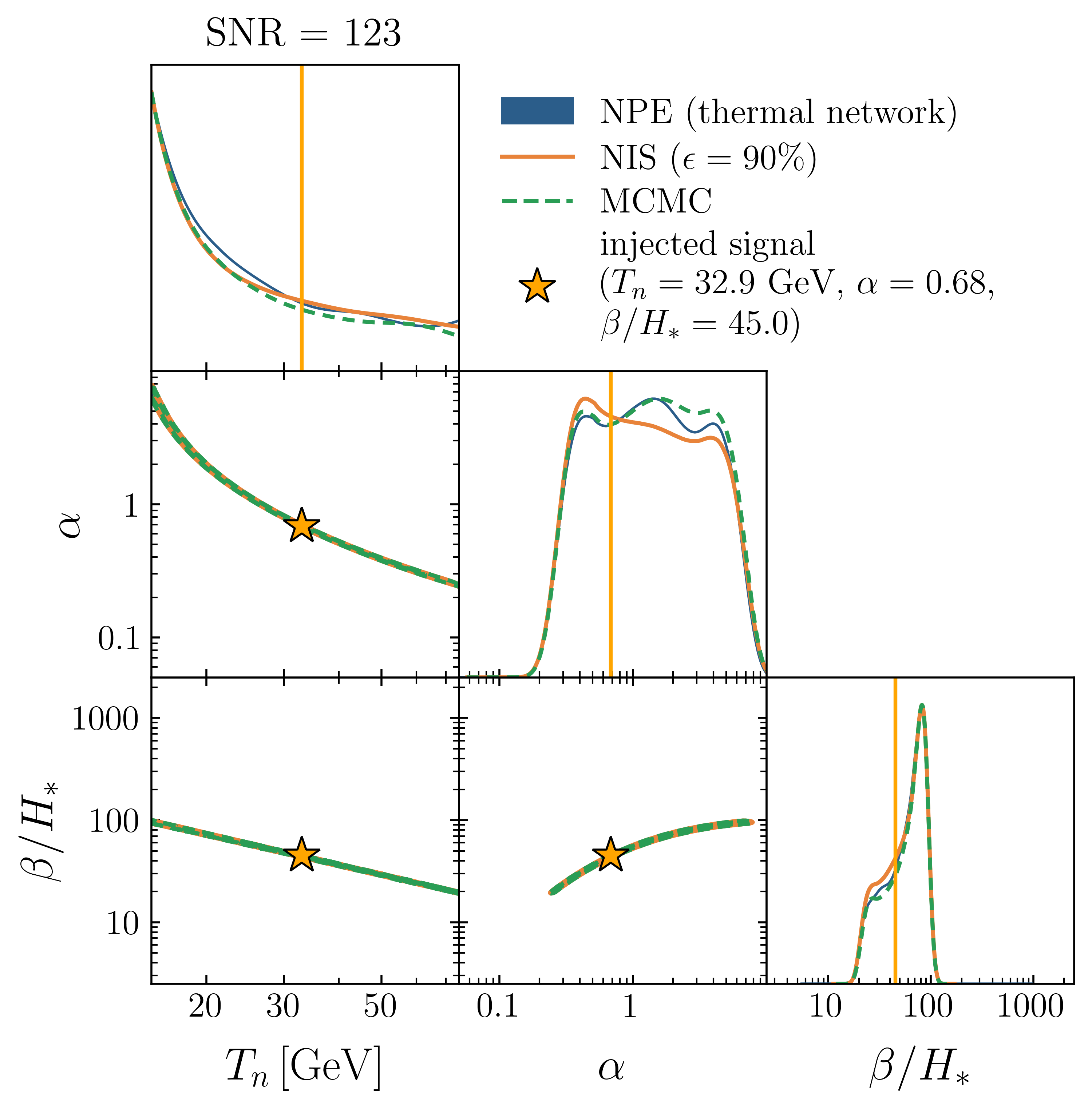}

  \caption{Top: One- and two-dimensional marginal distributions of
  the cxSM parameters $\{m_S, \lambda_S, \lambda_{HS}\}$ for three simulated
  LISA measurements with an injected signal of low (left), medium (center)
  and high (right) SNR, obtained with the amortized NPE (blue), NIS based on
  the NPE proposal (orange) and MCMC (green). The dashed black lines show the
  boundaries of the prior region (see \cref{sec:priors}).
  Bottom: One- and two-dimensional marginal distributions of the
  thermodynamic parameters $\{T_n, \alpha, \beta/H_*\}$ inferred from the
  same simulated measurements, with the amortized NPE of the thermodynamic
  parameters. The orange stars in the two-dimensional distributions and the
  orange vertical lines in the one-dimensional distributions mark the
  parameters of the injected SGWBs. The efficiencies $\epsilon$ of the
  importance sampling, defined in \cref{sec:sbi}, are given in the
  legends.}
  \label{fig:cxsm1}
\end{figure*}

We now turn to the reconstruction of the cxSM parameters
$\{m_S, \lambda_S, \lambda_{HS}\}$. The top row of \cref{fig:cxsm1} shows
their posterior distributions for three simulated LISA measurements with
low, medium and high SNR ($\mathrm{SNR} = 3$, $28$ and $123$), and the
bottom row shows the posteriors of the thermodynamic parameters inferred
from the same spectra. The shaded regions and the colored lines have the
same meaning as in \cref{fig:th1}, and the stars and the orange vertical
lines again indicate the parameters of the injected signals. In addition,
the dashed black lines show the boundaries of the prior region, i.e.~the part
of the parameter space within the ranges of $m_S$, $\lambda_S$ and
$\lambda_{HS}$ given in \cref{eq:cxsm_box} in which the transition is of first order
and the predicted signal is detectable with $\mathrm{SNR} > 1$
(see \cref{sec:priors}). As for the thermodynamic parameters, the injected
parameters lie within the $68\%$ credible regions in all three cases.

At low SNR, the posterior fills a broad band within the prior region, in
which the three cxSM parameters are positively correlated and only a combination
of them is constrained. With increasing SNR, the posterior contracts onto a
thin curved strip, which to a good approximation is a one-dimensional curve
in the three-dimensional parameter space. Along this curve, $m_S$,
$\lambda_S$ and $\lambda_{HS}$ increase simultaneously, such that a larger
singlet mass is compensated by larger couplings, and all points on the curve
predict very similar GW spectra.
The length of the strip, i.e.~the range of parameter values covered along the
degenerate direction, is set by the degeneracies and by the boundaries of the
prior region, and shrinks only slowly with increasing SNR. In contrast, the
thickness of the strip perpendicular to the degenerate direction decreases
rapidly with increasing SNR.

The NPE of the cxSM parameters (top row of \cref{fig:cxsm1}) performs
best for the wider posteriors.
For the low- and medium-SNR spectra, the medians agree with those of the NIS to within $0.1$ of the $68\%$ half-widths and the widths to within $3\%$, and the importance sampling reaches efficiencies of $86\%$ and $67\%$.
For the high-SNR spectrum (top right of \cref{fig:cxsm1}), the posterior
of the cxSM parameters is confined to a very thin strip. Here the NPE still
reproduces the location and the length of the strip, but it predicts a
thicker strip than the NIS. 
The marginal $68\%$ intervals of NPE and NIS
agree to within about $2\GeV$ in $m_S$, whereas the thickness of the strip
in $m_S$, as defined in \cref{sec:cost}, is $0.04\GeV$ for the
NPE and $0.02\GeV$ for the NIS. Accordingly, the
efficiency of the importance sampling drops to $5.8\%$.

This limitation of the NPE is related to where the high-SNR signals are
located in the parameter space. Signals with a large SNR arise for parameter
points close to the boundary beyond which the transition does not complete
and the universe remains trapped in the wrong phase.
Near this boundary, small changes of the cxSM parameters lead to large
changes of the thermodynamic parameters, and hence of the predicted spectrum.
As a consequence, the posterior turns into a very thin, essentially
one-dimensional curve that ends abruptly at the boundary, a feature that a smooth density estimator struggles to resolve exactly.
The NPE is therefore slightly conservative in this regime, and the importance sampling corrects the remaining difference with $10^{5}$ simulator calls
(see the discussion in \cref{sec:NIS} for details).
For the importance sampling, it is favorable that the NPE posterior is
wider and not narrower than the exact posterior. A proposal distribution that
is too wide only lowers the efficiency. A proposal distribution that is too
narrow, in contrast, would not cover parts of the posterior, which the
importance sampling cannot recover, such that the result would be biased.
The ability of the NPE to resolve such thin posteriors can likely be improved with more expressive flows and refined training schemes.

As for the thermodynamic parameters (see \cref{sec:results_thermal}),
the posterior of the cxSM parameters could in principle also be obtained
without the NPE, by drawing the cxSM parameters uniformly from the prior
region of \cref{sec:priors} and weighting each draw by its likelihood
of \cref{eq:binned_likelihood}.

For the three spectra of \cref{fig:cxsm1},
the efficiency of this direct sampling is of the order of $10^{-1}$,
$10^{-3}$ and $10^{-5}$ per point drawn from the prior region, respectively, and of the order of $10^{-3}$, $10^{-5}$ and $10^{-7}$ per point drawn uniformly in the box of \cref{eq:cxsm_box}, since only $0.6\%$ of these points lie in the prior region (see \cref{sec:priors}).
In contrast to the thermodynamic case, however,
here each draw requires the costly computation of the thermodynamic parameters
from the cxSM parameters described in \cref{sec:cxsm}. 

For the high-SNR spectrum, about $10^{7}$ such computations are needed for each effective sample, which makes the direct sampling computationally very expensive.
For the same reason, we restrict the MCMC sampling of the cxSM parameters to
the region of high likelihood determined beforehand with the GA, instead of
sampling the full prior region (see \cref{sec:mcmc}).

Alternatively, the posterior of the cxSM parameters can be obtained from
the NPE of the thermodynamic parameters, without training a separate network
for the cxSM parameters. This is possible because the predicted spectrum
depends on the cxSM parameters $\theta = \{m_S, \lambda_S, \lambda_{HS}\}$
only through the thermodynamic parameters
$\tau(\theta) = \{T_n, \alpha, \beta/H_*\}$. The likelihood of a cxSM
parameter point is therefore the likelihood of the corresponding
thermodynamic parameters. The NPE of the thermodynamic parameters, $q_\varphi(\tau \mid D)$,
approximates the posterior of the thermodynamic parameters, i.e.~the product
of this likelihood and the prior $\pi_\tau$ of \cref{eq:box_prior}. Dividing
by this prior yields the likelihood up to a constant, such that the
posterior of the cxSM parameters is given by
\begin{equation}
    p(\theta \mid D) \propto \pi(\theta)\,
    \frac{q_\varphi(\tau(\theta) \mid D)}{\pi_\tau(\tau(\theta))} \, ,
    \label{eq:reweight_thermal}
\end{equation}
with $\pi(\theta)$ the prior of the cxSM parameters. In practice, we apply
\cref{eq:reweight_thermal} to the points of the scan of the cxSM prior region
described in \cref{sec:priors}, for which the thermodynamic parameters have
already been computed. Each scan point is weighted by the ratio in
\cref{eq:reweight_thermal}, such that the analysis of a new spectrum requires
only evaluations of the NPE of the thermodynamic parameters, but no further computations of the
thermodynamic parameters. For the low- and medium-SNR spectra, the medians
of the one-dimensional posteriors obtained in this way (not shown in
\cref{fig:cxsm1}) differ from those of the NIS shown in the top row of
\cref{fig:cxsm1} by less than $0.04$ times the
half-width of the $68\%$ credible interval, and the widths of these intervals
agree to within $9\%$.

The precision of this approach is, however, limited by the number of
scan points. With the $5.5\times10^{4}$ scan points within the prior region
(the same points on which the NPE of the cxSM parameters is trained, see
\cref{sec:priors}), the ESS is of the order of $10^{4}$,
$10^{2}$ and only a few for the low-, medium- and high-SNR spectra of
\cref{fig:cxsm1}, respectively.
 The thin strip of the high-SNR posterior is thus covered by
only a few scan points with a significant weight. Resolving it would require
a considerably denser scan, which is again computationally expensive.

The bottom row of \cref{fig:cxsm1} shows the posteriors of the
thermodynamic parameters obtained from the same simulated measurements with
the NPE of the thermodynamic parameters, the corresponding NIS and the MCMC
analysis in the thermodynamic parameters. As for the spectra of
\cref{fig:th1}, the three methods agree very well. The injected parameters
lie within the $68\%$ credible regions, except for the low-SNR spectrum,
where they lie slightly outside the $68\%$ but within the $95\%$ credible
region in the plane of $\alpha$ and $\beta/H_*$. For correctly calibrated
posteriors, this is expected to occur in about one third of the cases.
The importance sampling
reaches efficiencies of $91\%$, $90\%$ and $90\%$ for the low-, medium- and
high-SNR spectra, respectively. In particular, for the high-SNR spectrum the
efficiency is much larger than the one found above for the cxSM parameters
with the same spectrum. The limitation of the NPE at high SNR discussed
above is therefore specific to the cxSM parameters. It does not arise from
the spectrum itself, but from the strong sensitivity of the thermodynamic
parameters to the cxSM parameters near the boundary of the region in which
the transition completes.

The NPE of the cxSM parameters and the NPE of the thermodynamic
parameters are trained independently. As a test of whether they describe
the same posterior, we use the samples of the NPE of the cxSM parameters as the proposal of
an importance sampling as in the NIS, but replace the exact likelihood by
the NPE of the thermodynamic parameters. The weight of each sample is then
$w(\theta) \propto q^{\tau}_\varphi(\tau(\theta) \mid D) /
q^{\theta}_\varphi(\theta \mid D)$, where $q^{\tau}_\varphi$ and
$q^{\theta}_\varphi$ denote the NPEs of the thermodynamic and the cxSM parameters,
respectively. This yields efficiencies of $86\%$, $72\%$ and $6.0\%$ for the
three spectra of \cref{fig:cxsm1}, compared with $86\%$, $67\%$ and $5.8\%$
for the NIS with the exact likelihood. Moreover, the medians of the thermodynamic
parameters computed from the samples of the NPE of the cxSM parameters agree with those of the
NIS to within $0.03$ in their decimal logarithms. The two networks are
therefore consistent with each other, and the NPE of the thermodynamic parameters can replace
the exact likelihood in the importance sampling without a noticeable loss of
accuracy.

\subsection{Inference of the bubble wall velocity}
\label{sec:vw}

We now treat the bubble wall velocity $v_w$ as a fourth free parameter
that has to be inferred from the data.
We apply a uniform prior in the range $v_w \in [0.1, 1]$ that covers deflagrations, hybrids and detonations.
For the SBI, we have 
trained two further networks
with the same architectures as the ones used in \cref{sec:results_thermal,sec:results_cxsm} (see details in \cref{sec:app_sbi}):
one for $\theta = \{T_n, \alpha, \beta/H_*, v_w\}$,
and one for $\theta = \{m_S, \lambda_S, \lambda_{HS}, v_w\}$.
Both networks were trained on a training set with
a total number of about
$10^5$ spectra.

\begin{figure}[t]
  \centering
  \includegraphics[width=0.32\textwidth]{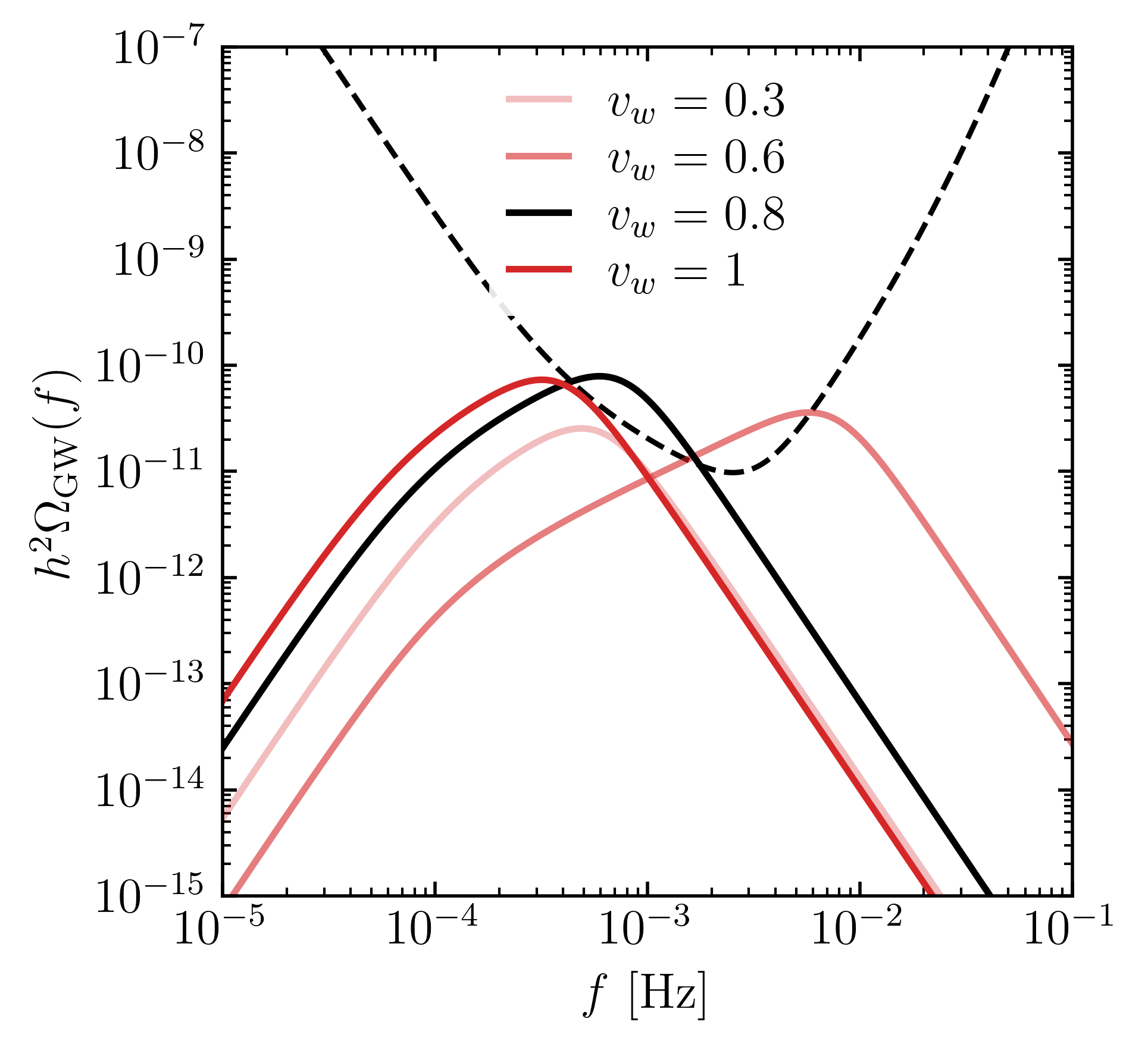}
  \caption{GW spectrum as a function of frequency for the bubble  wall velocities $v_w = 0.3$, $0.6$, $0.8$ (black) and $1$, with the other thermodynamic parameters kept fixed at $\alpha = 1$,
  $\beta/H_* = 100$ and $T_n = 60\GeV$. The black dashed curve shows the LISA
  noise spectrum $h^2\Omega_{\rm n}$.
  }
  \label{fig:vw_dep}
\end{figure}

In the template for the GW power spectrum~\cite{Caprini:2024hue},
$v_w$ enters through the mean bubble separation,
$H_* R_* \propto \max(v_w, c_s)/(\beta/H_*)$,
through the efficiency factor $\kappa(\alpha, v_w)$~\cite{Espinosa:2010hh},
and through the thickness of the sound shell, which sets the ratio of
the two break frequencies in the sound-wave template,
$f_2/f_1 \propto 1/\Delta_w$ with $\Delta_w = |v_w - c_s|/\max(v_w, c_s)$.
The effects of $v_w$ through $R_*$ and $\kappa$ shift the peak frequency
and change the amplitude, which can be compensated by changes in $\alpha$,
$\beta/H_*$ and $T_n$. The information one can infer on $v_w$ therefore stems mainly
from the spectral shape, i.e.~from $f_2/f_1$ and hence from $\Delta_w$.
A given value of $\Delta_w$ could be realized both by a subsonic
and by a supersonic wall velocity.
In such cases,
the posterior of $v_w$ has two branches.
At $v_w \sim c_s$ the actual spectrum depends on the hydrodynamics of hybrid walls~\cite{Gowling:2021gcy,Giese:2021dnw}, and the template does not provide a correct description of the GW spectra, so that constraints on $v_w$ in this range are not reliable.

\Cref{fig:vw_dep} shows how the GW spectrum changes with the bubble wall
velocity $v_w$, with the other thermodynamic parameters fixed to the values of the black curves in the top row of \cref{fig:analytic_GW_dep}.
In contrast to other macroscopic parameters,
the effect of $v_w$ on the spectrum is not monotonic.
For $v_w$ close to the speed of sound, e.g.~for $v_w = 0.6$, the sound shell is thin, such that the
two break frequencies are widely separated, and the spectrum becomes broad,
with its peak shifted to higher frequencies.
Further away from $c_s$, subsonic and supersonic wall velocities have similar values of $\Delta_w$ and therefore yield spectra of similar shape, which differ mainly in amplitude and peak frequency.

\begin{figure*}[t]
  \centering
  \includegraphics[width=0.48\textwidth]{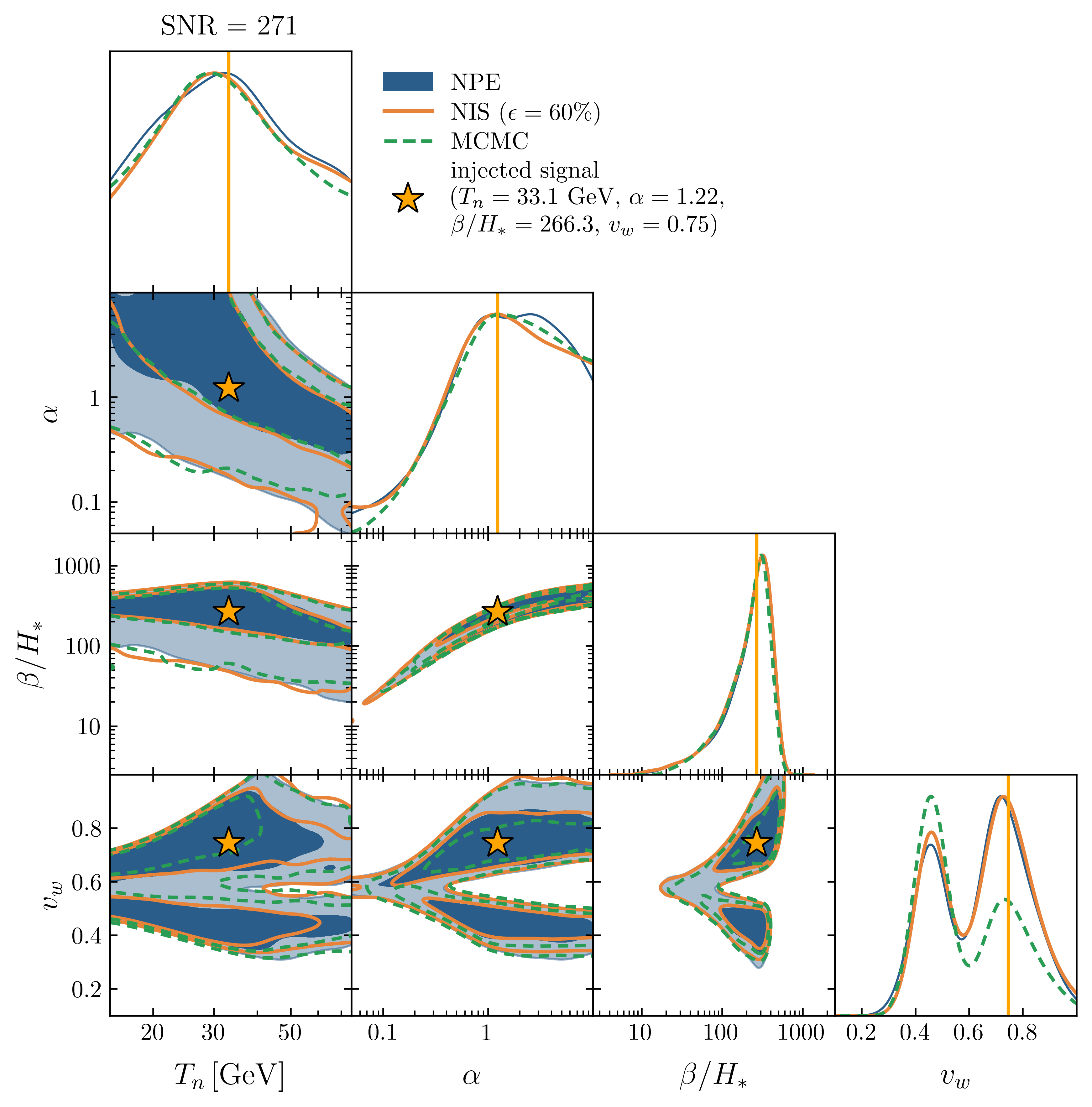}
  \includegraphics[width=0.48\textwidth]{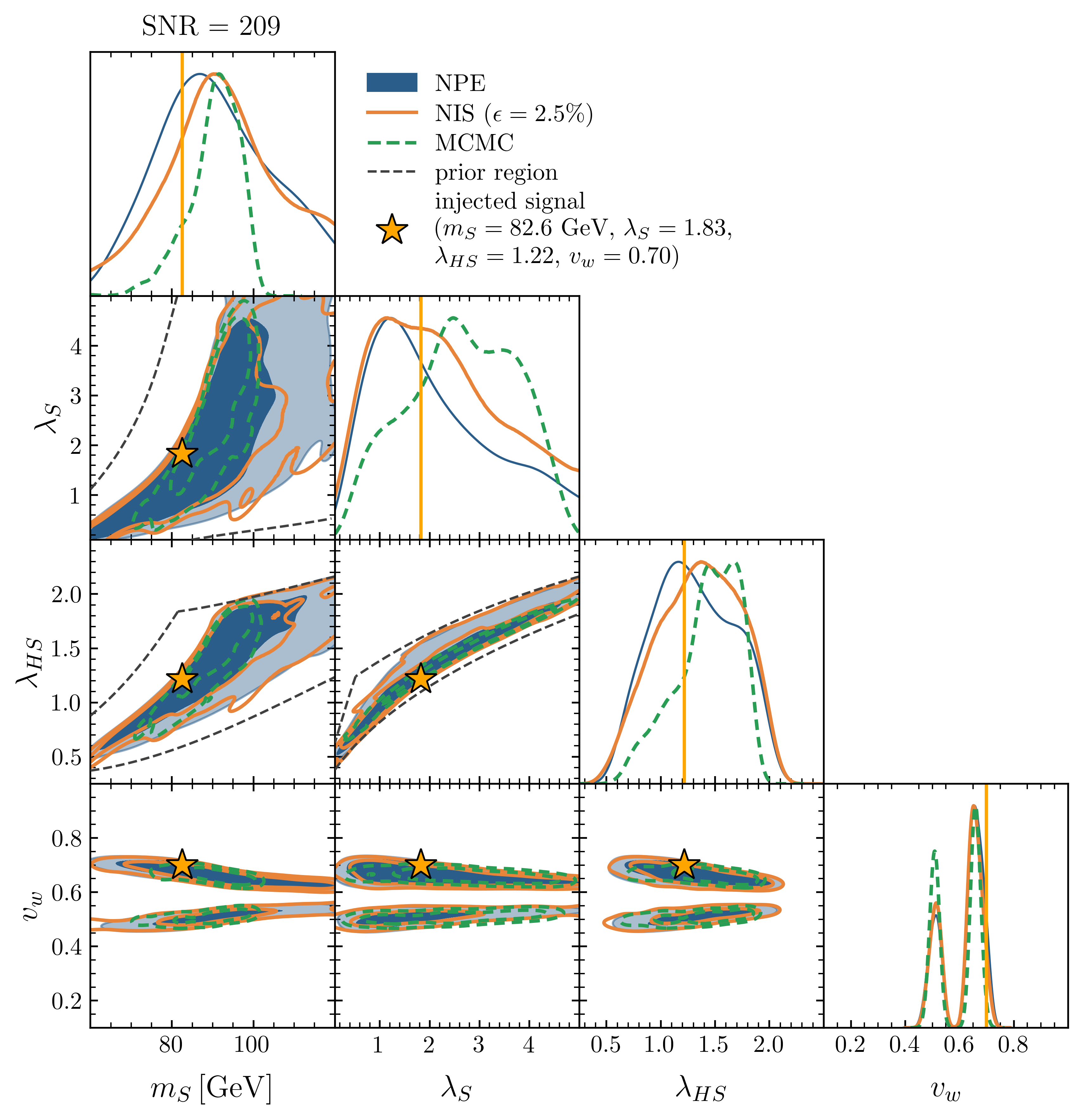}
  \caption{As in \cref{fig:th1}, but with the bubble wall velocity
    $v_w$ as an additional free parameter. Left: thermodynamic parameters
    $\{T_n, \alpha, \beta/H_*, v_w\}$ for a simulated LISA measurement
    with an injected signal of $\mathrm{SNR} = 271$. Right: cxSM
    parameters $\{m_S, \lambda_S, \lambda_{HS}, v_w\}$ for a simulated
    LISA measurement with an injected signal of $\mathrm{SNR} = 209$.}
  \label{fig:vw_posteriors}
\end{figure*}

The left panel of \cref{fig:vw_posteriors} shows, in the same way as
in \cref{fig:th1}, the one- and two-dimensional marginal posterior
distributions of
the thermodynamic parameters and $v_w$ for a strong signal with
$\mathrm{SNR} = 271$. The parameters of the injected signal
are $T_n = 33.1\GeV$, $\alpha = 1.22$,
$\beta/H_* = 266$ and $v_w = 0.75$.

The marginal posterior of $v_w$
has two peaks, one at the injected supersonic value and one at
$v_w \simeq 0.45$. The latter agrees with the subsonic wall velocity
$c_s(1-\Delta_w) \simeq 0.44$ that yields the same value of $\Delta_w$
as the injected one. The NPE and the NIS assign similar probabilities
to the two branches, such that the spectrum does not allow us to
distinguish between a subsonic and a supersonic wall. Within each
branch, $v_w$ is correlated with $T_n$, $\alpha$ and $\beta/H_*$, which
compensate the changes of the amplitude and of the peak frequency
caused by a change of $v_w$. As a consequence, $T_n$ and $\alpha$ are
only weakly constrained, extending over a large part of the prior
range, while $\beta/H_*$ is constrained from above,
$\beta / H_* \lesssim 500$.

The NPE agrees well with the NIS, which reaches an efficiency of
$\epsilon = 60\%$ for this spectrum.
The MCMC reproduces the location and the shape of both branches
of the posterior of $v_w$,
but assigns a larger probability to the subsonic branch. This
difference arises because the walkers of the MCMC rarely move between
the two well-separated branches, such that the relative weight of the
branches is not reliably sampled. The sampling of such multimodal
posteriors can be improved, for instance, by parallel
tempering~\cite{Swendsen:1986vqb,Earl:2005wvh,Vousden:2016eeu}, in
which additional chains sample a flattened version of the posterior
and can therefore move between the branches, or by nested
sampling~\cite{Skilling:2006gxv}, which
explores the full prior volume and therefore determines the relative
weights of separated modes more reliably. The NIS does not suffer from this
problem, since its samples are drawn independently from the NPE, which
covers both branches.

Beyond this individual spectrum, whose results are shown
in \cref{fig:vw_posteriors}, we test the calibration of the
network with free $v_w$ on $10^4$ simulated test spectra not used in
the training, following the procedure described in
\cref{sec:results_validation}. 
The $68\%$ credible regions contain the injected parameters for $69\%$ of these spectra, as expected for a
calibrated posterior.

The right panel of \cref{fig:vw_posteriors} shows, in the same way as
in \cref{fig:cxsm1}, the one- and two-dimensional marginal posterior
distributions of the cxSM parameters and $v_w$ for a strong signal with
$\mathrm{SNR} = 209$. The parameters of the injected signal are
$m_S = 82.6\GeV$, $\lambda_S = 1.83$, $\lambda_{HS} = 1.22$ and
$v_w = 0.70$. As for the thermodynamic parameters, the one-dimensional
marginal posterior of $v_w$ has two separate maxima, one at the supersonic
value $v_w \approx 0.66$, close to the injected value, and one at the
subsonic value $v_w \approx 0.50$. Both values correspond to approximately
the same sound-shell thickness $\Delta_w \approx 0.13$, such that the
spectrum again does not allow us to distinguish between a subsonic and a
supersonic wall.
In contrast to the thermodynamic parameters, however, each branch
is confined to a narrow interval of $v_w$. The posterior probability of
$v_w$ is concentrated in $0.47 \lesssim v_w \lesssim 0.53$ for the subsonic
branch and in $0.62 \lesssim v_w \lesssim 0.71$ for the supersonic branch,
whereas for the thermodynamic parameters (left panel of
\cref{fig:vw_posteriors}) the two branches extend over
$0.35 \lesssim v_w \lesssim 0.55$ and $0.6 \lesssim v_w \lesssim 1$,
respectively.

In \cref{sec:results_cxsm} we found that for a fixed and constant
bubble wall velocity,
$v_w = 0.8$, the posterior of a signal
with a comparable SNR is confined to a thin strip (see the right column
of \cref{fig:cxsm1}).
In contrast, here we find that the posterior fills a broad band
within the prior region.
The two-dimensional posterior
shown in the right panel of \cref{fig:vw_posteriors}
spans over values of $m_S$
from the lower to the upper bound of the prior range, and most of the prior
ranges of $\lambda_S$ and $\lambda_{HS}$.
As a consequence, a precise reconstruction
of the cxSM parameters is only possible if the bubble wall velocity is
known, for instance from a theoretical prediction
of $v_w$ within the model.

The interplay between $v_w$ and the cxSM parameters
differs from the one between $v_w$ and the thermodynamic parameters. In the
thermodynamic case (left panel of \cref{fig:vw_posteriors}),
a large change of $v_w$ can be compensated by
independent changes of $T_n$, $\alpha$ and $\beta/H_*$.
In the cxSM (right panel of \cref{fig:vw_posteriors}), the
thermodynamic parameters are functions of the three cxSM
parameters and cannot be varied independently. As a consequence of these
relations, only a small range of $v_w$ within each branch remains
compatible with the data, and $v_w$ can be inferred more precisely within
a given model than in the model-independent reconstruction of the
thermodynamic parameters.
This observation is in line with
Refs.~\cite{Gowling:2021gcy,Giese:2021dnw}, in which $v_w$ is well
determined once $T_n$ is fixed~\cite{Gowling:2021gcy}, or once $T_n$ and
$\beta/H_*$ are related to $\alpha$ by empirical
relations~\cite{Giese:2021dnw}.

The NPE reproduces the two branches of $v_w$ and the extent of the band
in the cxSM parameters found with the NIS, with small differences in the
one-dimensional distributions, for instance a peak of the distribution of
$m_S$ at slightly smaller values. The efficiency of the importance sampling
is only $\epsilon = 2.5\%$, such that the NIS contours are built from a
small number of effective samples and appear irregular. The MCMC analysis,
as in the left panel, assigns a larger probability to the subsonic branch
than the NIS. In addition, its posterior of the cxSM parameters is narrower
than the one of the NIS and does not extend to values of $m_S$ above about
$100\GeV$. We attribute these differences to an incomplete exploration of
the posterior by the walkers. For this scan, the sampling region cannot be
restricted with the GA (see \cref{sec:mcmc}), and the walkers have to
explore two narrow, well-separated branches of $v_w$ which each extend over
a broad band in the cxSM parameters. The methods for multimodal posteriors
discussed above for the thermodynamic parameters would also improve the
MCMC sampling in this case.

\subsection{Inference of the noise parameters}
\label{sec:noise}

To test the validity of treating the LISA noise amplitudes as fixed inputs with their fiducial values $A = 3\,{\rm fm\,s^{-2}/\sqrt{Hz}}$ and $P = 15\,{\rm pm/\sqrt{Hz}}$ (see \cref{sec:simulation} for discussion), we train two further networks, for the thermodynamic and for the cxSM parameters, in which $P$ and $A$ are inferred together with the physical parameters.
The training follows \cref{sec:app_sbi}, except that every training spectrum is simulated with its own noise amplitudes, drawn independently of the physical parameters from uniform priors within $\pm10\%$ of the fiducial values, $P \in [13.5, 16.5]\,{\rm pm/\sqrt{Hz}}$ and $A \in [2.7, 3.3]\,{\rm fm\,s^{-2}/\sqrt{Hz}}$.

\begin{figure*}[t]
  \centering
  \includegraphics[width=0.48\textwidth]{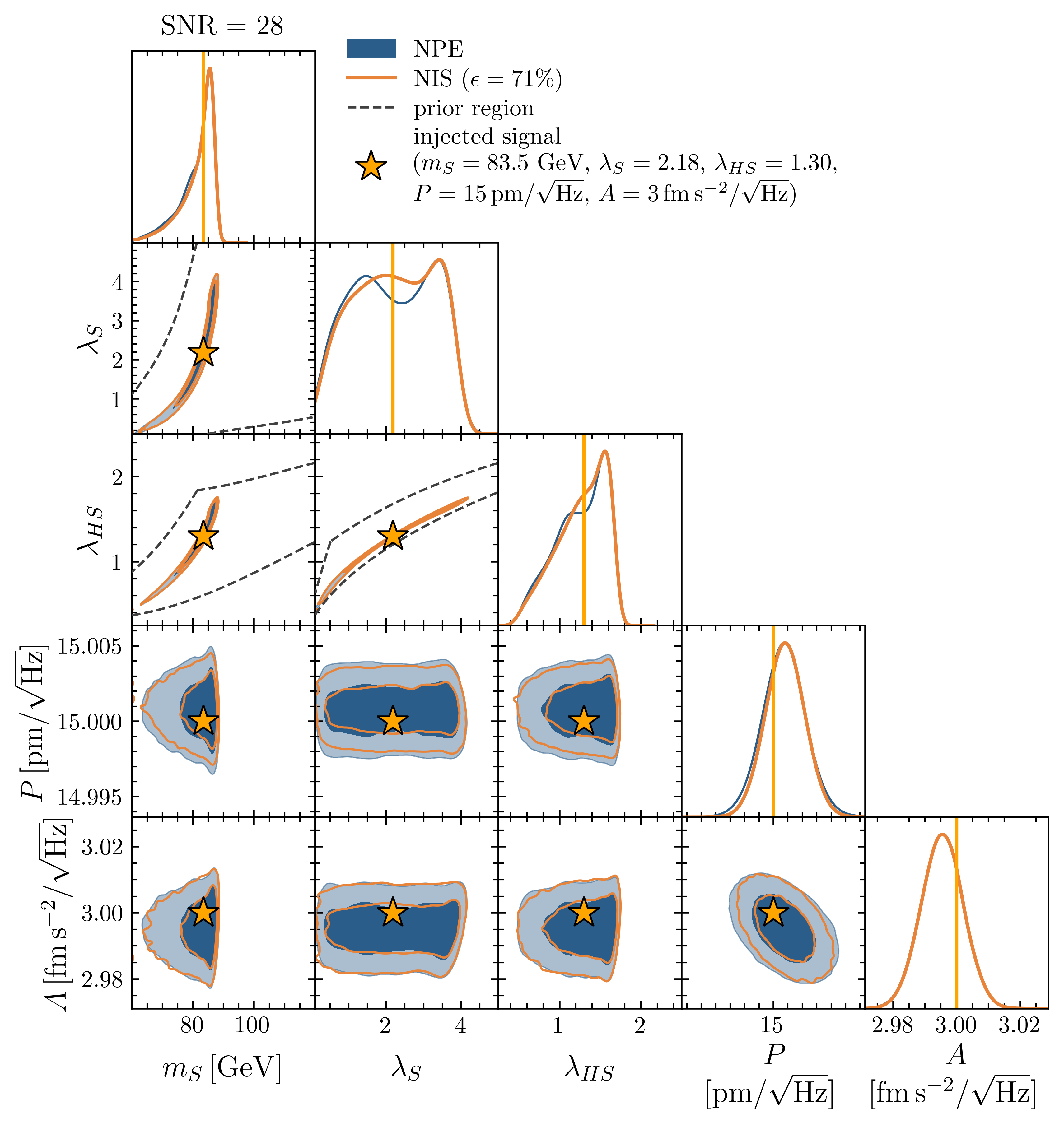}
  \includegraphics[width=0.48\textwidth]{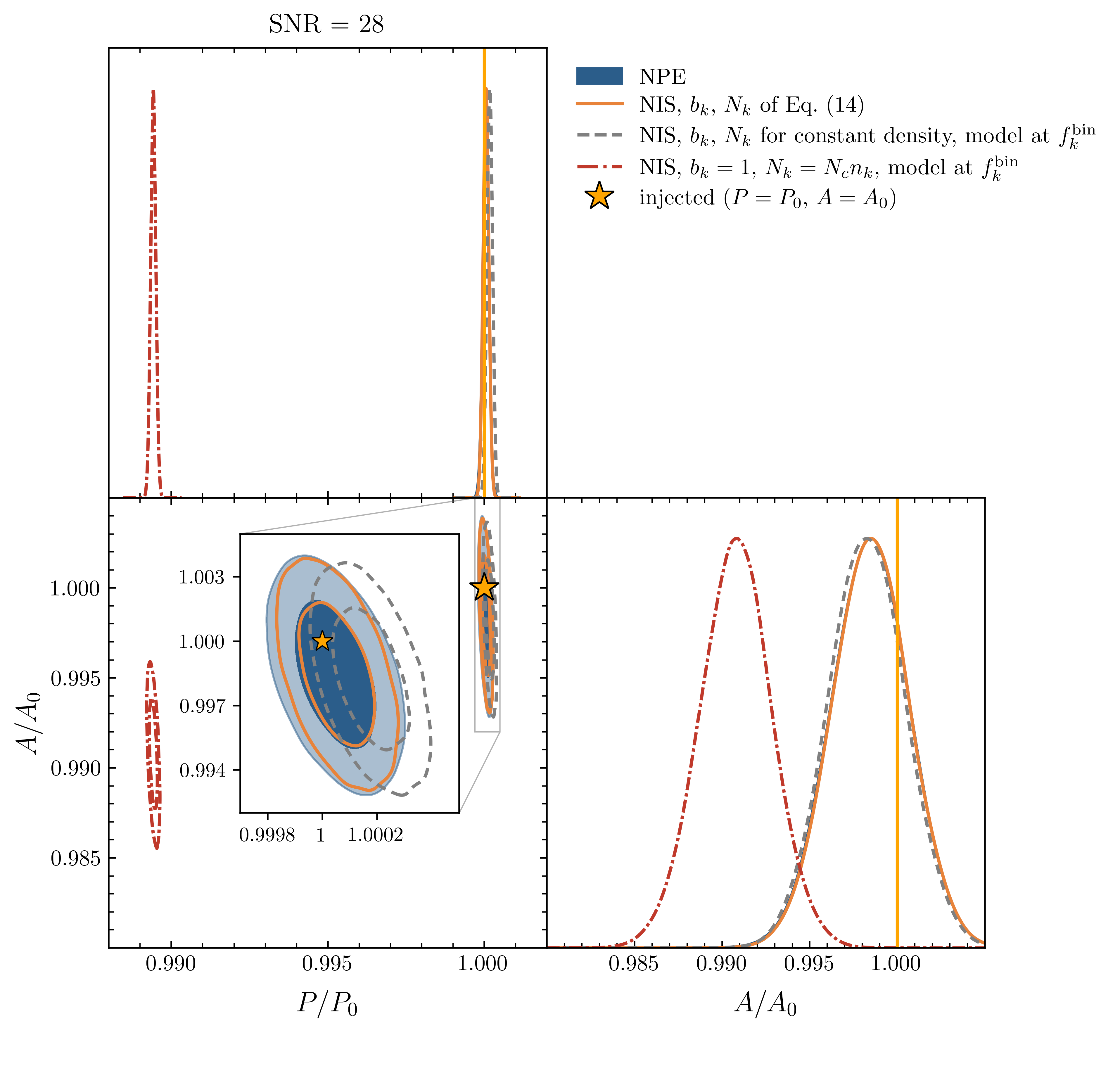}
  \caption{Posterior of the cxSM parameters and the noise amplitudes $P$ and $A$ for the medium-SNR spectrum of \cref{fig:cxsm1}, simulated with the fiducial noise amplitudes (\textit{left}). 
  Posterior of $P$ and $A$ for the same spectrum, with the cxSM parameters marginalized, from the NPE and from importance sampling with three versions of the binned likelihood: $b_k$ and $N_k$ of \cref{eq:bk_Nk}, $b_k$ and $N_k$ from a constant spectral density, and $b_k = 1$, $N_k = N_c n_k$ (\textit{right}).}
  \label{fig:APinf}
\end{figure*}

The left panel of \cref{fig:APinf} shows the posterior distribution for the medium-SNR cxSM spectrum of \cref{fig:cxsm1}.
The noise amplitudes are determined to about $0.01\%$ for $P$ and $0.2\%$ for $A$, far more precisely than the prior range, so that the prior has no effect on the result.
This reproduces the precision found for the full LISA pipeline in Ref.~\cite{Caprini:2024hue}, about $0.2\%$ for $A$, while $P$ is determined there about two times more precisely, as expected from their use of all three LISA's channels.
They are not correlated with the cxSM parameters, whose posterior agrees with that of the network with fixed noise.
The same holds for all six spectra of \cref{fig:th1,fig:cxsm1}, also when they are simulated with noise amplitudes offset from the fiducial values, which the networks recover within the mentioned precision.

The precision on $P$ makes the inference of the noise amplitudes a sensitive test of the binned likelihood discussed in \cref{sec:likelihood}.
Above about $10\,$mHz, where most of the degrees of freedom $N_k$ are, the spectrum is dominated by the OMS noise, whose amplitude scales as $P^2$.
A relative error $\delta$ in the predicted bin values $b_k \mathcal{M}_k$ at these frequencies is therefore absorbed by a shift of $P^2$ by about $\delta$.
The right panel of \cref{fig:APinf} illustrates this, where we reweight the importance-sampling proposals of the network with three versions of the binned likelihood.
With $b_k$ and $N_k$ of \cref{eq:bk_Nk}, the posterior is centered on the injected values and agrees with that of the NPE.
With $b_k$ and $N_k$ determined from simulations with a constant spectral density, which neglects the curvature of the spectrum within the bins, $P$ is shifted by $1.4\sigma$ for this spectrum.
With fixed noise weights, as used in Refs.~\cite{Flauger:2020qyi,Caprini:2024hue}, the binned spectrum would not be biased by the noise fluctuations, but the curvature effect would remain and evaluating the model at the bin frequency shifts $P$ by the same amount also for such a binning.
However, a comparison with the network would then require a network trained on spectra binned in the same way.
Without any correction, $b_k = 1$ and $N_k = N_c n_k$, the downward bias of the binned spectrum from the data-dependent weights shifts $P$ by about $1\%$, i.e.~by more than $100\sigma$, and $A$ by $3.5\sigma$.
In all three cases the posteriors of the GW parameters change by at most $0.2\sigma$, as the noise amplitudes absorb the error in the predicted noise level.

This supports the assumption of \cref{sec:simulation} that the noise amplitudes are constrained by the data independently of the GW signal.
In contrast, the networks trained with fixed $P$ and $A$ fail for spectra simulated with noise amplitudes that differ by only $1$--$2\%$ from their fiducial values.

This separation of noise and signal is achieved here with a
simplified data set.
LISA's time-delay interferometry (TDI) combines the
laser phases measured between its three spacecraft such that the dominant
laser frequency noise cancels, yielding the Michelson-like channels X, Y
and Z and from them the combinations A, E and T with uncorrelated
noise~\cite{Tinto:2020fcc}. 
We use only the X channel, in which signal
and noise enter together. 
In the full analysis, the noise amplitudes are
in addition constrained by the T channel, in which the GW signal largely
cancels~\cite{Flauger:2020qyi,Caprini:2024hue}, so that
their determination independently of the signal is expected to be even more robust.

\subsection{Validation and accuracy}
\label{sec:results_validation}

In order to test the coverage of the NPE over the whole prior, we perform an expected-coverage test~\cite{Hermans:2021rqv}, using, for each network, $10^4$ simulated test spectra that were not used in the training.
For each test spectrum, we compute the credible level of the highest-posterior-density region of the NPE whose boundary passes through the injected parameters, i.e.\ the posterior probability of all parameter points at which the posterior density is higher than at the injected parameters.
If the NPE is calibrated, these levels are uniformly distributed between zero and one, so that, for instance, $68\%$ of the test spectra have their injected parameters inside the $68\%$ credible region.
The possibility to verify the calibration over the whole prior in this
way is an important advantage of SBI. Since the NPE is amortized, it
provides the posteriors of all $10^4$ test spectra at negligible cost. The
same test for the MCMC analysis would require a separate MCMC run for each
test spectrum and would therefore be computationally prohibitive.

\begin{figure*}[t]
  \centering
  \includegraphics[width=0.32\textwidth]{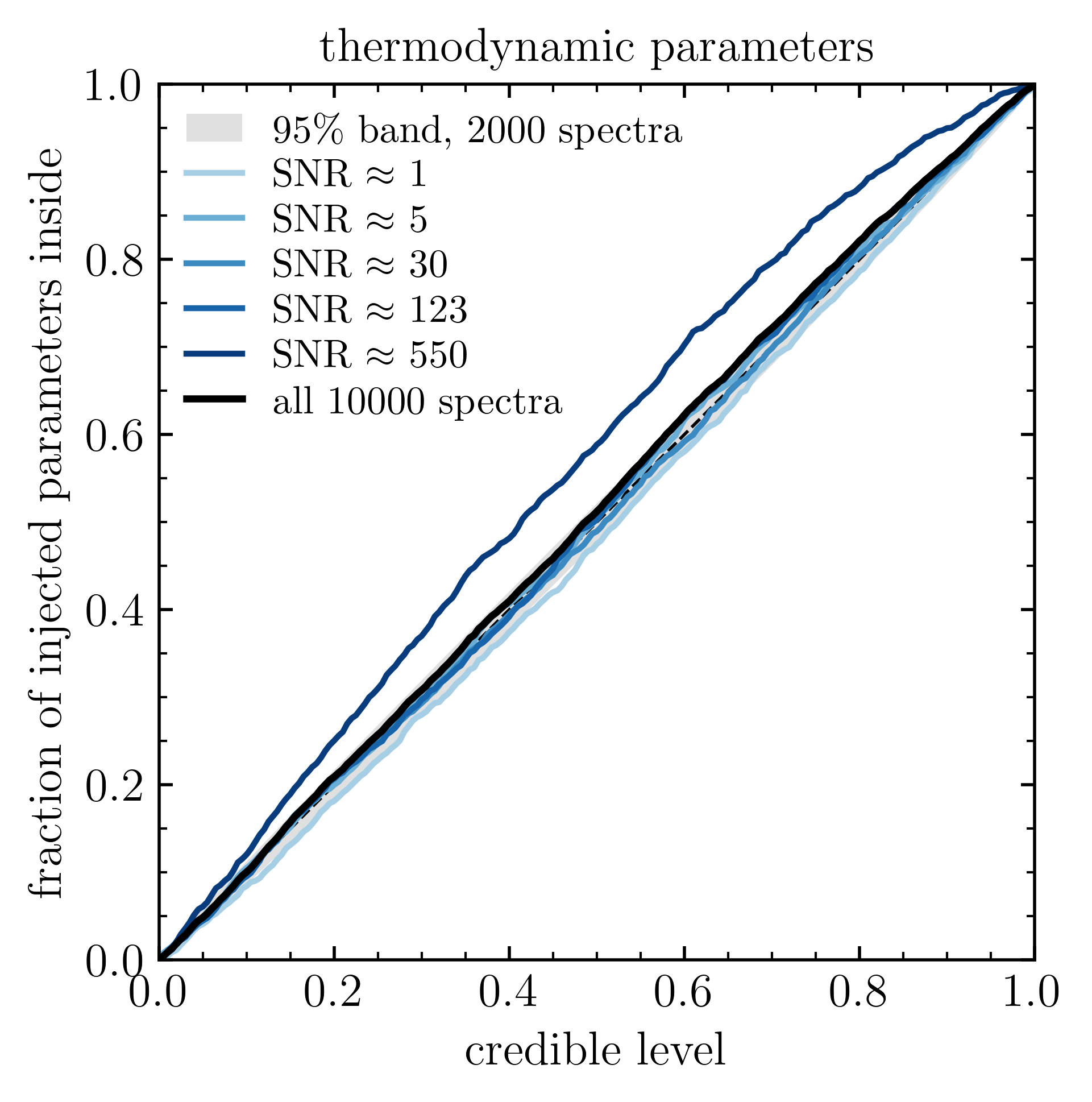}
  \includegraphics[width=0.32\textwidth]{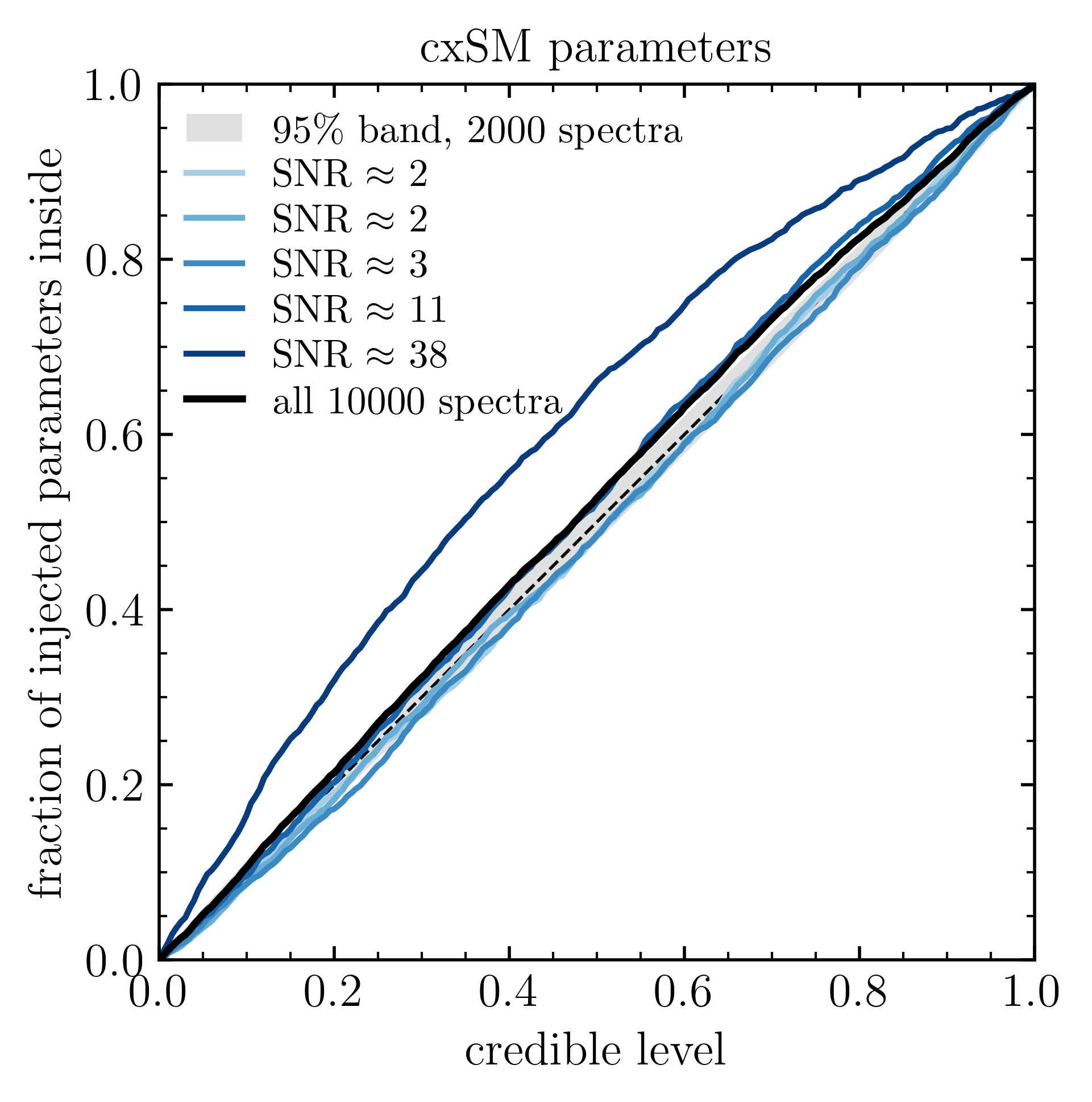}
  \caption{Calibration of the NPE for the thermodynamic \textit{(left)} and the cxSM \textit{(right)} parameters: fraction of test spectra whose injected parameters lie inside the highest-posterior-density region of a given credible level, for all test spectra (black) and for five equal groups ordered by the excess power of the data above the noise (see \cref{eq:excess}), labeled by their median SNR (blue). The gray bands represent the $95\%$ binomial range.}
  \label{fig:QQ}
\end{figure*}

In \cref{fig:QQ} we show, for every credible level, the fraction of test spectra whose injected parameters lie inside the region of that level (commonly called a coverage plot).

To see how the calibration depends on the signal strength, we order the test spectra by the excess power of the data over the noise,
\begin{equation}
    s^2 = \sum_k \left[ N_k \left( \frac{D^{\rm bin}_k}{b_k h^2 \Omega_{\rm n}(f^{\rm bin}_k)} - 1 \right)^2 - 1 \right] ,
    \label{eq:excess}
\end{equation}
with $b_k$ and $N_k$ given in \cref{eq:bk_Nk}, which vanishes on average for noise only and increases with the strength of the signal, and split them into five groups of $2000$ samples each.
Since $s^2$ is computed from the data alone, the grouping does not bias the test, unlike a grouping by the injected parameters, e.g.\ by the true SNR.

The left and right panels of \cref{fig:QQ} show the results for the
NPE of the thermodynamic parameters and of the cxSM parameters,
respectively. In each panel, the horizontal axis shows the credible level,
and the vertical axis shows the fraction of test spectra whose injected
parameters lie inside the credible region of that level. The black line
is obtained from all $10^4$ test spectra. Each of the five blue lines is
obtained only from the test spectra of one of the five groups. The shade of
blue ranges from light blue for the group with the smallest SNR to dark blue
for the group with the largest SNR, and each line is labeled by the median
SNR of its group. For a
calibrated NPE, all lines follow the diagonal (dashed line). A line above
the diagonal means that the credible regions contain the injected parameters
more often than their credible level states, i.e.\ that the posteriors of
the NPE are too broad (conservative). A line below the diagonal means that
the posteriors are too narrow (overconfident). The gray band shows the range
around the diagonal within which, at each credible level, the fraction
obtained from a group of 2000 test spectra lies with a probability of
$95\%$ for a calibrated NPE, as a result of the statistical fluctuations
due to the finite number of test spectra.

Overall, the $68\%$ credible regions contain the injected parameters for $70\%$ of the spectra with the thermodynamic network and for $71\%$ with the cxSM network.
The coverage is close to nominal in all groups except the one with the strongest signals, where it rises to $78\%$ and $81\%$, respectively.
Hence, the NPE is slightly conservative for the strongest signals. For
the cxSM parameters, this is in agreement with the high-SNR spectrum shown
in the top right panel of \cref{fig:cxsm1}. There, the two-dimensional
posterior regions of the NPE extend further in the direction perpendicular
to the degenerate direction than those of the NIS and MCMC
(see \cref{sec:results_cxsm} for details).

\begin{figure*}[t]
  \centering
  \includegraphics[width=0.32\textwidth]{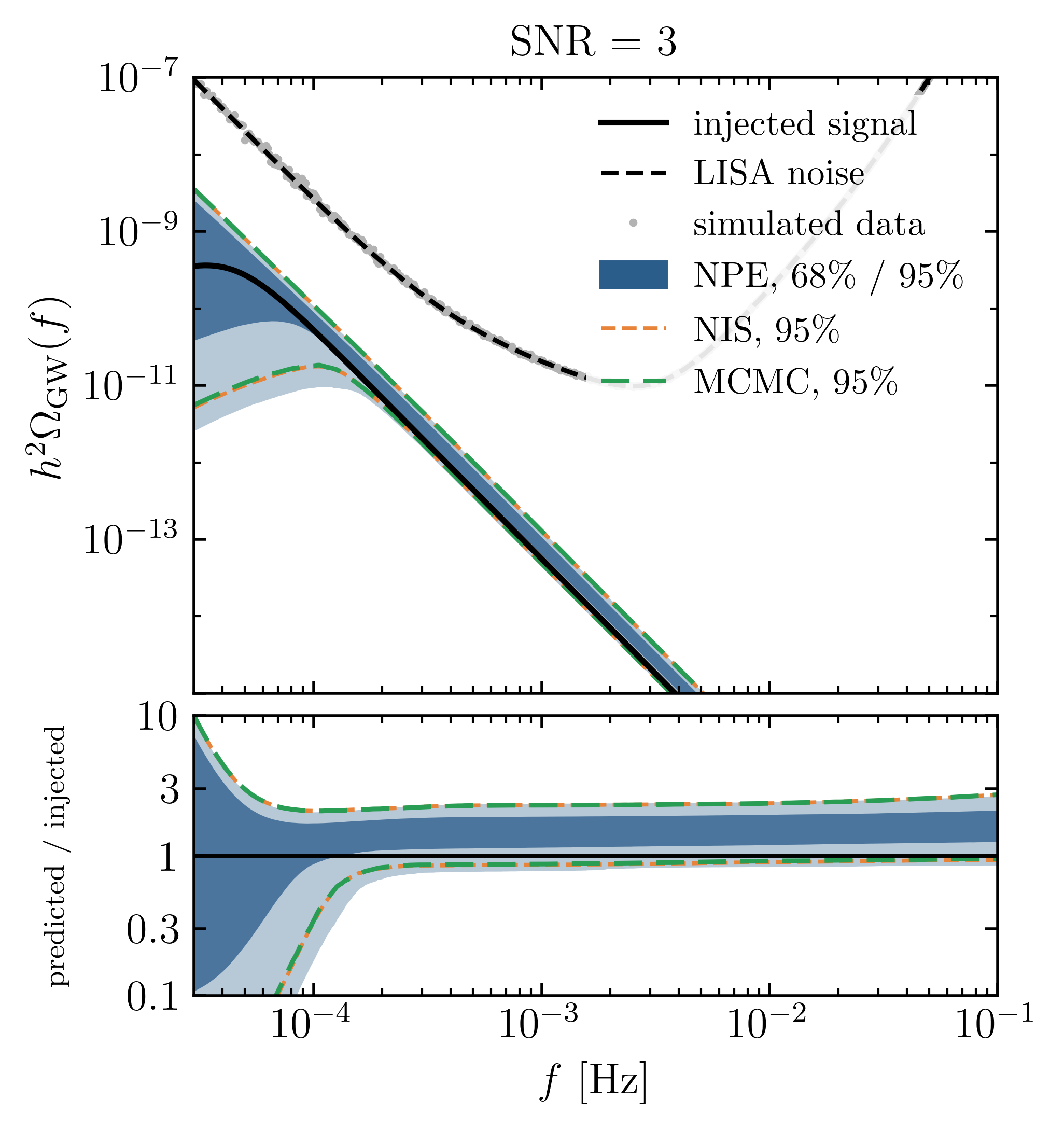}
  \includegraphics[width=0.32\textwidth]{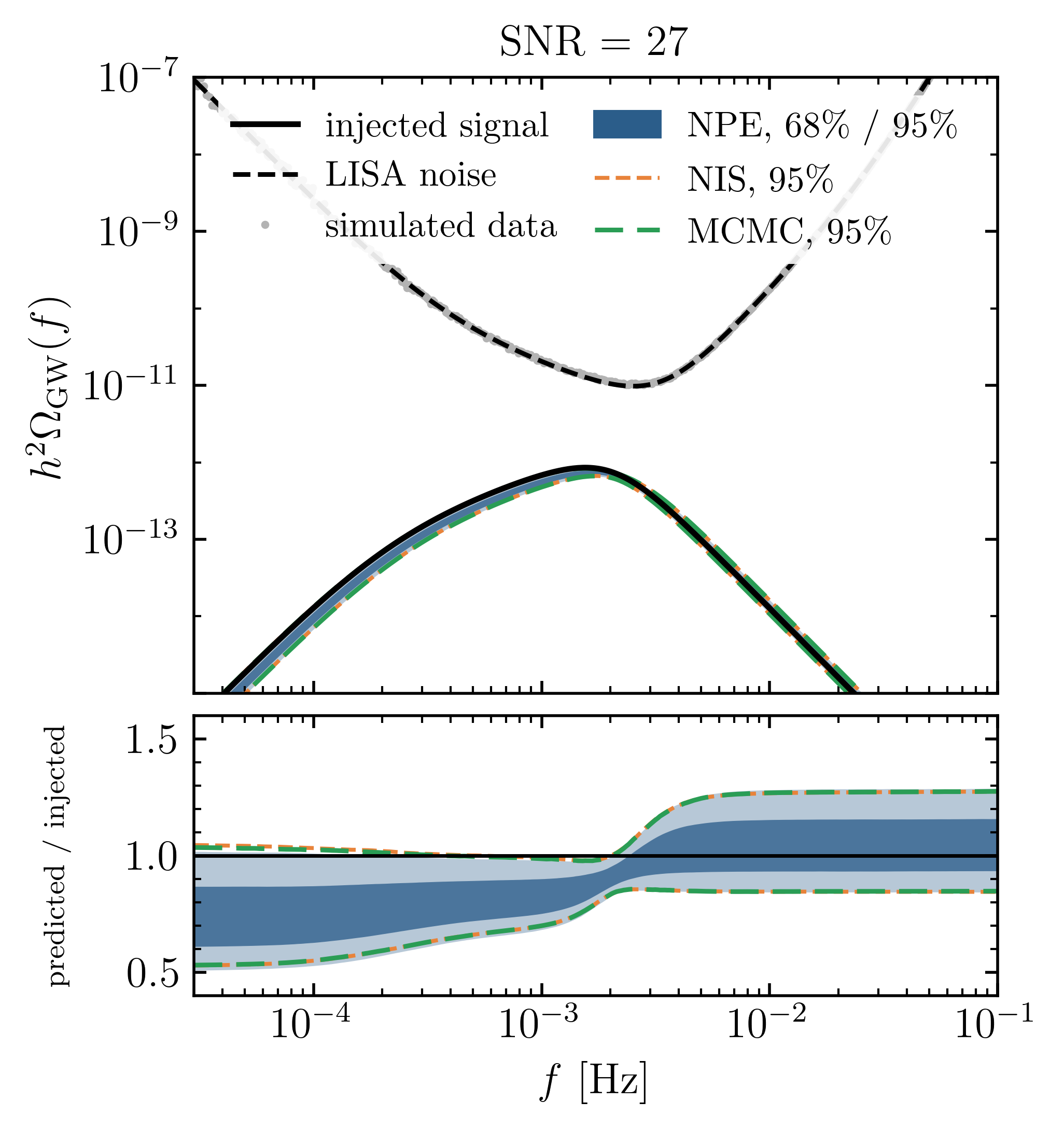}
  \includegraphics[width=0.32\textwidth]{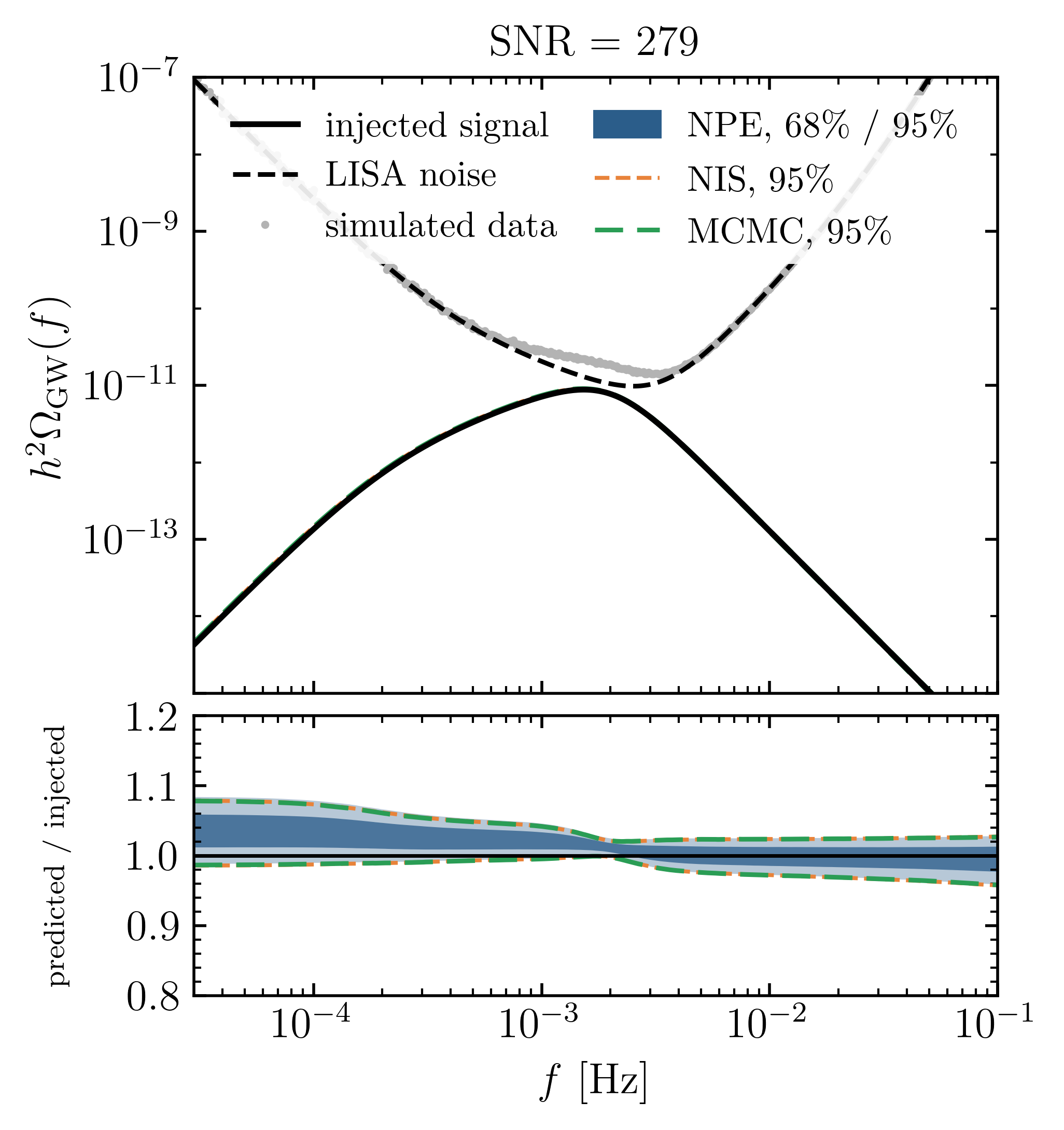}

  \includegraphics[width=0.32\textwidth]{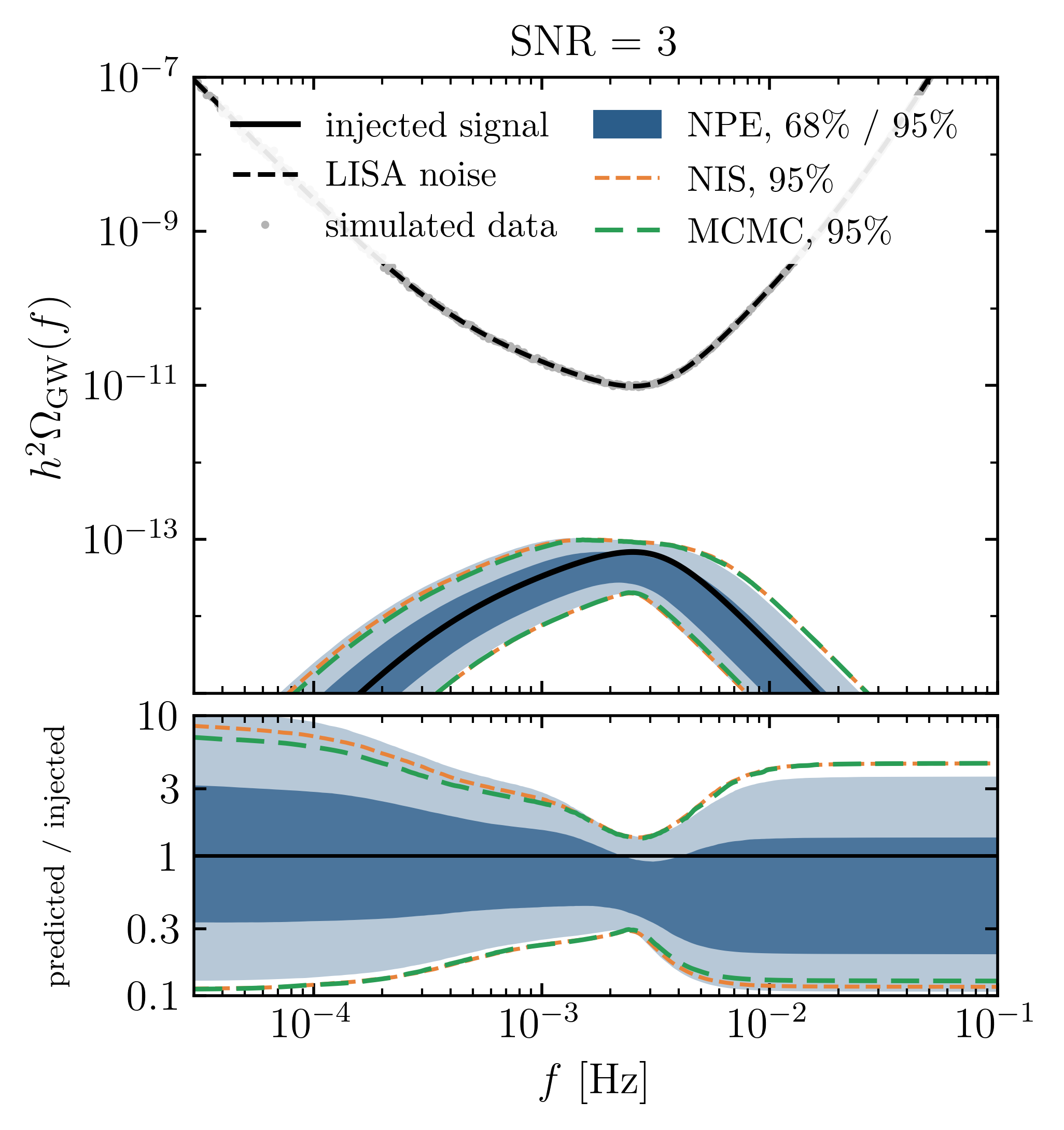}
  \includegraphics[width=0.32\textwidth]{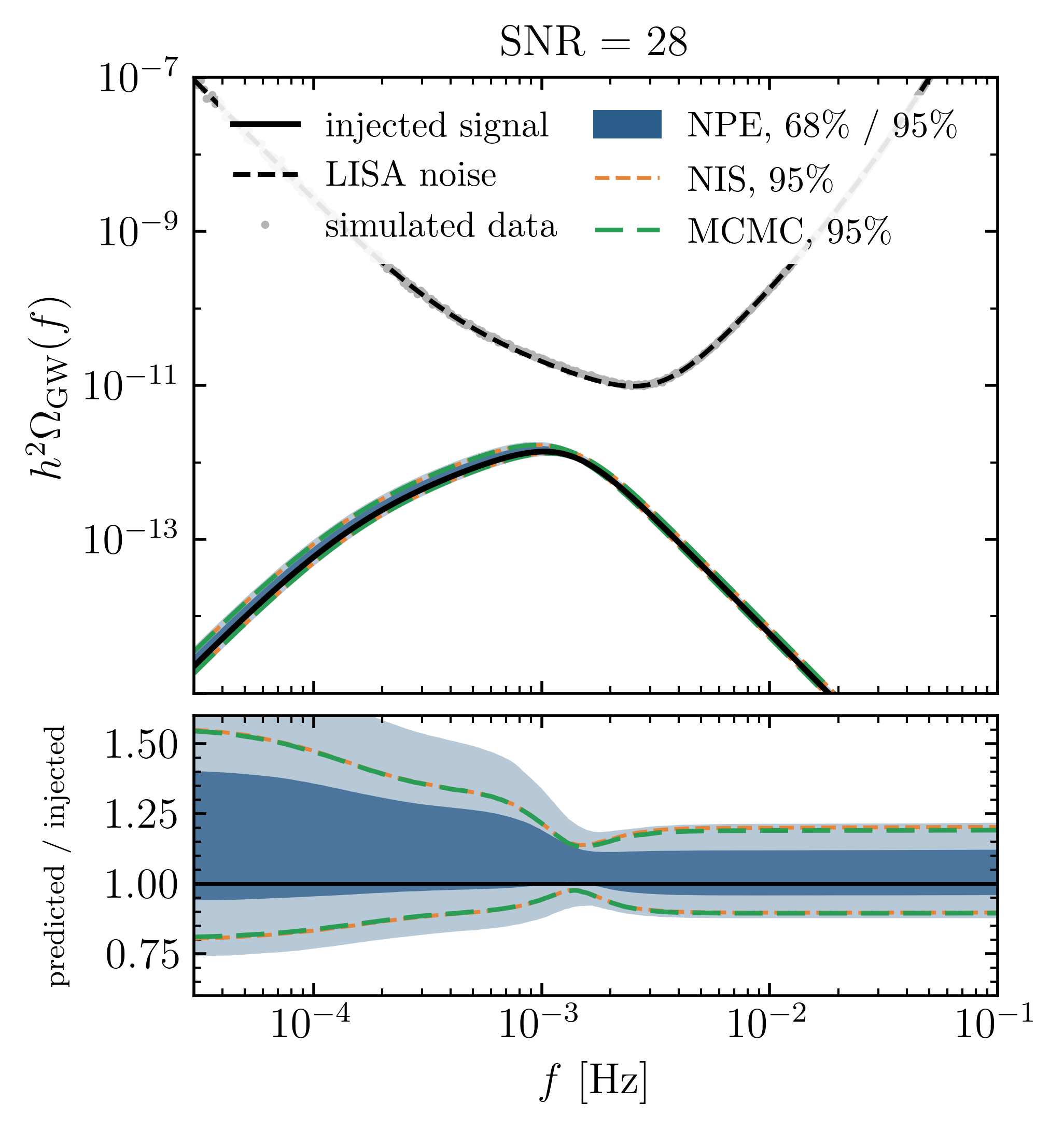}
  \includegraphics[width=0.32\textwidth]{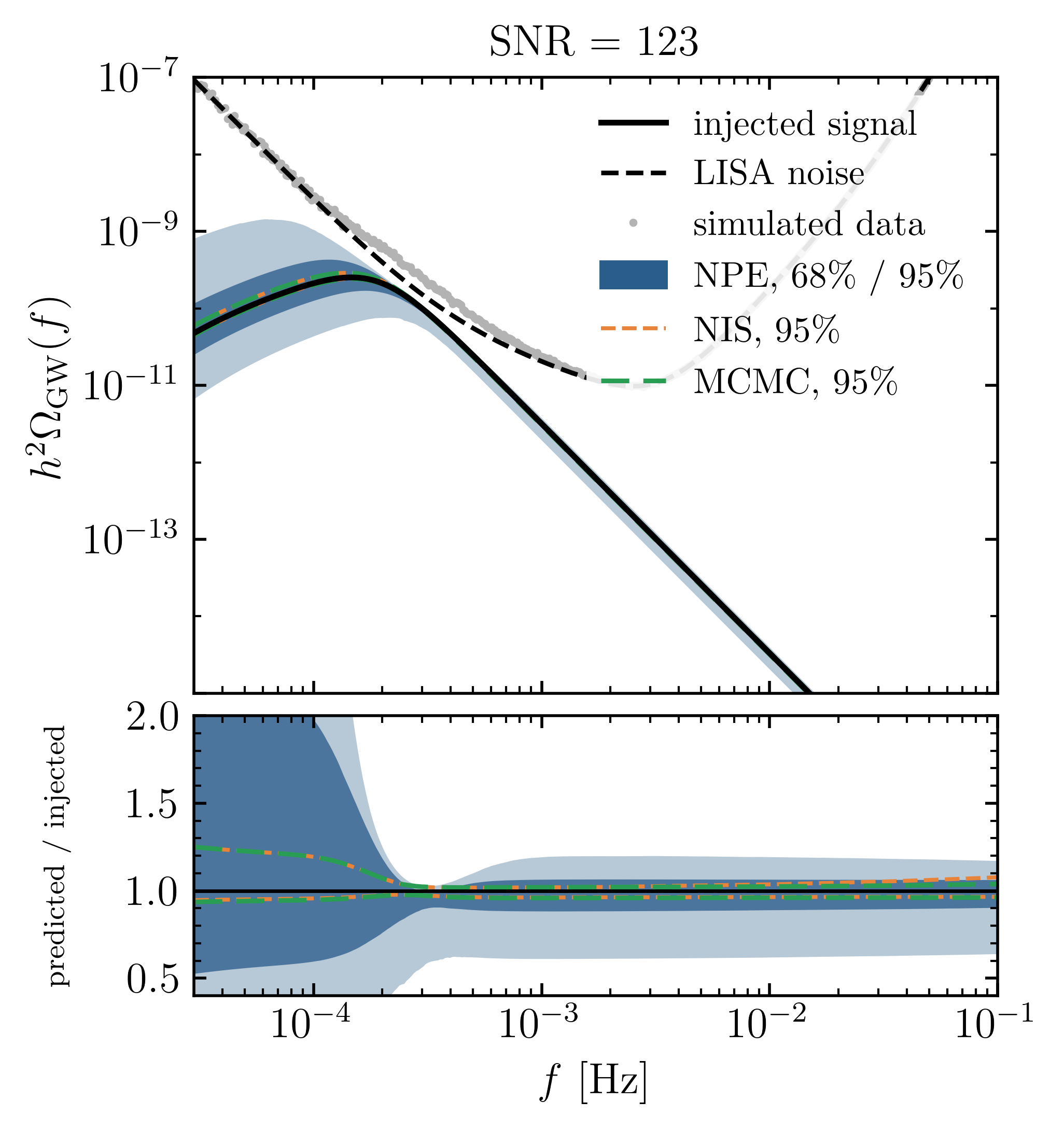}

  \caption{The reconstruction of the low-, medium- and high-SNR injected spectra of \cref{fig:th1} \textit{(top)} and \cref{fig:cxsm1} \textit{(bottom)}. Upper panels: GW signal predicted from the NPE posterior ($68\%$ and $95\%$ bands) and from the NIS and the MCMC ($95\%$, dashed), the injected signal, the simulated data and the LISA noise. Lower panels: ratio of the predicted to the injected signal.}
  \label{fig:ppc}
\end{figure*}

The coverage test checks the statistical properties of the posteriors, but not whether they describe the measured spectrum.
To this end,
we perform a signal reconstruction check: 
for parameters drawn from the posteriors determined by the NPE, NIS and MCMC, we compute the noise-free GW spectrum they predict and compare it with the injected noise-free signal.
The top row of \Cref{fig:ppc} shows the result of this check for the low-, medium- and
high-SNR spectra of \cref{fig:th1}, and the bottom row shows the
results for the spectra of \cref{fig:cxsm1}.
For each spectrum, we draw $10^4$ parameter points from each
posterior and evaluate the predicted noise-free spectrum at each bin
frequency. In the upper panel of each plot, the dark and light blue bands
show the central $68\%$ and $95\%$ intervals of the spectra predicted from
the NPE posterior at each frequency. The dashed orange and green lines show
the boundaries of the central $95\%$ intervals obtained from the NIS and the
MCMC posteriors, respectively. The black solid line shows the injected
signal, the gray dots show the simulated binned data divided by the bias factors, $D^{\rm bin}_k/b_k$ (see \cref{eq:bk_Nk}), and the black
dashed line shows the LISA noise spectrum. The lower panel of each plot
shows the ratio of the predicted spectra to the injected signal.

For the thermodynamic parameters (top row of \cref{fig:ppc}),
the predicted spectra agree with the injected signal to within $10$--$20\%$ around the peak frequency at an SNR of about $30$, and to within a few percent at an SNR of about $300$, even though the posteriors extend over a large range of the parameters along the degenerate directions.
For the low-SNR spectrum ($\mathrm{SNR} = 3$), the signal lies far
below the noise at all frequencies, and the predicted spectra are only
weakly constrained, in particular at the lowest frequencies around the
peak of the injected signal.
For all three spectra, the boundaries of the
central $95\%$ intervals obtained from the NIS and the MCMC posteriors
agree well with those of the $95\%$ band of the NPE.

For the cxSM parameters (bottom row of \cref{fig:ppc}), the spectra
predicted from the posteriors of the NPE, the NIS and the MCMC agree well
for the low- and medium-SNR spectra. For the medium-SNR spectrum, the
$95\%$ band of the NPE is only slightly broader than the central $95\%$
intervals obtained from the NIS and the MCMC posteriors, mainly below the
peak frequency. For the high-SNR spectrum, the NPE posterior extends
further in the direction perpendicular to the degenerate direction than the
NIS and the MCMC posteriors, as discussed for the top right panel of
\cref{fig:cxsm1} in \cref{sec:results_cxsm}. As a consequence, the $95\%$
band of the NPE is significantly broader than the central $95\%$ intervals
obtained from the NIS and the MCMC posteriors, which are confined to about $10\%$ around the injected signal.

When computing the bands of the NPE in \cref{fig:ppc}, we have to
take into account that not every NPE sample predicts a GW signal. The region
of the parameter space in which a first-order phase transition occurs ends
abruptly at a boundary, and the strongest signals arise close to this
boundary (see \cref{sec:results_cxsm}). Since the NPE density is a smooth
function of the parameters, a small fraction of the NPE probability lies
beyond this boundary. As a consequence, about $4\%$ of the NPE samples for the high-SNR cxSM
spectrum, and below $0.5\%$ for the low- and medium-SNR spectra,
correspond to parameter points for which no first-order phase transition
occurs, such that they predict no GW signal. These samples lie outside the
support of the prior, and we discard them when computing the bands, which
corresponds to restricting the NPE posterior to the support of the prior. In
the importance sampling, they receive zero weight.

In summary, although the posteriors in \cref{fig:th1,fig:cxsm1} extend
over large parameter ranges, the spectra predicted from them are confined to
narrow bands around the injected signal, as visible in \cref{fig:ppc}.
The large extent of the posteriors
therefore reflects the degeneracies in the map from the parameters to the GW
spectrum, and not an inaccurate reconstruction. For the same reason, the
signal reconstruction check cannot test whether the posterior probability is
distributed correctly along the degenerate directions. This is tested by the
coverage test and by the direct comparison of the posteriors of the NPE, the
NIS and the MCMC.

\subsection{Cost of the methods}
\label{sec:cost}

\begin{table*}[t]
  \centering
  \footnotesize
  \setlength{\tabcolsep}{6pt}
  \begin{tabular}{lcccccc}
    \toprule
    Method & One-time cost & \begin{tabular}{@{}c@{}}Cost per\\ spectrum\end{tabular} & Amortized & \begin{tabular}{@{}c@{}}Likelihood\\ needed\end{tabular} & ESS & \begin{tabular}{@{}c@{}}Peak signal,\\ $95\%$ CI\end{tabular} \\
    \midrule \noalign{\vskip 2pt}
    NPE                       & scan, $4$ h training      & $1$ s                                     & yes & no  & ---                                   & $0.88$--$1.33$ / $0.25$--$2.06$ \\ \noalign{\vskip 2pt}\arrayrulecolor{gray!50}\hline\arrayrulecolor{black}\noalign{\vskip 2pt}
    NIS                       & as NPE                      & \begin{tabular}{@{}c@{}}$10^{5}$ calls\\ ($26$\,core-h)\end{tabular} & no  & yes & $3.7\times10^{4}$ / $8.7\times10^{2}$ & $0.94$--$1.21$ / $0.96$--$1.14$ \\ \noalign{\vskip 2pt}\arrayrulecolor{gray!50}\hline\arrayrulecolor{black}\noalign{\vskip 2pt}
    \begin{tabular}{@{}l@{}}Thermodynamic\\ reweighting\end{tabular} & scan, $2$ h training      & $10$ s                                    & yes & no  & $78$ / $2$                            & $0.95$--$1.23$ / --- \\ \noalign{\vskip 2pt}\arrayrulecolor{gray!50}\hline\arrayrulecolor{black}\noalign{\vskip 2pt}
    Direct sampling           & scan                        & $10$ s                                    & yes & yes & $71$ / $2$                            & $0.94$--$1.21$ / --- \\ \noalign{\vskip 2pt}\arrayrulecolor{gray!50}\hline\arrayrulecolor{black}\noalign{\vskip 2pt}
    MCMC                      & ---                         & \begin{tabular}{@{}c@{}}$4.5\times10^{6}$ calls\\ ($10^{3}$\,core-h)\end{tabular} & no  & yes & $1.0\times10^{3}$ / $9.7\times10^{2}$ & $0.95$--$1.21$ / $0.96$--$1.14$ \\
    \noalign{\vskip 2pt}\bottomrule
  \end{tabular}
  \caption{Cost and accuracy of the reconstruction of the cxSM parameters for the medium- and high-SNR spectra of \cref{fig:cxsm1} (separated by a slash): amortized neural posterior estimation (NPE), the NPE as proposal for importance sampling with the exact likelihood (NIS), reweighting of the prior scan with the thermodynamic NPE (thermodynamic reweighting) or with the likelihood (direct sampling), and MCMC.
    The scan of the prior region ($9.3\times10^{6}$ simulator calls, about $5\times10^{2}$ core-hours) provides the training set of the cxSM NPE and the $5.5\times10^{4}$ points used for the reweighting and the direct sampling; the thermodynamic NPE is trained on spectra simulated directly from the thermodynamic parameters.
    A simulator call in the region of first-order transitions takes about $1$ core-second. The MCMC cost includes the GA runs that define its sampling region.
    Amortized: a new spectrum needs no further simulator calls. Likelihood needed: the method evaluates the likelihood of \cref{eq:binned_likelihood}.
    The last column gives the central $95\%$ credible interval (CI) of the reconstructed signal at the peak frequency, relative to the injected signal (see \cref{fig:ppc}); with an ESS of order one, the scan-based methods do not resolve the high-SNR posterior.}
  \label{tab:cost}
\end{table*}

The inference methods compared in \cref{sec:results} differ not only
in the accuracy of the reconstructed posteriors, but also in their
computational cost. This cost determines whether a method can be applied
to a large number of measured spectra, as required for instance for the
coverage test of \cref{sec:results_validation}.
The main computational cost in our scans stems from the
evaluation of the simulator, i.e.~of the map from the cxSM
parameters $\{m_S,\lambda_S,\lambda_{HS}\}$ to the
thermodynamic parameters $\{T_n, \alpha, \beta/H_*\}$.
Each evaluation takes about $1$ s on
a single CPU core.
The total number of simulator calls determines the cost of every method.
The sampling methods introduced in
\cref{sec:mcmc,sec:sbi} differ
in where their computational cost is incurred.
Since the likelihood-based MCMC calls the simulator for every
sampled parameter point, its full computational cost arises anew for
each injected signal.
In contrast, the NPE is trained only once on the simulations of
the prior scan, and then the posterior of every new injected signal is
obtained amortized without further simulator calls.
The NIS lies in between, because it evaluates the exact likelihood with
the NPE posterior as proposal, and therefore requires new simulator
calls for each injected signal.

\Cref{tab:cost} compares the computational cost and the accuracy of
the different methods for the medium- and high-SNR cxSM spectra of
\cref{fig:cxsm1}. In addition to the NPE, the NIS and the MCMC, it
contains the two methods based on the scan of the cxSM prior region
discussed in \cref{sec:results_cxsm}. These are the reweighting of the
scan points with the NPE of the thermodynamic parameters according to
\cref{eq:reweight_thermal} (thermodynamic reweighting), and the direct
sampling, in which each scan point is weighted by its likelihood. Since
the thermodynamic parameters of the scan points are computed only once,
neither of the two methods requires further simulator calls for a new
spectrum.
All timings were obtained on a laptop with an eight-core Apple M1 Pro processor, six performance and two efficiency cores, on which the networks were also trained.
We characterize the accuracy of a posterior by two quantities. The
ESS of \cref{eq:ess} gives the number of effectively independent samples
with which a method resolves the posterior. 
For the MCMC, the ESS is the number of samples after burn-in divided by the integrated autocorrelation time of the chains and multiplied by the efficiency of the reweighting to the uniform prior (see \cref{sec:app_mcmc_ga}). 
We compute the autocorrelation time from the autocorrelation function averaged over the walkers and take the largest value among the three cxSM parameters.

In addition, we give the central $95\%$ credible interval of the GW
signal at the peak frequency predicted from the posterior samples,
relative to the injected signal (see \cref{fig:ppc}). This interval
is indicative of 
how precisely the reconstructed parameters determine the observed
spectrum.

All methods except the MCMC share the one-time cost of the scan of
the prior region, which comprises $9.3\times10^{6}$ simulator calls, or
about $5\times10^{2}$ core-hours. Since most scan points do not feature a
first-order transition and are therefore evaluated quickly, the average
cost per call in the scan is smaller than in the region of first-order
transitions.
Once this cost is paid, the NPE yields the posterior of a new spectrum
in about a second, without further simulator calls and with any number
of independent samples. This makes the coverage tests over $10^{4}$
test spectra in \cref{sec:results_validation} possible.

The NPE is less accurate than the exact methods, in particular at high
SNR. 
There, the NPE strip is about twice as thick as the exact one
($0.04\GeV$ compared with $0.02\GeV$ in $m_S$, 
defined as the half-width of the central $68\%$ credible
interval of the deviation of $m_S$ from a smooth surface $m_S^{\rm ridge}(\lambda_S, \lambda_{HS})$ fitted to the NIS posterior), and the $95\%$ credible interval of
the peak signal is $0.25$--$2.06$, compared with $0.96$--$1.14$ for the
NIS and the MCMC.
The NIS removes this difference at the cost of $10^{5}$ simulator calls
per spectrum, or $26$ core-hours (about four hours on the laptop), with
an ESS of $3.7\times10^{4}$ and $8.7\times10^{2}$ for the medium- and
high-SNR spectra, respectively.
The thermodynamic reweighting and the direct sampling are almost as fast
as the NPE for a new spectrum, and at medium SNR they reproduce the
$95\%$ credible interval of the peak signal of the NIS and the MCMC. Their
resolution is, however, fixed by the density of the scan. Only about
$70$ to $80$ effective samples remain at medium SNR, and about two at high
SNR. With only two effective samples, the $95\%$ credible interval of the peak signal
cannot be determined, which is why the last column of \cref{tab:cost}
contains no entry for these two methods at high SNR. The NIS and the MCMC,
in contrast, constrain the peak signal to $0.96$--$1.14$ of the injected
one.

The MCMC reaches the same accuracy as the NIS, with an ESS of about
$10^{3}$ and $95\%$ credible intervals of the peak signal that agree to within $0.01$.
However, it
requires about $4.5\times10^{6}$ simulator calls per spectrum, including
the GA runs that define its sampling region, corresponding to about
$10^{3}$ core-hours. This is about $40$ times the cost of the NIS and
twice the cost of the one-time scan. Since this cost arises anew for every
spectrum, the coverage test over $10^{4}$ test spectra of
\cref{sec:results_validation} would require about $10^{7}$ core-hours
with the MCMC.
The NPE thus provides the fastest approximate posterior, and, as the
proposal of the NIS, the most efficient route to the exact one.

  \section{Conclusions}
\label{sec:conclusions}

The inference of the parameters of an underlying model from observations is a key step 
in confronting theory with Nature, and efficient tools for this inference are therefore of high importance.
A cosmological first-order phase transition, such as a first-order
electroweak phase transition in an extension of the SM,
would source a SGWB that could be
observed with LISA, offering the opportunity to probe BSM physics. 
In this work, we have studied SBI of
the parameters characterizing such a phase transition from a simulated LISA
measurement of the SGWB. We have applied SBI to the thermodynamic parameters
$\{T_n, \alpha, \beta/H_*\}$ and to the parameters
$\{m_S, \lambda_S, \lambda_{HS}\}$ of the cxSM. In both cases, the bubble wall velocity $v_w$ is
either fixed or jointly inferred with the other parameters. By direct
comparison with importance sampling using the exact likelihood and with a
likelihood-based MCMC analysis, we have shown
that NPE reliably infers the thermodynamic
parameters, the parameters of the cxSM, the bubble wall velocity, as well as
the LISA noise amplitudes.

We have demonstrated that SBI possesses several advantages over
likelihood-based sampling methods, the main ones being the amortization of
the trained network and the fact that no explicit expression for the
likelihood has to be derived or evaluated, since the posterior is
learned directly from simulations.
The computational cost of SBI arises only once, from the simulation of
the training data, which for the cxSM requires a scan of a sparse prior
region, and from the training of the network.
Once trained, the NPE returns the posterior of a new injected spectrum in about a second.
In contrast, a likelihood-based analysis such as the MCMC has to be
repeated from scratch for every new measured spectrum and requires a large
number of likelihood evaluations, each of which involves, for the cxSM, the
computation of the thermodynamic parameters from the model parameters.

Moreover, we have shown that
NPE provides an efficient proposal for importance sampling with the exact likelihood.
If the cxSM parameters are drawn uniformly from the prior, the
efficiency of the importance sampling is only $10^{-3}$--$10^{-7}$,
decreasing for a higher SNR. If they are drawn from
the NPE posterior instead, the efficiency is $5$--$90\%$.
For the thermodynamic parameters the NPE performs even better, with efficiencies of $80$--$95\%$ that 
remain high also for the strongest signals.
Thus, NIS, in which parameter points are
drawn from the NPE posterior and reweighted with the exact likelihood,
combines the speed of the network with the accuracy of a likelihood-based
analysis.
Similarly, samples from the NPE posterior could be used as starting
points for the walkers of an MCMC analysis. This would shorten the burn-in
phase and help the walkers to explore all regions of a multimodal posterior.

For the cxSM, small changes of the BSM parameters lead to large changes
of the predicted GW spectrum near the boundary beyond
which the transition does not complete, which is precisely where the
strongest signals arise. As a consequence, we have found that the posteriors
obtained from measurements with a large SNR are confined to thin, curved
strips in the parameter space. The length of these strips is set by the
degeneracies of the GW spectrum and shrinks only slowly with increasing SNR.
We have found that the NPE identifies these strips well, but provides a
conservative estimate of their thickness, which can, however, be
systematically improved with the NIS.
The good agreement between NPE and likelihood-based MCMC suggests 
that the parameter reconstruction is primarily limited by intrinsic degeneracies in 
the GW spectral templates rather than by the inference methodology.

We have found that, when the bubble wall velocity is inferred jointly
with the thermodynamic parameters, a change of $v_w$ can be largely compensated by
independent changes of $T_n$, $\alpha$ and $\beta/H_*$, such that $v_w$ is
only weakly constrained. Within the cxSM, the thermodynamic parameters are
functions of the model parameters and cannot be varied independently. As a
consequence, when $v_w$ is inferred jointly with the cxSM parameters, it is
more strongly constrained. Hence, an interpretation of the SGWB within
a specific model allows for a more precise reconstruction of $v_w$ than
a model-independent reconstruction in terms of the thermodynamic parameters.
Since a reliable theoretical prediction of $v_w$ requires a model-dependent
computation of the bubble wall dynamics, which is notoriously difficult,
measuring $v_w$ from the SGWB within a given model may be a more feasible
alternative.

Finally, we have demonstrated that inferring the noise amplitudes $P$ and $A$ together with the physical parameters determines them to $0.01\%$ and $0.2\%$, without correlation with the cxSM parameters, whereas networks trained at fixed noise fail already for offsets of $1$--$2\%$ in the injected spectra.

We expect SBI to be even more advantageous for models with more
parameters, such as more complicated scalar extensions of the SM, whose
posteriors have non-trivial regions and shapes, and for the simultaneous
inference of several GW signals of different origins, including
astrophysical foregrounds. In addition, SBI makes it possible to include
theoretical uncertainties of the GW prediction systematically and to study
their impact on the inferred model parameters, which we leave for future
work.
The inferred posteriors may moreover be combined with those of collider measurements and astronomical data, providing global constraints on new physics.

In summary, the amortized NPE provides reliable posteriors at negligible
computational cost per data set, and the NIS allows them to be verified and
corrected with the exact likelihood at a moderate additional cost. This
makes systematic analyses of a large number of SGWB measurements feasible,
both for LISA and for other future GW observatories.

  \section*{Acknowledgments}
We thank Mikael Chala for useful discussions and comments on the paper.
T.B.~and K.R.~acknowledge support
of the Spanish Agencia
Estatal de Investigaci\'on through
the grant IFT Centro de Excelencia Severo Ochoa CEX2020-001007-S
funded by MCIN/AEI/10.13039/501100011033. 
The work of A.D.~is supported by the European Research Council under
grant agreement n. 101230200.
The project that gave rise to these
results received the support of a
fellowship from the ``la Caixa''
Foundation (ID 100010434). The
fellowship code is LCF/BQ/PI24/12040018.
We acknowledge the use of \texttt{Claude Opus v.5.5} for assistance in
developing the code and in the writing of this paper.

  \appendix
  \crefalias{section}{appendix}
\crefname{appendix}{Appendix}{Appendices}

\section{Details on MCMC method}
\label{sec:app_mcmc}

This appendix collects the technical details of the MCMC analysis of \cref{sec:mcmc}. 
We discuss the sampler and its settings, and the preparation of the sampling region for the cxSM parameters.

\subsection{Sampler and settings}
\label{sec:app_mcmc_sampler}

In the affine-invariant ensemble sampling introduced in \cref{sec:mcmc},
each walker is updated with the so-called stretch move~\cite{Goodman:2010dyf},
which proposes a new
position on the straight line through the current position of the walker and
the position of another, randomly chosen walker.
As a consequence, the performance of the sampler is unchanged under linear
transformations of the parameters.

In order to reduce the runtime, we evaluate
the likelihood for many walkers in parallel on many CPU cores. To this end, we use the parallel
version of the stretch move~\cite{Foreman-Mackey:2012any}, in which the
ensemble is split into two
halves, and all walkers of one half are updated simultaneously using the
positions of the walkers of the other half. 
Since the positions used to
construct the proposals remain fixed while a half is updated, no walker is
moved based on the position of a walker that is updated at the same time.
This avoids race conditions between the parallel updates, which would
otherwise bias the sampled distribution away from the posterior.

We evolve 200 walkers for $10^4$ steps, both for the thermodynamic and for
the cxSM parameters. The first half of each chain is discarded as burn-in,
such that the posterior is estimated from the remaining
$200 \times 5 \times 10^{3} = 10^{6}$ samples.

\subsection{Sampling region for the cxSM parameters}
\label{sec:app_mcmc_ga}

Since only about $0.6\%$ of the points in the box of \cref{eq:cxsm_box}
predict a detectable signal (see \cref{sec:priors}), the region of high likelihood
used for the cxSM scans of \cref{sec:mcmc} is determined beforehand by
maximizing the likelihood 200 times with the GA implemented in \texttt{evortran}.

Each of these runs starts from
an independent random population of 500 parameter points within the box of
\cref{eq:cxsm_box}, which is evolved for at most 25 generations. A run
terminates once the $\ln \mathcal{L}$ of its best-fit point is so close to
the $\ln \mathcal{L}$ of the injected parameter point that the remaining
difference can be attributed to the experimental noise. Each run thus yields
one best-fit point. Due to the degeneracies among the cxSM parameters, the 200
best-fit points do not coincide but are distributed along the region of high
likelihood. In the plane of $\ln \lambda_{HS}$ and $\ln \lambda_S$, they lie
to a very good approximation on a straight line,
\begin{equation}
    \ln \lambda_S^{\rm fit}(\lambda_{HS}) = a \ln \lambda_{HS} + c \, ,
    \label{eq:lams_fit}
\end{equation}
whose slope $a$ and intercept $c$ we determine with a least-squares fit to
the best-fit points. We denote by $\sigma_{\rm res}$ the standard deviation of
the residuals $\ln \lambda_S - \ln \lambda_S^{\rm fit}(\lambda_{HS})$ of the
best-fit points.
Instead of $\lambda_S$, the MCMC then samples the deviation
$\delta = \ln \lambda_S - \ln \lambda_S^{\rm fit}(\lambda_{HS})$ from the
fitted line, together with $m_S$ and $\ln \lambda_{HS}$. The walkers are
restricted to $|\delta| \leq 10\, \sigma_{\rm res}$, and to the ranges of
$m_S$ and $\lambda_{HS}$ of \cref{eq:cxsm_box}.
The walkers are initialized uniformly within this region, and the prior is
uniform in $m_S$, $\ln \lambda_{HS}$ and $\delta \propto \ln \lambda_S$,
which corresponds to a prior
$\pi(m_S, \lambda_S, \lambda_{HS}) \propto 1/(\lambda_S \lambda_{HS})$ in the
cxSM parameters. In order to compare with the SBI, which is based on the
prior that is uniform in $\lambda_S$ and $\lambda_{HS}$ (see
\cref{sec:priors}), we reweight each MCMC sample by
$\lambda_S \lambda_{HS}$.

\section{Details on SBI method}
\label{sec:app_sbi}

This appendix collects the architecture and the training details of the networks used in \cref{sec:sbi}, together with the comparisons on which the choices are based.

\subsection{Embedding network}
\label{sec:app_embedding}

The data from $571$ frequency bins of \cref{eq:binning_def} is compressed down to $64$ summary features by an embedding network that is trained together with the neural flow.
We have compared two embedding architectures with similar numbers of parameters, as well as a flow without an embedding network.

The fully connected network (MLP) maps the bin values through $571 \to 256 \to 128 \to 64$ layers and has $1.9\times10^{5}$ parameters.
The convolutional network (CNN) acts along the frequency direction and consists of four convolutional layers with $16$, $32$, $32$ and $32$ channels with the kernel sizes $7$, $5$, $5$ and $3$, each followed by a SiLU activation and a max-pooling step that halves the resolution, so that the $571$ bins are reduced to $35$.
These are followed by two $1120 \to 128 \to 64$ layers, for a total of $1.6\times10^{5}$ parameters.

Trained on the same simulated pairs, the CNN improves the log-likelihood of the validation set by about $1.2$ per spectrum when the cxSM parameters are inferred, and by about $0.4$ for the thermodynamic parameters. 
Without an embedding network, the flow has to process all $571$ bins directly, which requires a larger network, gives wider posteriors and lowers the validation log-likelihood by about $3.0$ per spectrum compared with the CNN.
Given these observations, we note that employing the embedding of the input features
therefore
directly improves the output results.
We perform the SBI in \cref{sec:sbi} with the CNN embedding network, described above.

The CNN is, however, more expensive to train than the MLP, with the full training taking about ten times longer.
The evaluation of a trained network is negligible in both cases, so this cost is paid once and does not affect the per-spectrum times of \cref{tab:cost}.

\subsection{Flow and training}
\label{sec:app_flow}

\begin{figure*}[t]
  \centering
  \includegraphics[width=0.4\textwidth]{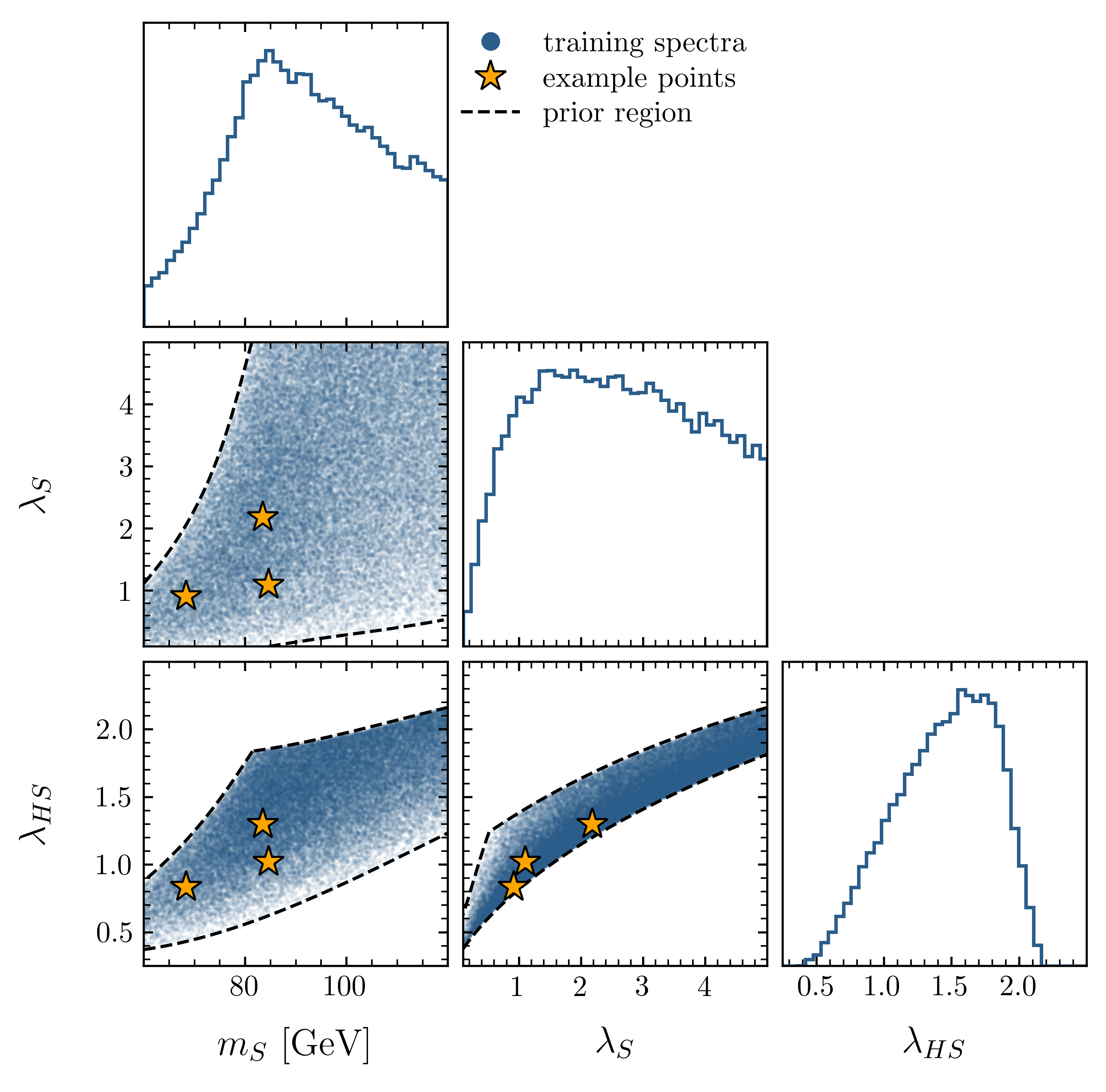}
  \includegraphics[width=0.4\textwidth]{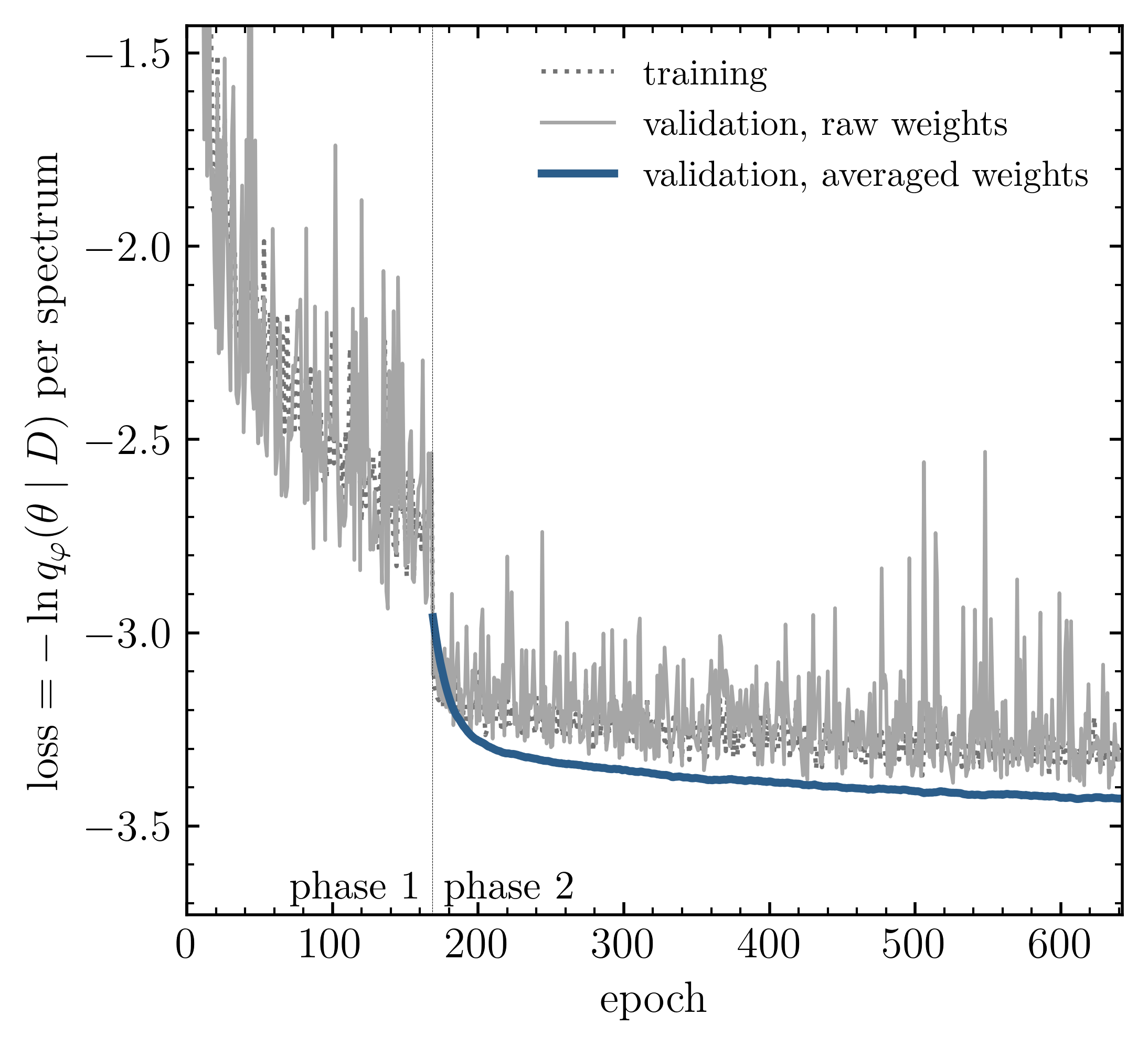}
  \caption{Training data for the cxSM NPE, drawn uniformly over the prior box and kept if they give a first-order phase transition with $\mathrm{SNR} > 1$, together with the boundary of the resulting prior region (dashed).
  The stars mark the example spectra of \cref{fig:cxsm1} \textit{(left)}. Training curve of the cxSM NPE used in \cref{sec:results} \textit{(right)}.}
  \label{fig:training}
\end{figure*}

The conditional density of \cref{eq:npe} is represented by a neural spline flow~\cite{Durkan:2019nsf} of ten coupling transforms with $128$ hidden features each, conditioned on the $64$ summary features encoding the measurement.
The parameters and the bin values are standardized to zero mean and unit variance over the training set before training, so that no further rescaling is applied inside the network, and the prior, described in \cref{sec:priors}, is formulated in the same standardized variables.

The weights are optimized with
adaptive moment estimation~(Adam)~\cite{Kingma:2014vow}
with a batch size of $512$, first with a learning rate of $5\times10^{-4}$ until the validation loss stops improving, and then, starting from the best weights, with $10^{-4}$ while keeping an exponential moving average (EMA) of the weights, $\bar\varphi' = 0.999\,\bar\varphi + 0.001\,\varphi$~\cite{Polyak:1992,moralesbrotons2024exponentialmovingaverageweights}.
The averaged network with the lowest validation loss is used in the analysis.
This second phase is needed because the posteriors are narrow strips along the degenerate directions, whose width is sensitive to the fluctuations of the weights during the optimization.
The two phases improve the validation log-likelihood by about $0.6$ per spectrum compared with training at a single constant learning rate, and the averaging removes the fluctuations between epochs.

To prevent the network from memorizing the specific noise realizations of the spectra in the training set, we draw a new noise realization at each training epoch.
As the noise is multiplicative (see \cref{sec:likelihood}), we proceed by multiplying the stored noise-free binned values of each training data point by a binned noise, simulated with the nominal LISA noise spectrum, normalized to unit mean in every bin, at no additional evaluations of the simulator.
This closes the gap between the calibration of the network on the training and on the test spectra.

We set the dropout rate to zero in all networks reported here.
A rate as small as $0.15$ in the embedding network leaves the training and validation losses almost unchanged, and yet widens the posteriors by up to a factor of two across the degenerate directions of \cref{sec:simulation}.
Similarly, small or zero dropout rates have been found to perform best for the flexible posterior parameterizations in Ref.~\cite{WagnerCarena:2020tbd}.

The network for the cxSM parameters is trained on about
$5\times10^{4}$ simulated spectra (see \cref{fig:training}), while the one for the thermodynamic parameters on about $9\times10^{4}$, with validation sets of
about $5\times10^{3}$ spectra.
The comparisons between networks and the coverage tests are carried out on
$10^{4}$
additional simulated spectra per network.
We use the \texttt{sbi} package~\cite{TejeroCantero:2020apm} for the SBI analysis, and \texttt{GetDist}~\cite{Lewis:2019xzd} for the posterior contours.

\subsection{Neural Importance Sampling}
\label{sec:NIS}

In \cref{sec:results} the NIS posteriors are obtained by calculating the weights per \cref{eq:is_weights} with the binned likelihood of \cref{eq:binned_likelihood}, and the same prior as for the training of the corresponding NPE (see \cref{sec:priors}).
The proposal for the sampling is a mixture of the NPE posterior ($60\%$), a widened NPE, obtained by drawing the base distribution of the neural flow with a standard deviation of $1.5$ instead of $1$ ($30\%$), and a uniform distribution over the prior support ($10\%$).
We introduce the two broader components to cover regions in which the NPE might underestimate the posterior, while keeping the density of the proposal known exactly.
We draw $N = 10^{5}$ proposals for each network posterior, for which every proposal requires an evaluation of the simulator.

The efficiency $\epsilon = \mathrm{ESS}/N$ reported in the figures refers to the NPE alone as the proposal, i.e.\ to the weights $w_i \propto \mathcal{L}(\theta_i)\,\pi(\theta_i)/q_\varphi(\theta_i \mid D)$ of samples $\theta_i$ drawn from the NPE, as in Ref.~\cite{Dax:2022pxd}.
It thus measures the quality of the NPE, and equals one for an exact posterior, whereas the mixture proposals used for the NIS contours have a lower efficiency by construction.
It is obtained from the draws of the NPE component of the mixture, which are exact samples of the network and require no additional evaluations of the simulator.
Samples outside the support of the prior, such as those without a first-order transition, enter with zero weight and thus lower $\epsilon$.

  \bibliography{lit,lit_extra}

\begin{thebibliography}{126}%
\makeatletter
\providecommand \@ifxundefined [1]{%
 \@ifx{#1\undefined}
}%
\providecommand \@ifnum [1]{%
 \ifnum #1\expandafter \@firstoftwo
 \else \expandafter \@secondoftwo
 \fi
}%
\providecommand \@ifx [1]{%
 \ifx #1\expandafter \@firstoftwo
 \else \expandafter \@secondoftwo
 \fi
}%
\providecommand \natexlab [1]{#1}%
\providecommand \enquote  [1]{``#1''}%
\providecommand \bibnamefont  [1]{#1}%
\providecommand \bibfnamefont [1]{#1}%
\providecommand \citenamefont [1]{#1}%
\providecommand \href@noop [0]{\@secondoftwo}%
\providecommand \href [0]{\begingroup \@sanitize@url \@href}%
\providecommand \@href[1]{\@@startlink{#1}\@@href}%
\providecommand \@@href[1]{\endgroup#1\@@endlink}%
\providecommand \@sanitize@url [0]{\catcode `\\12\catcode `\$12\catcode
  `\&12\catcode `\#12\catcode `\^12\catcode `\_12\catcode `\%12\relax}%
\providecommand \@@startlink[1]{}%
\providecommand \@@endlink[0]{}%
\providecommand \url  [0]{\begingroup\@sanitize@url \@url }%
\providecommand \@url [1]{\endgroup\@href {#1}{\urlprefix }}%
\providecommand \urlprefix  [0]{URL }%
\providecommand \Eprint [0]{\href }%
\providecommand \doibase [0]{https://doi.org/}%
\providecommand \selectlanguage [0]{\@gobble}%
\providecommand \bibinfo  [0]{\@secondoftwo}%
\providecommand \bibfield  [0]{\@secondoftwo}%
\providecommand \translation [1]{[#1]}%
\providecommand \BibitemOpen [0]{}%
\providecommand \bibitemStop [0]{}%
\providecommand \bibitemNoStop [0]{.\EOS\space}%
\providecommand \EOS [0]{\spacefactor3000\relax}%
\providecommand \BibitemShut  [1]{\csname bibitem#1\endcsname}%
\let\auto@bib@innerbib\@empty
\bibitem [{\citenamefont {Witten}(1984)}]{Witten:1984rs}%
  \BibitemOpen
  \bibfield  {author} {\bibinfo {author} {\bibfnamefont {E.}~\bibnamefont
  {Witten}},\ }\bibfield  {title} {\emph {\bibinfo {title} {{Cosmic Separation
  of Phases}}},\ }\href {https://doi.org/10.1103/PhysRevD.30.272} {\bibfield
  {journal} {\bibinfo  {journal} {Phys. Rev. D}\ }\textbf {\bibinfo {volume}
  {30}},\ \bibinfo {pages} {272} (\bibinfo {year} {1984})}\BibitemShut
  {NoStop}%
\bibitem [{\citenamefont {Hogan}(1986)}]{Hogan:1986qda}%
  \BibitemOpen
  \bibfield  {author} {\bibinfo {author} {\bibfnamefont {C.~J.}\ \bibnamefont
  {Hogan}},\ }\bibfield  {title} {\emph {\bibinfo {title} {{Gravitational
  radiation from cosmological phase transitions}}},\ }\href@noop {} {\bibfield
  {journal} {\bibinfo  {journal} {Mon. Not. Roy. Astron. Soc.}\ }\textbf
  {\bibinfo {volume} {218}},\ \bibinfo {pages} {629} (\bibinfo {year}
  {1986})}\BibitemShut {NoStop}%
\bibitem [{\citenamefont {Kosowsky}\ \emph {et~al.}(1992)\citenamefont
  {Kosowsky}, \citenamefont {Turner},\ and\ \citenamefont
  {Watkins}}]{Kosowsky:1992rz}%
  \BibitemOpen
  \bibfield  {author} {\bibinfo {author} {\bibfnamefont {A.}~\bibnamefont
  {Kosowsky}}, \bibinfo {author} {\bibfnamefont {M.~S.}\ \bibnamefont
  {Turner}},\ and\ \bibinfo {author} {\bibfnamefont {R.}~\bibnamefont
  {Watkins}},\ }\bibfield  {title} {\emph {\bibinfo {title} {{Gravitational
  Waves from First Order Cosmological Phase Transitions}}},\ }\href
  {https://doi.org/10.1103/PhysRevLett.69.2026} {\bibfield  {journal} {\bibinfo
   {journal} {Phys. Rev. Lett.}\ }\textbf {\bibinfo {volume} {69}},\ \bibinfo
  {pages} {2026} (\bibinfo {year} {1992})}\BibitemShut {NoStop}%
\bibitem [{\citenamefont {Kamionkowski}\ \emph {et~al.}(1994)\citenamefont
  {Kamionkowski}, \citenamefont {Kosowsky},\ and\ \citenamefont
  {Turner}}]{Kamionkowski:1993fg}%
  \BibitemOpen
  \bibfield  {author} {\bibinfo {author} {\bibfnamefont {M.}~\bibnamefont
  {Kamionkowski}}, \bibinfo {author} {\bibfnamefont {A.}~\bibnamefont
  {Kosowsky}},\ and\ \bibinfo {author} {\bibfnamefont {M.~S.}\ \bibnamefont
  {Turner}},\ }\bibfield  {title} {\emph {\bibinfo {title} {{Gravitational
  radiation from first order phase transitions}}},\ }\href
  {https://doi.org/10.1103/PhysRevD.49.2837} {\bibfield  {journal} {\bibinfo
  {journal} {Phys. Rev. D}\ }\textbf {\bibinfo {volume} {49}},\ \bibinfo
  {pages} {2837} (\bibinfo {year} {1994})},\ \Eprint
  {https://arxiv.org/abs/astro-ph/9310044} {arXiv:astro-ph/9310044}
  \BibitemShut {NoStop}%
\bibitem [{\citenamefont {Grojean}\ and\ \citenamefont
  {Servant}(2007)}]{Grojean:2006bp}%
  \BibitemOpen
  \bibfield  {author} {\bibinfo {author} {\bibfnamefont {C.}~\bibnamefont
  {Grojean}}\ and\ \bibinfo {author} {\bibfnamefont {G.}~\bibnamefont
  {Servant}},\ }\bibfield  {title} {\emph {\bibinfo {title} {{Gravitational
  Waves from Phase Transitions at the Electroweak Scale and Beyond}}},\ }\href
  {https://doi.org/10.1103/PhysRevD.75.043507} {\bibfield  {journal} {\bibinfo
  {journal} {Phys. Rev. D}\ }\textbf {\bibinfo {volume} {75}},\ \bibinfo
  {pages} {043507} (\bibinfo {year} {2007})},\ \Eprint
  {https://arxiv.org/abs/hep-ph/0607107} {arXiv:hep-ph/0607107} \BibitemShut
  {NoStop}%
\bibitem [{\citenamefont {Agazie}\ \emph {et~al.}(2023)\citenamefont {Agazie}
  \emph {et~al.}}]{NANOGrav:2023gor}%
  \BibitemOpen
  \bibfield  {author} {\bibinfo {author} {\bibfnamefont {G.}~\bibnamefont
  {Agazie}} \emph {et~al.} (\bibinfo {collaboration} {NANOGrav}),\ }\bibfield
  {title} {\emph {\bibinfo {title} {{The NANOGrav 15 yr Data Set: Evidence for
  a Gravitational-wave Background}}},\ }\href
  {https://doi.org/10.3847/2041-8213/acdac6} {\bibfield  {journal} {\bibinfo
  {journal} {Astrophys. J. Lett.}\ }\textbf {\bibinfo {volume} {951}},\
  \bibinfo {pages} {L8} (\bibinfo {year} {2023})},\ \Eprint
  {https://arxiv.org/abs/2306.16213} {arXiv:2306.16213 [astro-ph.HE]}
  \BibitemShut {NoStop}%
\bibitem [{\citenamefont {Antoniadis}\ \emph {et~al.}(2023)\citenamefont
  {Antoniadis} \emph {et~al.}}]{EPTA:2023fyk}%
  \BibitemOpen
  \bibfield  {author} {\bibinfo {author} {\bibfnamefont {J.}~\bibnamefont
  {Antoniadis}} \emph {et~al.} (\bibinfo {collaboration} {EPTA, InPTA}),\
  }\bibfield  {title} {\emph {\bibinfo {title} {{The second data release from
  the European Pulsar Timing Array - III. Search for gravitational wave
  signals}}},\ }\href {https://doi.org/10.1051/0004-6361/202346844} {\bibfield
  {journal} {\bibinfo  {journal} {Astron. Astrophys.}\ }\textbf {\bibinfo
  {volume} {678}},\ \bibinfo {pages} {A50} (\bibinfo {year} {2023})},\ \Eprint
  {https://arxiv.org/abs/2306.16214} {arXiv:2306.16214 [astro-ph.HE]}
  \BibitemShut {NoStop}%
\bibitem [{\citenamefont {Amaro-Seoane}\ \emph {et~al.}(2017)\citenamefont
  {Amaro-Seoane} \emph {et~al.}}]{LISA:2017pwj}%
  \BibitemOpen
  \bibfield  {author} {\bibinfo {author} {\bibfnamefont {P.}~\bibnamefont
  {Amaro-Seoane}} \emph {et~al.} (\bibinfo {collaboration} {LISA}),\ }\bibfield
   {title} {\emph {\bibinfo {title} {{Laser Interferometer Space Antenna}}},\
  }\href@noop {} {\  (\bibinfo {year} {2017})},\ \Eprint
  {https://arxiv.org/abs/1702.00786} {arXiv:1702.00786 [astro-ph.IM]}
  \BibitemShut {NoStop}%
\bibitem [{\citenamefont {Colpi}\ \emph {et~al.}(2024)\citenamefont {Colpi}
  \emph {et~al.}}]{LISA:2024hlh}%
  \BibitemOpen
  \bibfield  {author} {\bibinfo {author} {\bibfnamefont {M.}~\bibnamefont
  {Colpi}} \emph {et~al.} (\bibinfo {collaboration} {LISA}),\ }\bibfield
  {title} {\emph {\bibinfo {title} {{LISA Definition Study Report}}},\
  }\href@noop {} {\  (\bibinfo {year} {2024})},\ \Eprint
  {https://arxiv.org/abs/2402.07571} {arXiv:2402.07571 [astro-ph.CO]}
  \BibitemShut {NoStop}%
\bibitem [{\citenamefont {Hu}\ and\ \citenamefont {Wu}(2017)}]{Hu:2017mde}%
  \BibitemOpen
  \bibfield  {author} {\bibinfo {author} {\bibfnamefont {W.-R.}\ \bibnamefont
  {Hu}}\ and\ \bibinfo {author} {\bibfnamefont {Y.-L.}\ \bibnamefont {Wu}},\
  }\bibfield  {title} {\emph {\bibinfo {title} {{The Taiji Program in Space for
  gravitational wave physics and the nature of gravity}}},\ }\href
  {https://doi.org/10.1093/nsr/nwx116} {\bibfield  {journal} {\bibinfo
  {journal} {Natl. Sci. Rev.}\ }\textbf {\bibinfo {volume} {4}},\ \bibinfo
  {pages} {685} (\bibinfo {year} {2017})}\BibitemShut {NoStop}%
\bibitem [{\citenamefont {Luo}\ \emph {et~al.}(2016)\citenamefont {Luo} \emph
  {et~al.}}]{TianQin:2015yph}%
  \BibitemOpen
  \bibfield  {author} {\bibinfo {author} {\bibfnamefont {J.}~\bibnamefont
  {Luo}} \emph {et~al.} (\bibinfo {collaboration} {TianQin}),\ }\bibfield
  {title} {\emph {\bibinfo {title} {{TianQin: a space-borne gravitational wave
  detector}}},\ }\href {https://doi.org/10.1088/0264-9381/33/3/035010}
  {\bibfield  {journal} {\bibinfo  {journal} {Class. Quant. Grav.}\ }\textbf
  {\bibinfo {volume} {33}},\ \bibinfo {pages} {035010} (\bibinfo {year}
  {2016})},\ \Eprint {https://arxiv.org/abs/1512.02076} {arXiv:1512.02076
  [astro-ph.IM]} \BibitemShut {NoStop}%
\bibitem [{\citenamefont {Caprini}\ \emph {et~al.}(2016)\citenamefont {Caprini}
  \emph {et~al.}}]{Caprini:2015zlo}%
  \BibitemOpen
  \bibfield  {author} {\bibinfo {author} {\bibfnamefont {C.}~\bibnamefont
  {Caprini}} \emph {et~al.},\ }\bibfield  {title} {\emph {\bibinfo {title}
  {{Science with the space-based interferometer eLISA. II: Gravitational waves
  from cosmological phase transitions}}},\ }\href
  {https://doi.org/10.1088/1475-7516/2016/04/001} {\bibfield  {journal}
  {\bibinfo  {journal} {JCAP}\ }\textbf {\bibinfo {volume} {04}},\ \bibinfo
  {pages} {001}},\ \Eprint {https://arxiv.org/abs/1512.06239} {arXiv:1512.06239
  [astro-ph.CO]} \BibitemShut {NoStop}%
\bibitem [{\citenamefont {Caprini}\ \emph {et~al.}(2020)\citenamefont {Caprini}
  \emph {et~al.}}]{Caprini:2019egz}%
  \BibitemOpen
  \bibfield  {author} {\bibinfo {author} {\bibfnamefont {C.}~\bibnamefont
  {Caprini}} \emph {et~al.},\ }\bibfield  {title} {\emph {\bibinfo {title}
  {{Detecting gravitational waves from cosmological phase transitions with
  LISA: an update}}},\ }\href {https://doi.org/10.1088/1475-7516/2020/03/024}
  {\bibfield  {journal} {\bibinfo  {journal} {JCAP}\ }\textbf {\bibinfo
  {volume} {03}},\ \bibinfo {pages} {024}},\ \Eprint
  {https://arxiv.org/abs/1910.13125} {arXiv:1910.13125 [astro-ph.CO]}
  \BibitemShut {NoStop}%
\bibitem [{\citenamefont {Seto}\ \emph {et~al.}(2001)\citenamefont {Seto},
  \citenamefont {Kawamura},\ and\ \citenamefont {Nakamura}}]{Seto:2001qf}%
  \BibitemOpen
  \bibfield  {author} {\bibinfo {author} {\bibfnamefont {N.}~\bibnamefont
  {Seto}}, \bibinfo {author} {\bibfnamefont {S.}~\bibnamefont {Kawamura}},\
  and\ \bibinfo {author} {\bibfnamefont {T.}~\bibnamefont {Nakamura}},\
  }\bibfield  {title} {\emph {\bibinfo {title} {{Possibility of direct
  measurement of the acceleration of the universe using 0.1-Hz band laser
  interferometer gravitational wave antenna in space}}},\ }\href
  {https://doi.org/10.1103/PhysRevLett.87.221103} {\bibfield  {journal}
  {\bibinfo  {journal} {Phys. Rev. Lett.}\ }\textbf {\bibinfo {volume} {87}},\
  \bibinfo {pages} {221103} (\bibinfo {year} {2001})},\ \Eprint
  {https://arxiv.org/abs/astro-ph/0108011} {arXiv:astro-ph/0108011}
  \BibitemShut {NoStop}%
\bibitem [{\citenamefont {Kawamura}\ \emph {et~al.}(2021)\citenamefont
  {Kawamura} \emph {et~al.}}]{Kawamura:2020pcg}%
  \BibitemOpen
  \bibfield  {author} {\bibinfo {author} {\bibfnamefont {S.}~\bibnamefont
  {Kawamura}} \emph {et~al.},\ }\bibfield  {title} {\emph {\bibinfo {title}
  {{Current status of space gravitational wave antenna DECIGO and B-DECIGO}}},\
  }\href {https://doi.org/10.1093/ptep/ptab019} {\bibfield  {journal} {\bibinfo
   {journal} {PTEP}\ }\textbf {\bibinfo {volume} {2021}},\ \bibinfo {pages}
  {05A105} (\bibinfo {year} {2021})},\ \Eprint
  {https://arxiv.org/abs/2006.13545} {arXiv:2006.13545 [gr-qc]} \BibitemShut
  {NoStop}%
\bibitem [{\citenamefont {Crowder}\ and\ \citenamefont
  {Cornish}(2005)}]{Crowder:2005nr}%
  \BibitemOpen
  \bibfield  {author} {\bibinfo {author} {\bibfnamefont {J.}~\bibnamefont
  {Crowder}}\ and\ \bibinfo {author} {\bibfnamefont {N.~J.}\ \bibnamefont
  {Cornish}},\ }\bibfield  {title} {\emph {\bibinfo {title} {{Beyond LISA:
  Exploring future gravitational wave missions}}},\ }\href
  {https://doi.org/10.1103/PhysRevD.72.083005} {\bibfield  {journal} {\bibinfo
  {journal} {Phys. Rev. D}\ }\textbf {\bibinfo {volume} {72}},\ \bibinfo
  {pages} {083005} (\bibinfo {year} {2005})},\ \Eprint
  {https://arxiv.org/abs/gr-qc/0506015} {arXiv:gr-qc/0506015} \BibitemShut
  {NoStop}%
\bibitem [{\citenamefont {Punturo}\ \emph {et~al.}(2010)\citenamefont {Punturo}
  \emph {et~al.}}]{Punturo:2010zz}%
  \BibitemOpen
  \bibfield  {author} {\bibinfo {author} {\bibfnamefont {M.}~\bibnamefont
  {Punturo}} \emph {et~al.},\ }\bibfield  {title} {\emph {\bibinfo {title}
  {{The Einstein Telescope: A third-generation gravitational wave
  observatory}}},\ }\href {https://doi.org/10.1088/0264-9381/27/19/194002}
  {\bibfield  {journal} {\bibinfo  {journal} {Class. Quant. Grav.}\ }\textbf
  {\bibinfo {volume} {27}},\ \bibinfo {pages} {194002} (\bibinfo {year}
  {2010})}\BibitemShut {NoStop}%
\bibitem [{\citenamefont {Maggiore}\ \emph {et~al.}(2020)\citenamefont
  {Maggiore} \emph {et~al.}}]{ET:2019dnz}%
  \BibitemOpen
  \bibfield  {author} {\bibinfo {author} {\bibfnamefont {M.}~\bibnamefont
  {Maggiore}} \emph {et~al.} (\bibinfo {collaboration} {ET}),\ }\bibfield
  {title} {\emph {\bibinfo {title} {{Science Case for the Einstein
  Telescope}}},\ }\href {https://doi.org/10.1088/1475-7516/2020/03/050}
  {\bibfield  {journal} {\bibinfo  {journal} {JCAP}\ }\textbf {\bibinfo
  {volume} {03}},\ \bibinfo {pages} {050}},\ \Eprint
  {https://arxiv.org/abs/1912.02622} {arXiv:1912.02622 [astro-ph.CO]}
  \BibitemShut {NoStop}%
\bibitem [{\citenamefont {Reitze}\ \emph {et~al.}(2019)\citenamefont {Reitze}
  \emph {et~al.}}]{Reitze:2019iox}%
  \BibitemOpen
  \bibfield  {author} {\bibinfo {author} {\bibfnamefont {D.}~\bibnamefont
  {Reitze}} \emph {et~al.},\ }\bibfield  {title} {\emph {\bibinfo {title}
  {{Cosmic Explorer: The U.S. Contribution to Gravitational-Wave Astronomy
  beyond LIGO}}},\ }\href@noop {} {\bibfield  {journal} {\bibinfo  {journal}
  {Bull. Am. Astron. Soc.}\ }\textbf {\bibinfo {volume} {51}},\ \bibinfo
  {pages} {035} (\bibinfo {year} {2019})},\ \Eprint
  {https://arxiv.org/abs/1907.04833} {arXiv:1907.04833 [astro-ph.IM]}
  \BibitemShut {NoStop}%
\bibitem [{\citenamefont {Kajantie}\ \emph
  {et~al.}(1996{\natexlab{a}})\citenamefont {Kajantie}, \citenamefont {Laine},
  \citenamefont {Rummukainen},\ and\ \citenamefont
  {Shaposhnikov}}]{Kajantie:1996mn}%
  \BibitemOpen
  \bibfield  {author} {\bibinfo {author} {\bibfnamefont {K.}~\bibnamefont
  {Kajantie}}, \bibinfo {author} {\bibfnamefont {M.}~\bibnamefont {Laine}},
  \bibinfo {author} {\bibfnamefont {K.}~\bibnamefont {Rummukainen}},\ and\
  \bibinfo {author} {\bibfnamefont {M.~E.}\ \bibnamefont {Shaposhnikov}},\
  }\bibfield  {title} {\emph {\bibinfo {title} {{Is there a~ hot electroweak
  phase transition at $m_H \gtrsim m_W$?}}},\ }\href
  {https://doi.org/10.1103/PhysRevLett.77.2887} {\bibfield  {journal} {\bibinfo
   {journal} {Phys. Rev. Lett.}\ }\textbf {\bibinfo {volume} {77}},\ \bibinfo
  {pages} {2887} (\bibinfo {year} {1996}{\natexlab{a}})},\ \Eprint
  {https://arxiv.org/abs/hep-ph/9605288} {arXiv:hep-ph/9605288} \BibitemShut
  {NoStop}%
\bibitem [{\citenamefont {Csikor}\ \emph {et~al.}(1999)\citenamefont {Csikor},
  \citenamefont {Fodor},\ and\ \citenamefont {Heitger}}]{Csikor:1998eu}%
  \BibitemOpen
  \bibfield  {author} {\bibinfo {author} {\bibfnamefont {F.}~\bibnamefont
  {Csikor}}, \bibinfo {author} {\bibfnamefont {Z.}~\bibnamefont {Fodor}},\ and\
  \bibinfo {author} {\bibfnamefont {J.}~\bibnamefont {Heitger}},\ }\bibfield
  {title} {\emph {\bibinfo {title} {{Endpoint of the hot electroweak phase
  transition}}},\ }\href {https://doi.org/10.1103/PhysRevLett.82.21} {\bibfield
   {journal} {\bibinfo  {journal} {Phys. Rev. Lett.}\ }\textbf {\bibinfo
  {volume} {82}},\ \bibinfo {pages} {21} (\bibinfo {year} {1999})},\ \Eprint
  {https://arxiv.org/abs/hep-ph/9809291} {arXiv:hep-ph/9809291} \BibitemShut
  {NoStop}%
\bibitem [{\citenamefont {D'Onofrio}\ and\ \citenamefont
  {Rummukainen}(2016)}]{DOnofrio:2015gop}%
  \BibitemOpen
  \bibfield  {author} {\bibinfo {author} {\bibfnamefont {M.}~\bibnamefont
  {D'Onofrio}}\ and\ \bibinfo {author} {\bibfnamefont {K.}~\bibnamefont
  {Rummukainen}},\ }\bibfield  {title} {\emph {\bibinfo {title} {{Standard
  model cross-over on the lattice}}},\ }\href
  {https://doi.org/10.1103/PhysRevD.93.025003} {\bibfield  {journal} {\bibinfo
  {journal} {Phys. Rev. D}\ }\textbf {\bibinfo {volume} {93}},\ \bibinfo
  {pages} {025003} (\bibinfo {year} {2016})},\ \Eprint
  {https://arxiv.org/abs/1508.07161} {arXiv:1508.07161 [hep-ph]} \BibitemShut
  {NoStop}%
\bibitem [{\citenamefont {Aoki}\ \emph {et~al.}(2006)\citenamefont {Aoki},
  \citenamefont {Endrodi}, \citenamefont {Fodor}, \citenamefont {Katz},\ and\
  \citenamefont {Szabo}}]{Aoki:2006we}%
  \BibitemOpen
  \bibfield  {author} {\bibinfo {author} {\bibfnamefont {Y.}~\bibnamefont
  {Aoki}}, \bibinfo {author} {\bibfnamefont {G.}~\bibnamefont {Endrodi}},
  \bibinfo {author} {\bibfnamefont {Z.}~\bibnamefont {Fodor}}, \bibinfo
  {author} {\bibfnamefont {S.~D.}\ \bibnamefont {Katz}},\ and\ \bibinfo
  {author} {\bibfnamefont {K.~K.}\ \bibnamefont {Szabo}},\ }\bibfield  {title}
  {\emph {\bibinfo {title} {{The Order of the quantum chromodynamics transition
  predicted by the standard model of particle physics}}},\ }\href
  {https://doi.org/10.1038/nature05120} {\bibfield  {journal} {\bibinfo
  {journal} {Nature}\ }\textbf {\bibinfo {volume} {443}},\ \bibinfo {pages}
  {675} (\bibinfo {year} {2006})},\ \Eprint
  {https://arxiv.org/abs/hep-lat/0611014} {arXiv:hep-lat/0611014} \BibitemShut
  {NoStop}%
\bibitem [{\citenamefont {Sakharov}(1967)}]{Sakharov:1967dj}%
  \BibitemOpen
  \bibfield  {author} {\bibinfo {author} {\bibfnamefont {A.~D.}\ \bibnamefont
  {Sakharov}},\ }\bibfield  {title} {\emph {\bibinfo {title} {{Violation of CP
  Invariance, C asymmetry, and baryon asymmetry of the universe}}},\ }\href
  {https://doi.org/10.1070/PU1991v034n05ABEH002497} {\bibfield  {journal}
  {\bibinfo  {journal} {Pisma Zh. Eksp. Teor. Fiz.}\ }\textbf {\bibinfo
  {volume} {5}},\ \bibinfo {pages} {32} (\bibinfo {year} {1967})}\BibitemShut
  {NoStop}%
\bibitem [{\citenamefont {Kuzmin}\ \emph {et~al.}(1985)\citenamefont {Kuzmin},
  \citenamefont {Rubakov},\ and\ \citenamefont {Shaposhnikov}}]{Kuzmin:1985mm}%
  \BibitemOpen
  \bibfield  {author} {\bibinfo {author} {\bibfnamefont {V.~A.}\ \bibnamefont
  {Kuzmin}}, \bibinfo {author} {\bibfnamefont {V.~A.}\ \bibnamefont
  {Rubakov}},\ and\ \bibinfo {author} {\bibfnamefont {M.~E.}\ \bibnamefont
  {Shaposhnikov}},\ }\bibfield  {title} {\emph {\bibinfo {title} {{On the
  Anomalous Electroweak Baryon Number Nonconservation in the Early
  Universe}}},\ }\href {https://doi.org/10.1016/0370-2693(85)91028-7}
  {\bibfield  {journal} {\bibinfo  {journal} {Phys. Lett. B}\ }\textbf
  {\bibinfo {volume} {155}},\ \bibinfo {pages} {36} (\bibinfo {year}
  {1985})}\BibitemShut {NoStop}%
\bibitem [{\citenamefont {Morrissey}\ and\ \citenamefont
  {Ramsey-Musolf}(2012)}]{Morrissey:2012db}%
  \BibitemOpen
  \bibfield  {author} {\bibinfo {author} {\bibfnamefont {D.~E.}\ \bibnamefont
  {Morrissey}}\ and\ \bibinfo {author} {\bibfnamefont {M.~J.}\ \bibnamefont
  {Ramsey-Musolf}},\ }\bibfield  {title} {\emph {\bibinfo {title} {{Electroweak
  baryogenesis}}},\ }\href {https://doi.org/10.1088/1367-2630/14/12/125003}
  {\bibfield  {journal} {\bibinfo  {journal} {New J. Phys.}\ }\textbf {\bibinfo
  {volume} {14}},\ \bibinfo {pages} {125003} (\bibinfo {year} {2012})},\
  \Eprint {https://arxiv.org/abs/1206.2942} {arXiv:1206.2942 [hep-ph]}
  \BibitemShut {NoStop}%
\bibitem [{\citenamefont {Cline}(2006)}]{Cline:2006ts}%
  \BibitemOpen
  \bibfield  {author} {\bibinfo {author} {\bibfnamefont {J.~M.}\ \bibnamefont
  {Cline}},\ }in\ \href@noop {} {\emph {\bibinfo {booktitle} {{Les Houches
  Summer School - Session 86: Particle Physics and Cosmology: The Fabric of
  Spacetime}}}}\ (\bibinfo {year} {2006})\ \Eprint
  {https://arxiv.org/abs/hep-ph/0609145} {arXiv:hep-ph/0609145} \BibitemShut
  {NoStop}%
\bibitem [{\citenamefont {Profumo}\ \emph {et~al.}(2007)\citenamefont
  {Profumo}, \citenamefont {Ramsey-Musolf},\ and\ \citenamefont
  {Shaughnessy}}]{Profumo:2007wc}%
  \BibitemOpen
  \bibfield  {author} {\bibinfo {author} {\bibfnamefont {S.}~\bibnamefont
  {Profumo}}, \bibinfo {author} {\bibfnamefont {M.~J.}\ \bibnamefont
  {Ramsey-Musolf}},\ and\ \bibinfo {author} {\bibfnamefont {G.}~\bibnamefont
  {Shaughnessy}},\ }\bibfield  {title} {\emph {\bibinfo {title} {{Singlet Higgs
  phenomenology and the electroweak phase transition}}},\ }\href
  {https://doi.org/10.1088/1126-6708/2007/08/010} {\bibfield  {journal}
  {\bibinfo  {journal} {JHEP}\ }\textbf {\bibinfo {volume} {08}},\ \bibinfo
  {pages} {010}},\ \Eprint {https://arxiv.org/abs/0705.2425} {arXiv:0705.2425
  [hep-ph]} \BibitemShut {NoStop}%
\bibitem [{\citenamefont {Espinosa}\ \emph {et~al.}(2012)\citenamefont
  {Espinosa}, \citenamefont {Konstandin},\ and\ \citenamefont
  {Riva}}]{Espinosa:2011ax}%
  \BibitemOpen
  \bibfield  {author} {\bibinfo {author} {\bibfnamefont {J.~R.}\ \bibnamefont
  {Espinosa}}, \bibinfo {author} {\bibfnamefont {T.}~\bibnamefont
  {Konstandin}},\ and\ \bibinfo {author} {\bibfnamefont {F.}~\bibnamefont
  {Riva}},\ }\bibfield  {title} {\emph {\bibinfo {title} {{Strong Electroweak
  Phase Transitions in the Standard Model with a Singlet}}},\ }\href
  {https://doi.org/10.1016/j.nuclphysb.2011.09.010} {\bibfield  {journal}
  {\bibinfo  {journal} {Nucl. Phys. B}\ }\textbf {\bibinfo {volume} {854}},\
  \bibinfo {pages} {592} (\bibinfo {year} {2012})},\ \Eprint
  {https://arxiv.org/abs/1107.5441} {arXiv:1107.5441 [hep-ph]} \BibitemShut
  {NoStop}%
\bibitem [{\citenamefont {Aad}\ \emph {et~al.}(2022)\citenamefont {Aad} \emph
  {et~al.}}]{ATLAS:2022vkf}%
  \BibitemOpen
  \bibfield  {author} {\bibinfo {author} {\bibfnamefont {G.}~\bibnamefont
  {Aad}} \emph {et~al.} (\bibinfo {collaboration} {ATLAS}),\ }\bibfield
  {title} {\emph {\bibinfo {title} {{A detailed map of Higgs boson interactions
  by the ATLAS experiment ten years after the discovery}}},\ }\href
  {https://doi.org/10.1038/s41586-022-04893-w} {\bibfield  {journal} {\bibinfo
  {journal} {Nature}\ }\textbf {\bibinfo {volume} {607}},\ \bibinfo {pages}
  {52} (\bibinfo {year} {2022})},\ \bibinfo {note} {[Erratum: Nature 612, E24
  (2022)]},\ \Eprint {https://arxiv.org/abs/2207.00092} {arXiv:2207.00092
  [hep-ex]} \BibitemShut {NoStop}%
\bibitem [{\citenamefont {Tumasyan}\ \emph {et~al.}(2022)\citenamefont
  {Tumasyan} \emph {et~al.}}]{CMS:2022dwd}%
  \BibitemOpen
  \bibfield  {author} {\bibinfo {author} {\bibfnamefont {A.}~\bibnamefont
  {Tumasyan}} \emph {et~al.} (\bibinfo {collaboration} {CMS}),\ }\bibfield
  {title} {\emph {\bibinfo {title} {{A portrait of the Higgs boson by the CMS
  experiment ten years after the discovery.}}},\ }\href
  {https://doi.org/10.1038/s41586-022-04892-x} {\bibfield  {journal} {\bibinfo
  {journal} {Nature}\ }\textbf {\bibinfo {volume} {607}},\ \bibinfo {pages}
  {60} (\bibinfo {year} {2022})},\ \bibinfo {note} {[Erratum: Nature 623,
  (2023)]},\ \Eprint {https://arxiv.org/abs/2207.00043} {arXiv:2207.00043
  [hep-ex]} \BibitemShut {NoStop}%
\bibitem [{\citenamefont {Caprini}\ \emph {et~al.}(2024)\citenamefont
  {Caprini}, \citenamefont {Jinno}, \citenamefont {Lewicki}, \citenamefont
  {Madge}, \citenamefont {Merchand}, \citenamefont {Nardini}, \citenamefont
  {Pieroni}, \citenamefont {Roper~Pol},\ and\ \citenamefont
  {Vaskonen}}]{Caprini:2024hue}%
  \BibitemOpen
  \bibfield  {author} {\bibinfo {author} {\bibfnamefont {C.}~\bibnamefont
  {Caprini}}, \bibinfo {author} {\bibfnamefont {R.}~\bibnamefont {Jinno}},
  \bibinfo {author} {\bibfnamefont {M.}~\bibnamefont {Lewicki}}, \bibinfo
  {author} {\bibfnamefont {E.}~\bibnamefont {Madge}}, \bibinfo {author}
  {\bibfnamefont {M.}~\bibnamefont {Merchand}}, \bibinfo {author}
  {\bibfnamefont {G.}~\bibnamefont {Nardini}}, \bibinfo {author} {\bibfnamefont
  {M.}~\bibnamefont {Pieroni}}, \bibinfo {author} {\bibfnamefont
  {A.}~\bibnamefont {Roper~Pol}},\ and\ \bibinfo {author} {\bibfnamefont
  {V.}~\bibnamefont {Vaskonen}} (\bibinfo {collaboration} {LISA Cosmology
  Working Group}),\ }\bibfield  {title} {\emph {\bibinfo {title}
  {{Gravitational waves from first-order phase transitions in LISA:
  reconstruction pipeline and physics interpretation}}},\ }\href
  {https://doi.org/10.1088/1475-7516/2024/10/020} {\bibfield  {journal}
  {\bibinfo  {journal} {JCAP}\ }\textbf {\bibinfo {volume} {10}},\ \bibinfo
  {pages} {020}},\ \Eprint {https://arxiv.org/abs/2403.03723} {arXiv:2403.03723
  [astro-ph.CO]} \BibitemShut {NoStop}%
\bibitem [{\citenamefont {Coleman}(1977)}]{Coleman:1977py}%
  \BibitemOpen
  \bibfield  {author} {\bibinfo {author} {\bibfnamefont {S.~R.}\ \bibnamefont
  {Coleman}},\ }\bibfield  {title} {\emph {\bibinfo {title} {{The Fate of the
  False Vacuum. 1. Semiclassical Theory}}},\ }\href
  {https://doi.org/10.1103/PhysRevD.15.2929} {\bibfield  {journal} {\bibinfo
  {journal} {Phys. Rev. D}\ }\textbf {\bibinfo {volume} {15}},\ \bibinfo
  {pages} {2929} (\bibinfo {year} {1977})},\ \bibinfo {note} {[Erratum:
  Phys.Rev.D 16, 1248 (1977)]}\BibitemShut {NoStop}%
\bibitem [{\citenamefont {Caprini}\ \emph {et~al.}(2019)\citenamefont
  {Caprini}, \citenamefont {Figueroa}, \citenamefont {Flauger}, \citenamefont
  {Nardini}, \citenamefont {Peloso}, \citenamefont {Pieroni}, \citenamefont
  {Ricciardone},\ and\ \citenamefont {Tasinato}}]{Caprini:2019pxz}%
  \BibitemOpen
  \bibfield  {author} {\bibinfo {author} {\bibfnamefont {C.}~\bibnamefont
  {Caprini}}, \bibinfo {author} {\bibfnamefont {D.~G.}\ \bibnamefont
  {Figueroa}}, \bibinfo {author} {\bibfnamefont {R.}~\bibnamefont {Flauger}},
  \bibinfo {author} {\bibfnamefont {G.}~\bibnamefont {Nardini}}, \bibinfo
  {author} {\bibfnamefont {M.}~\bibnamefont {Peloso}}, \bibinfo {author}
  {\bibfnamefont {M.}~\bibnamefont {Pieroni}}, \bibinfo {author} {\bibfnamefont
  {A.}~\bibnamefont {Ricciardone}},\ and\ \bibinfo {author} {\bibfnamefont
  {G.}~\bibnamefont {Tasinato}},\ }\bibfield  {title} {\emph {\bibinfo {title}
  {{Reconstructing the spectral shape of a stochastic gravitational wave
  background with LISA}}},\ }\href
  {https://doi.org/10.1088/1475-7516/2019/11/017} {\bibfield  {journal}
  {\bibinfo  {journal} {JCAP}\ }\textbf {\bibinfo {volume} {11}},\ \bibinfo
  {pages} {017}},\ \Eprint {https://arxiv.org/abs/1906.09244} {arXiv:1906.09244
  [astro-ph.CO]} \BibitemShut {NoStop}%
\bibitem [{\citenamefont {Gowling}\ \emph {et~al.}(2023)\citenamefont
  {Gowling}, \citenamefont {Hindmarsh}, \citenamefont {Hooper},\ and\
  \citenamefont {Torrado}}]{Gowling:2022pzb}%
  \BibitemOpen
  \bibfield  {author} {\bibinfo {author} {\bibfnamefont {C.}~\bibnamefont
  {Gowling}}, \bibinfo {author} {\bibfnamefont {M.}~\bibnamefont {Hindmarsh}},
  \bibinfo {author} {\bibfnamefont {D.~C.}\ \bibnamefont {Hooper}},\ and\
  \bibinfo {author} {\bibfnamefont {J.}~\bibnamefont {Torrado}},\ }\bibfield
  {title} {\emph {\bibinfo {title} {{Reconstructing physical parameters from
  template gravitational wave spectra at LISA: first order phase
  transitions}}},\ }\href {https://doi.org/10.1088/1475-7516/2023/04/061}
  {\bibfield  {journal} {\bibinfo  {journal} {JCAP}\ }\textbf {\bibinfo
  {volume} {04}},\ \bibinfo {pages} {061}},\ \Eprint
  {https://arxiv.org/abs/2209.13551} {arXiv:2209.13551 [astro-ph.CO]}
  \BibitemShut {NoStop}%
\bibitem [{\citenamefont
  {Biek{\"o}tter}(2026{\natexlab{a}})}]{Biekotter:2025gkp}%
  \BibitemOpen
  \bibfield  {author} {\bibinfo {author} {\bibfnamefont {T.}~\bibnamefont
  {Biek{\"o}tter}},\ }\bibfield  {title} {\emph {\bibinfo {title} {{evortran: A
  modern Fortran package for genetic algorithms with applications from LHC data
  fitting to LISA signal reconstruction}}},\ }\href
  {https://doi.org/10.21468/SciPostPhysCodeb.64} {\bibfield  {journal}
  {\bibinfo  {journal} {SciPost Phys. Codeb.}\ }\textbf {\bibinfo {volume}
  {64}},\ \bibinfo {pages} {1} (\bibinfo {year} {2026}{\natexlab{a}})},\
  \Eprint {https://arxiv.org/abs/2507.06082} {arXiv:2507.06082 [hep-ph]}
  \BibitemShut {NoStop}%
\bibitem [{\citenamefont {Biek{\"o}tter}(2025{\natexlab{a}})}]{evortran}%
  \BibitemOpen
  \bibfield  {author} {\bibinfo {author} {\bibfnamefont {T.}~\bibnamefont
  {Biek{\"o}tter}},\ }\href@noop {} {\bibinfo {title} {{\textit{evortran -- A
  modern Fortran package for genetic algorithms}}}},\ \bibinfo {howpublished}
  {\url{https://gitlab.com/thomas.biekoetter/evortran}} (\bibinfo {year}
  {2025}{\natexlab{a}})\BibitemShut {NoStop}%
\bibitem [{\citenamefont {Goodman}\ and\ \citenamefont
  {Weare}(2010)}]{Goodman:2010dyf}%
  \BibitemOpen
  \bibfield  {author} {\bibinfo {author} {\bibfnamefont {J.}~\bibnamefont
  {Goodman}}\ and\ \bibinfo {author} {\bibfnamefont {J.}~\bibnamefont
  {Weare}},\ }\bibfield  {title} {\emph {\bibinfo {title} {{Ensemble samplers
  with affine invariance}}},\ }\href {https://doi.org/10.2140/camcos.2010.5.65}
  {\bibfield  {journal} {\bibinfo  {journal} {Commun. Appl. Math. Comput. Sc.}\
  }\textbf {\bibinfo {volume} {5}},\ \bibinfo {pages} {65} (\bibinfo {year}
  {2010})}\BibitemShut {NoStop}%
\bibitem [{\citenamefont {Foreman-Mackey}\ \emph {et~al.}(2013)\citenamefont
  {Foreman-Mackey}, \citenamefont {Hogg}, \citenamefont {Lang},\ and\
  \citenamefont {Goodman}}]{Foreman-Mackey:2012any}%
  \BibitemOpen
  \bibfield  {author} {\bibinfo {author} {\bibfnamefont {D.}~\bibnamefont
  {Foreman-Mackey}}, \bibinfo {author} {\bibfnamefont {D.~W.}\ \bibnamefont
  {Hogg}}, \bibinfo {author} {\bibfnamefont {D.}~\bibnamefont {Lang}},\ and\
  \bibinfo {author} {\bibfnamefont {J.}~\bibnamefont {Goodman}},\ }\bibfield
  {title} {\emph {\bibinfo {title} {{emcee: The MCMC Hammer}}},\ }\href
  {https://doi.org/10.1086/670067} {\bibfield  {journal} {\bibinfo  {journal}
  {Publ. Astron. Soc. Pac.}\ }\textbf {\bibinfo {volume} {125}},\ \bibinfo
  {pages} {306} (\bibinfo {year} {2013})},\ \Eprint
  {https://arxiv.org/abs/1202.3665} {arXiv:1202.3665 [astro-ph.IM]}
  \BibitemShut {NoStop}%
\bibitem [{\citenamefont {Biek{\"o}tter}(2025{\natexlab{b}})}]{copa}%
  \BibitemOpen
  \bibfield  {author} {\bibinfo {author} {\bibfnamefont {T.}~\bibnamefont
  {Biek{\"o}tter}},\ }\href@noop {} {\bibinfo {title} {{\textit{copa -- Chains
  for Optimization and Probabilistic Analysis}}}},\ \bibinfo {howpublished}
  {\url{https://github.com/thomasbiekoetter/copa}} (\bibinfo {year}
  {2025}{\natexlab{b}})\BibitemShut {NoStop}%
\bibitem [{\citenamefont {Cranmer}\ \emph {et~al.}(2020)\citenamefont
  {Cranmer}, \citenamefont {Brehmer},\ and\ \citenamefont
  {Louppe}}]{Cranmer:2019eaq}%
  \BibitemOpen
  \bibfield  {author} {\bibinfo {author} {\bibfnamefont {K.}~\bibnamefont
  {Cranmer}}, \bibinfo {author} {\bibfnamefont {J.}~\bibnamefont {Brehmer}},\
  and\ \bibinfo {author} {\bibfnamefont {G.}~\bibnamefont {Louppe}},\
  }\bibfield  {title} {\emph {\bibinfo {title} {{The frontier of
  simulation-based inference}}},\ }\href
  {https://doi.org/10.1073/pnas.1912789117} {\bibfield  {journal} {\bibinfo
  {journal} {Proc. Nat. Acad. Sci.}\ }\textbf {\bibinfo {volume} {117}},\
  \bibinfo {pages} {30055} (\bibinfo {year} {2020})},\ \Eprint
  {https://arxiv.org/abs/1911.01429} {arXiv:1911.01429 [stat.ML]} \BibitemShut
  {NoStop}%
\bibitem [{\citenamefont {Papamakarios}\ \emph {et~al.}(2021)\citenamefont
  {Papamakarios}, \citenamefont {Nalisnick}, \citenamefont {Rezende},
  \citenamefont {Mohamed},\ and\ \citenamefont
  {Lakshminarayanan}}]{Papamakarios:2019fms}%
  \BibitemOpen
  \bibfield  {author} {\bibinfo {author} {\bibfnamefont {G.}~\bibnamefont
  {Papamakarios}}, \bibinfo {author} {\bibfnamefont {E.}~\bibnamefont
  {Nalisnick}}, \bibinfo {author} {\bibfnamefont {D.~J.}\ \bibnamefont
  {Rezende}}, \bibinfo {author} {\bibfnamefont {S.}~\bibnamefont {Mohamed}},\
  and\ \bibinfo {author} {\bibfnamefont {B.}~\bibnamefont {Lakshminarayanan}},\
  }\bibfield  {title} {\emph {\bibinfo {title} {{Normalizing Flows for
  Probabilistic Modeling and Inference}}},\ }\href
  {https://doi.org/10.5555/3546258.3546315} {\bibfield  {journal} {\bibinfo
  {journal} {J. Machine Learning Res.}\ }\textbf {\bibinfo {volume} {22}},\
  \bibinfo {pages} {2617} (\bibinfo {year} {2021})},\ \Eprint
  {https://arxiv.org/abs/1912.02762} {arXiv:1912.02762 [stat.ML]} \BibitemShut
  {NoStop}%
\bibitem [{\citenamefont {Brehmer}\ and\ \citenamefont
  {Cranmer}(2020)}]{Brehmer:2020cvb}%
  \BibitemOpen
  \bibfield  {author} {\bibinfo {author} {\bibfnamefont {J.}~\bibnamefont
  {Brehmer}}\ and\ \bibinfo {author} {\bibfnamefont {K.}~\bibnamefont
  {Cranmer}},\ }\bibfield  {title} {\emph {\bibinfo {title} {{Simulation-based
  inference methods for particle physics}}},\ }\href@noop {} {\  (\bibinfo
  {year} {2020})},\ \Eprint {https://arxiv.org/abs/2010.06439}
  {arXiv:2010.06439 [hep-ph]} \BibitemShut {NoStop}%
\bibitem [{\citenamefont {Aad}\ \emph {et~al.}(2025{\natexlab{a}})\citenamefont
  {Aad} \emph {et~al.}}]{ATLAS:2024jry}%
  \BibitemOpen
  \bibfield  {author} {\bibinfo {author} {\bibfnamefont {G.}~\bibnamefont
  {Aad}} \emph {et~al.} (\bibinfo {collaboration} {ATLAS}),\ }\bibfield
  {title} {\emph {\bibinfo {title} {{Measurement of off-shell Higgs boson
  production in the $H^*\rightarrow ZZ\rightarrow 4\ell$ decay channel using a
  neural simulation-based inference technique in 13{\,}TeV pp collisions with
  the ATLAS detector}}},\ }\href {https://doi.org/10.1088/1361-6633/adcd9a}
  {\bibfield  {journal} {\bibinfo  {journal} {Rept. Prog. Phys.}\ }\textbf
  {\bibinfo {volume} {88}},\ \bibinfo {pages} {057803} (\bibinfo {year}
  {2025}{\natexlab{a}})},\ \Eprint {https://arxiv.org/abs/2412.01548}
  {arXiv:2412.01548 [hep-ex]} \BibitemShut {NoStop}%
\bibitem [{\citenamefont {Chatterjee}\ \emph {et~al.}(2025)\citenamefont
  {Chatterjee}, \citenamefont {Choudhury}, \citenamefont {Mitra}, \citenamefont
  {Mondal},\ and\ \citenamefont {Mondal}}]{Chatterjee:2025gej}%
  \BibitemOpen
  \bibfield  {author} {\bibinfo {author} {\bibfnamefont {A.}~\bibnamefont
  {Chatterjee}}, \bibinfo {author} {\bibfnamefont {A.}~\bibnamefont
  {Choudhury}}, \bibinfo {author} {\bibfnamefont {S.}~\bibnamefont {Mitra}},
  \bibinfo {author} {\bibfnamefont {A.}~\bibnamefont {Mondal}},\ and\ \bibinfo
  {author} {\bibfnamefont {S.}~\bibnamefont {Mondal}},\ }\bibfield  {title}
  {\emph {\bibinfo {title} {{Exploring the BSM parameter space with neural
  network aided Simulation-Based Inference}}},\ }\href
  {https://doi.org/10.1007/JHEP12(2025)138} {\bibfield  {journal} {\bibinfo
  {journal} {JHEP}\ }\textbf {\bibinfo {volume} {12}},\ \bibinfo {pages}
  {138}},\ \Eprint {https://arxiv.org/abs/2502.11928} {arXiv:2502.11928
  [hep-ph]} \BibitemShut {NoStop}%
\bibitem [{\citenamefont {Aad}\ \emph {et~al.}(2025{\natexlab{b}})\citenamefont
  {Aad} \emph {et~al.}}]{ATLAS:2025clx}%
  \BibitemOpen
  \bibfield  {author} {\bibinfo {author} {\bibfnamefont {G.}~\bibnamefont
  {Aad}} \emph {et~al.} (\bibinfo {collaboration} {ATLAS}),\ }\bibfield
  {title} {\emph {\bibinfo {title} {{An implementation of neural
  simulation-based inference for parameter estimation in ATLAS}}},\ }\href
  {https://doi.org/10.1088/1361-6633/add370} {\bibfield  {journal} {\bibinfo
  {journal} {Rept. Prog. Phys.}\ }\textbf {\bibinfo {volume} {88}},\ \bibinfo
  {pages} {067801} (\bibinfo {year} {2025}{\natexlab{b}})},\ \Eprint
  {https://arxiv.org/abs/2412.01600} {arXiv:2412.01600 [physics.data-an]}
  \BibitemShut {NoStop}%
\bibitem [{\citenamefont {Bahl}\ and\ \citenamefont
  {Das}(2026)}]{Bahl:2026zdd}%
  \BibitemOpen
  \bibfield  {author} {\bibinfo {author} {\bibfnamefont {H.}~\bibnamefont
  {Bahl}}\ and\ \bibinfo {author} {\bibfnamefont {R.}~\bibnamefont {Das}},\
  }\bibfield  {title} {\emph {\bibinfo {title} {{Sequential SBI for Resonance
  Searches at the LHC}}},\ }\href@noop {} {\  (\bibinfo {year} {2026})},\
  \Eprint {https://arxiv.org/abs/2610.30245} {arXiv:2610.30245 [hep-ph]}
  \BibitemShut {NoStop}%
\bibitem [{\citenamefont {Dax}\ \emph {et~al.}(2021)\citenamefont {Dax},
  \citenamefont {Green}, \citenamefont {Gair}, \citenamefont {Macke},
  \citenamefont {Buonanno},\ and\ \citenamefont {Sch{\"o}lkopf}}]{Dax:2021tsq}%
  \BibitemOpen
  \bibfield  {author} {\bibinfo {author} {\bibfnamefont {M.}~\bibnamefont
  {Dax}}, \bibinfo {author} {\bibfnamefont {S.~R.}\ \bibnamefont {Green}},
  \bibinfo {author} {\bibfnamefont {J.}~\bibnamefont {Gair}}, \bibinfo {author}
  {\bibfnamefont {J.~H.}\ \bibnamefont {Macke}}, \bibinfo {author}
  {\bibfnamefont {A.}~\bibnamefont {Buonanno}},\ and\ \bibinfo {author}
  {\bibfnamefont {B.}~\bibnamefont {Sch{\"o}lkopf}},\ }\bibfield  {title}
  {\emph {\bibinfo {title} {{Real-Time Gravitational Wave Science with Neural
  Posterior Estimation}}},\ }\href
  {https://doi.org/10.1103/PhysRevLett.127.241103} {\bibfield  {journal}
  {\bibinfo  {journal} {Phys. Rev. Lett.}\ }\textbf {\bibinfo {volume} {127}},\
  \bibinfo {pages} {241103} (\bibinfo {year} {2021})},\ \Eprint
  {https://arxiv.org/abs/2106.12594} {arXiv:2106.12594 [gr-qc]} \BibitemShut
  {NoStop}%
\bibitem [{\citenamefont {Dax}\ \emph {et~al.}(2023)\citenamefont {Dax},
  \citenamefont {Green}, \citenamefont {Gair}, \citenamefont {P{\"u}rrer},
  \citenamefont {Wildberger}, \citenamefont {Macke}, \citenamefont {Buonanno},\
  and\ \citenamefont {Sch{\"o}lkopf}}]{Dax:2022pxd}%
  \BibitemOpen
  \bibfield  {author} {\bibinfo {author} {\bibfnamefont {M.}~\bibnamefont
  {Dax}}, \bibinfo {author} {\bibfnamefont {S.~R.}\ \bibnamefont {Green}},
  \bibinfo {author} {\bibfnamefont {J.}~\bibnamefont {Gair}}, \bibinfo {author}
  {\bibfnamefont {M.}~\bibnamefont {P{\"u}rrer}}, \bibinfo {author}
  {\bibfnamefont {J.}~\bibnamefont {Wildberger}}, \bibinfo {author}
  {\bibfnamefont {J.~H.}\ \bibnamefont {Macke}}, \bibinfo {author}
  {\bibfnamefont {A.}~\bibnamefont {Buonanno}},\ and\ \bibinfo {author}
  {\bibfnamefont {B.}~\bibnamefont {Sch{\"o}lkopf}},\ }\bibfield  {title}
  {\emph {\bibinfo {title} {{Neural Importance Sampling for Rapid and Reliable
  Gravitational-Wave Inference}}},\ }\href
  {https://doi.org/10.1103/PhysRevLett.130.171403} {\bibfield  {journal}
  {\bibinfo  {journal} {Phys. Rev. Lett.}\ }\textbf {\bibinfo {volume} {130}},\
  \bibinfo {pages} {171403} (\bibinfo {year} {2023})},\ \Eprint
  {https://arxiv.org/abs/2210.05686} {arXiv:2210.05686 [gr-qc]} \BibitemShut
  {NoStop}%
\bibitem [{\citenamefont {Bhardwaj}\ \emph {et~al.}(2023)\citenamefont
  {Bhardwaj}, \citenamefont {Alvey}, \citenamefont {Miller}, \citenamefont
  {Nissanke},\ and\ \citenamefont {Weniger}}]{Bhardwaj:2023xph}%
  \BibitemOpen
  \bibfield  {author} {\bibinfo {author} {\bibfnamefont {U.}~\bibnamefont
  {Bhardwaj}}, \bibinfo {author} {\bibfnamefont {J.}~\bibnamefont {Alvey}},
  \bibinfo {author} {\bibfnamefont {B.~K.}\ \bibnamefont {Miller}}, \bibinfo
  {author} {\bibfnamefont {S.}~\bibnamefont {Nissanke}},\ and\ \bibinfo
  {author} {\bibfnamefont {C.}~\bibnamefont {Weniger}},\ }\bibfield  {title}
  {\emph {\bibinfo {title} {{Sequential simulation-based inference for
  gravitational wave signals}}},\ }\href
  {https://doi.org/10.1103/PhysRevD.108.042004} {\bibfield  {journal} {\bibinfo
   {journal} {Phys. Rev. D}\ }\textbf {\bibinfo {volume} {108}},\ \bibinfo
  {pages} {042004} (\bibinfo {year} {2023})},\ \Eprint
  {https://arxiv.org/abs/2304.02035} {arXiv:2304.02035 [gr-qc]} \BibitemShut
  {NoStop}%
\bibitem [{\citenamefont {V{\'\i}lchez}\ and\ \citenamefont
  {Sopuerta}(2025)}]{Vilchez:2024qnw}%
  \BibitemOpen
  \bibfield  {author} {\bibinfo {author} {\bibfnamefont {I.~M.}\ \bibnamefont
  {V{\'\i}lchez}}\ and\ \bibinfo {author} {\bibfnamefont {C.~F.}\ \bibnamefont
  {Sopuerta}},\ }\bibfield  {title} {\emph {\bibinfo {title} {{Efficient
  Massive Black Hole Binary parameter estimation for LISA using Sequential
  Neural Likelihood}}},\ }\href {https://doi.org/10.1088/1475-7516/2025/04/022}
  {\bibfield  {journal} {\bibinfo  {journal} {JCAP}\ }\textbf {\bibinfo
  {volume} {04}},\ \bibinfo {pages} {022}},\ \Eprint
  {https://arxiv.org/abs/2406.00565} {arXiv:2406.00565 [gr-qc]} \BibitemShut
  {NoStop}%
\bibitem [{\citenamefont {Spadaro}\ \emph {et~al.}(2026)\citenamefont
  {Spadaro}, \citenamefont {Gair}, \citenamefont {Gerosa}, \citenamefont
  {Green}, \citenamefont {Buscicchio}, \citenamefont {Gupte}, \citenamefont
  {Tenorio}, \citenamefont {Clyne}, \citenamefont {P{\"u}rrer},\ and\
  \citenamefont {Korsakova}}]{Spadaro:2026evb}%
  \BibitemOpen
  \bibfield  {author} {\bibinfo {author} {\bibfnamefont {A.}~\bibnamefont
  {Spadaro}}, \bibinfo {author} {\bibfnamefont {J.}~\bibnamefont {Gair}},
  \bibinfo {author} {\bibfnamefont {D.}~\bibnamefont {Gerosa}}, \bibinfo
  {author} {\bibfnamefont {S.~R.}\ \bibnamefont {Green}}, \bibinfo {author}
  {\bibfnamefont {R.}~\bibnamefont {Buscicchio}}, \bibinfo {author}
  {\bibfnamefont {N.}~\bibnamefont {Gupte}}, \bibinfo {author} {\bibfnamefont
  {R.}~\bibnamefont {Tenorio}}, \bibinfo {author} {\bibfnamefont
  {S.}~\bibnamefont {Clyne}}, \bibinfo {author} {\bibfnamefont
  {M.}~\bibnamefont {P{\"u}rrer}},\ and\ \bibinfo {author} {\bibfnamefont
  {N.}~\bibnamefont {Korsakova}},\ }\bibfield  {title} {\emph {\bibinfo {title}
  {{Accurate and efficient simulation-based inference for massive black-hole
  binaries with LISA}}},\ }\href {https://doi.org/10.1103/mmyl-wdgq} {\bibfield
   {journal} {\bibinfo  {journal} {Phys. Rev. D}\ }\textbf {\bibinfo {volume}
  {114}},\ \bibinfo {pages} {023038} (\bibinfo {year} {2026})},\ \Eprint
  {https://arxiv.org/abs/2603.20431} {arXiv:2603.20431 [astro-ph.HE]}
  \BibitemShut {NoStop}%
\bibitem [{\citenamefont {Srinivasan}\ \emph {et~al.}(2025)\citenamefont
  {Srinivasan}, \citenamefont {Barausse}, \citenamefont {Korsakova},\ and\
  \citenamefont {Trotta}}]{Srinivasan:2025etu}%
  \BibitemOpen
  \bibfield  {author} {\bibinfo {author} {\bibfnamefont {R.}~\bibnamefont
  {Srinivasan}}, \bibinfo {author} {\bibfnamefont {E.}~\bibnamefont
  {Barausse}}, \bibinfo {author} {\bibfnamefont {N.}~\bibnamefont
  {Korsakova}},\ and\ \bibinfo {author} {\bibfnamefont {R.}~\bibnamefont
  {Trotta}},\ }\bibfield  {title} {\emph {\bibinfo {title} {{Simulation-based
  population inference of LISA{\textquoteright}s Galactic binaries: Bypassing
  the global fit}}},\ }\href {https://doi.org/10.1103/shym-w46f} {\bibfield
  {journal} {\bibinfo  {journal} {Phys. Rev. D}\ }\textbf {\bibinfo {volume}
  {112}},\ \bibinfo {pages} {103043} (\bibinfo {year} {2025})},\ \Eprint
  {https://arxiv.org/abs/2506.22543} {arXiv:2506.22543 [astro-ph.GA]}
  \BibitemShut {NoStop}%
\bibitem [{\citenamefont {Alvey}\ \emph {et~al.}(2024)\citenamefont {Alvey},
  \citenamefont {Bhardwaj}, \citenamefont {Domcke}, \citenamefont {Pieroni},\
  and\ \citenamefont {Weniger}}]{Alvey:2023npw}%
  \BibitemOpen
  \bibfield  {author} {\bibinfo {author} {\bibfnamefont {J.}~\bibnamefont
  {Alvey}}, \bibinfo {author} {\bibfnamefont {U.}~\bibnamefont {Bhardwaj}},
  \bibinfo {author} {\bibfnamefont {V.}~\bibnamefont {Domcke}}, \bibinfo
  {author} {\bibfnamefont {M.}~\bibnamefont {Pieroni}},\ and\ \bibinfo {author}
  {\bibfnamefont {C.}~\bibnamefont {Weniger}},\ }\bibfield  {title} {\emph
  {\bibinfo {title} {{Simulation-based inference for stochastic gravitational
  wave background data analysis}}},\ }\href
  {https://doi.org/10.1103/PhysRevD.109.083008} {\bibfield  {journal} {\bibinfo
   {journal} {Phys. Rev. D}\ }\textbf {\bibinfo {volume} {109}},\ \bibinfo
  {pages} {083008} (\bibinfo {year} {2024})},\ \Eprint
  {https://arxiv.org/abs/2309.07954} {arXiv:2309.07954 [gr-qc]} \BibitemShut
  {NoStop}%
\bibitem [{\citenamefont {Dimitriou}\ \emph {et~al.}(2024)\citenamefont
  {Dimitriou}, \citenamefont {Figueroa},\ and\ \citenamefont
  {Zaldivar}}]{Dimitriou:2023knw}%
  \BibitemOpen
  \bibfield  {author} {\bibinfo {author} {\bibfnamefont {A.}~\bibnamefont
  {Dimitriou}}, \bibinfo {author} {\bibfnamefont {D.~G.}\ \bibnamefont
  {Figueroa}},\ and\ \bibinfo {author} {\bibfnamefont {B.}~\bibnamefont
  {Zaldivar}},\ }\bibfield  {title} {\emph {\bibinfo {title} {{Fast
  likelihood-free reconstruction of gravitational wave backgrounds}}},\ }\href
  {https://doi.org/10.1088/1475-7516/2024/09/032} {\bibfield  {journal}
  {\bibinfo  {journal} {JCAP}\ }\textbf {\bibinfo {volume} {09}},\ \bibinfo
  {pages} {032}},\ \Eprint {https://arxiv.org/abs/2309.08430} {arXiv:2309.08430
  [astro-ph.CO]} \BibitemShut {NoStop}%
\bibitem [{\citenamefont {Dimitriou}\ \emph {et~al.}(2026)\citenamefont
  {Dimitriou}, \citenamefont {Figueroa}, \citenamefont {Simakachorn},
  \citenamefont {Stomberg},\ and\ \citenamefont
  {Zaldivar}}]{Dimitriou:2026agw}%
  \BibitemOpen
  \bibfield  {author} {\bibinfo {author} {\bibfnamefont {A.}~\bibnamefont
  {Dimitriou}}, \bibinfo {author} {\bibfnamefont {D.~G.}\ \bibnamefont
  {Figueroa}}, \bibinfo {author} {\bibfnamefont {P.}~\bibnamefont
  {Simakachorn}}, \bibinfo {author} {\bibfnamefont {I.}~\bibnamefont
  {Stomberg}},\ and\ \bibinfo {author} {\bibfnamefont {B.}~\bibnamefont
  {Zaldivar}},\ }\bibfield  {title} {\emph {\bibinfo {title} {{Cosmic string
  gravitational wave backgrounds at LISA: II. Reconstruction of conventional
  signals over astrophysical foregrounds}}},\ }\href@noop {} {\  (\bibinfo
  {year} {2026})},\ \Eprint {https://arxiv.org/abs/2607.15219}
  {arXiv:2607.15219 [astro-ph.CO]} \BibitemShut {NoStop}%
\bibitem [{\citenamefont {Lemke}\ \emph {et~al.}(2026)\citenamefont {Lemke},
  \citenamefont {Mitridate}, \citenamefont {Konstandin}, \citenamefont
  {Pieroni},\ and\ \citenamefont {Alvey}}]{Lemke:2026gxj}%
  \BibitemOpen
  \bibfield  {author} {\bibinfo {author} {\bibfnamefont {A.-M.}\ \bibnamefont
  {Lemke}}, \bibinfo {author} {\bibfnamefont {A.}~\bibnamefont {Mitridate}},
  \bibinfo {author} {\bibfnamefont {T.}~\bibnamefont {Konstandin}}, \bibinfo
  {author} {\bibfnamefont {M.}~\bibnamefont {Pieroni}},\ and\ \bibinfo {author}
  {\bibfnamefont {J.}~\bibnamefont {Alvey}},\ }\bibfield  {title} {\emph
  {\bibinfo {title} {{Detecting Gravitational-Wave Anisotropies with
  Simulation-Based Inference}}},\ }\href@noop {} {\  (\bibinfo {year}
  {2026})},\ \Eprint {https://arxiv.org/abs/2605.24082} {arXiv:2605.24082
  [astro-ph.CO]} \BibitemShut {NoStop}%
\bibitem [{\citenamefont {Barger}\ \emph {et~al.}(2009)\citenamefont {Barger},
  \citenamefont {Langacker}, \citenamefont {McCaskey}, \citenamefont
  {Ramsey-Musolf},\ and\ \citenamefont {Shaughnessy}}]{Barger:2008jx}%
  \BibitemOpen
  \bibfield  {author} {\bibinfo {author} {\bibfnamefont {V.}~\bibnamefont
  {Barger}}, \bibinfo {author} {\bibfnamefont {P.}~\bibnamefont {Langacker}},
  \bibinfo {author} {\bibfnamefont {M.}~\bibnamefont {McCaskey}}, \bibinfo
  {author} {\bibfnamefont {M.}~\bibnamefont {Ramsey-Musolf}},\ and\ \bibinfo
  {author} {\bibfnamefont {G.}~\bibnamefont {Shaughnessy}},\ }\bibfield
  {title} {\emph {\bibinfo {title} {{Complex Singlet Extension of the Standard
  Model}}},\ }\href {https://doi.org/10.1103/PhysRevD.79.015018} {\bibfield
  {journal} {\bibinfo  {journal} {Phys. Rev. D}\ }\textbf {\bibinfo {volume}
  {79}},\ \bibinfo {pages} {015018} (\bibinfo {year} {2009})},\ \Eprint
  {https://arxiv.org/abs/0811.0393} {arXiv:0811.0393 [hep-ph]} \BibitemShut
  {NoStop}%
\bibitem [{\citenamefont {Gonderinger}\ \emph {et~al.}(2012)\citenamefont
  {Gonderinger}, \citenamefont {Lim},\ and\ \citenamefont
  {Ramsey-Musolf}}]{Gonderinger:2012rd}%
  \BibitemOpen
  \bibfield  {author} {\bibinfo {author} {\bibfnamefont {M.}~\bibnamefont
  {Gonderinger}}, \bibinfo {author} {\bibfnamefont {H.}~\bibnamefont {Lim}},\
  and\ \bibinfo {author} {\bibfnamefont {M.~J.}\ \bibnamefont
  {Ramsey-Musolf}},\ }\bibfield  {title} {\emph {\bibinfo {title} {{Complex
  Scalar Singlet Dark Matter: Vacuum Stability and Phenomenology}}},\ }\href
  {https://doi.org/10.1103/PhysRevD.86.043511} {\bibfield  {journal} {\bibinfo
  {journal} {Phys. Rev. D}\ }\textbf {\bibinfo {volume} {86}},\ \bibinfo
  {pages} {043511} (\bibinfo {year} {2012})},\ \Eprint
  {https://arxiv.org/abs/1202.1316} {arXiv:1202.1316 [hep-ph]} \BibitemShut
  {NoStop}%
\bibitem [{\citenamefont {Biek{\"o}tter}\ \emph {et~al.}(2025)\citenamefont
  {Biek{\"o}tter}, \citenamefont {Dashko}, \citenamefont {L{\"o}schner},\ and\
  \citenamefont {Weiglein}}]{Biekotter:2025npc}%
  \BibitemOpen
  \bibfield  {author} {\bibinfo {author} {\bibfnamefont {T.}~\bibnamefont
  {Biek{\"o}tter}}, \bibinfo {author} {\bibfnamefont {A.}~\bibnamefont
  {Dashko}}, \bibinfo {author} {\bibfnamefont {M.}~\bibnamefont
  {L{\"o}schner}},\ and\ \bibinfo {author} {\bibfnamefont {G.}~\bibnamefont
  {Weiglein}},\ }\bibfield  {title} {\emph {\bibinfo {title} {{Perturbative
  aspects of the electroweak phase transition with a complex singlet and
  implications for gravitational wave predictions}}},\ }\href@noop {} {\
  (\bibinfo {year} {2025})},\ \Eprint {https://arxiv.org/abs/2511.14831}
  {arXiv:2511.14831 [hep-ph]} \BibitemShut {NoStop}%
\bibitem [{\citenamefont {Lewicki}\ \emph {et~al.}(2024)\citenamefont
  {Lewicki}, \citenamefont {Merchand}, \citenamefont {Sagunski}, \citenamefont
  {Schicho},\ and\ \citenamefont {Schmitt}}]{Lewicki:2024xan}%
  \BibitemOpen
  \bibfield  {author} {\bibinfo {author} {\bibfnamefont {M.}~\bibnamefont
  {Lewicki}}, \bibinfo {author} {\bibfnamefont {M.}~\bibnamefont {Merchand}},
  \bibinfo {author} {\bibfnamefont {L.}~\bibnamefont {Sagunski}}, \bibinfo
  {author} {\bibfnamefont {P.}~\bibnamefont {Schicho}},\ and\ \bibinfo {author}
  {\bibfnamefont {D.}~\bibnamefont {Schmitt}},\ }\bibfield  {title} {\emph
  {\bibinfo {title} {{Impact of theoretical uncertainties on model parameter
  reconstruction from GW signals sourced by cosmological phase transitions}}},\
  }\href {https://doi.org/10.1103/PhysRevD.110.023538} {\bibfield  {journal}
  {\bibinfo  {journal} {Phys. Rev. D}\ }\textbf {\bibinfo {volume} {110}},\
  \bibinfo {pages} {023538} (\bibinfo {year} {2024})},\ \Eprint
  {https://arxiv.org/abs/2403.03769} {arXiv:2403.03769 [hep-ph]} \BibitemShut
  {NoStop}%
\bibitem [{\citenamefont {Ashoorioon}\ and\ \citenamefont
  {Konstandin}(2009)}]{Ashoorioon:2009nf}%
  \BibitemOpen
  \bibfield  {author} {\bibinfo {author} {\bibfnamefont {A.}~\bibnamefont
  {Ashoorioon}}\ and\ \bibinfo {author} {\bibfnamefont {T.}~\bibnamefont
  {Konstandin}},\ }\bibfield  {title} {\emph {\bibinfo {title} {{Strong
  electroweak phase transitions without collider traces}}},\ }\href
  {https://doi.org/10.1088/1126-6708/2009/07/086} {\bibfield  {journal}
  {\bibinfo  {journal} {JHEP}\ }\textbf {\bibinfo {volume} {07}},\ \bibinfo
  {pages} {086}},\ \Eprint {https://arxiv.org/abs/0904.0353} {arXiv:0904.0353
  [hep-ph]} \BibitemShut {NoStop}%
\bibitem [{\citenamefont {Curtin}\ \emph {et~al.}(2014)\citenamefont {Curtin},
  \citenamefont {Meade},\ and\ \citenamefont {Yu}}]{Curtin:2014jma}%
  \BibitemOpen
  \bibfield  {author} {\bibinfo {author} {\bibfnamefont {D.}~\bibnamefont
  {Curtin}}, \bibinfo {author} {\bibfnamefont {P.}~\bibnamefont {Meade}},\ and\
  \bibinfo {author} {\bibfnamefont {C.-T.}\ \bibnamefont {Yu}},\ }\bibfield
  {title} {\emph {\bibinfo {title} {{Testing Electroweak Baryogenesis with
  Future Colliders}}},\ }\href {https://doi.org/10.1007/JHEP11(2014)127}
  {\bibfield  {journal} {\bibinfo  {journal} {JHEP}\ }\textbf {\bibinfo
  {volume} {11}},\ \bibinfo {pages} {127}},\ \Eprint
  {https://arxiv.org/abs/1409.0005} {arXiv:1409.0005 [hep-ph]} \BibitemShut
  {NoStop}%
\bibitem [{\citenamefont {Chiang}\ \emph {et~al.}(2018)\citenamefont {Chiang},
  \citenamefont {Ramsey-Musolf},\ and\ \citenamefont
  {Senaha}}]{Chiang:2017nmu}%
  \BibitemOpen
  \bibfield  {author} {\bibinfo {author} {\bibfnamefont {C.-W.}\ \bibnamefont
  {Chiang}}, \bibinfo {author} {\bibfnamefont {M.~J.}\ \bibnamefont
  {Ramsey-Musolf}},\ and\ \bibinfo {author} {\bibfnamefont {E.}~\bibnamefont
  {Senaha}},\ }\bibfield  {title} {\emph {\bibinfo {title} {{Standard Model
  with a Complex Scalar Singlet: Cosmological Implications and Theoretical
  Considerations}}},\ }\href {https://doi.org/10.1103/PhysRevD.97.015005}
  {\bibfield  {journal} {\bibinfo  {journal} {Phys. Rev. D}\ }\textbf {\bibinfo
  {volume} {97}},\ \bibinfo {pages} {015005} (\bibinfo {year} {2018})},\
  \Eprint {https://arxiv.org/abs/1707.09960} {arXiv:1707.09960 [hep-ph]}
  \BibitemShut {NoStop}%
\bibitem [{\citenamefont {Aalbers}\ \emph {et~al.}(2023)\citenamefont {Aalbers}
  \emph {et~al.}}]{LZ:2022lsv}%
  \BibitemOpen
  \bibfield  {author} {\bibinfo {author} {\bibfnamefont {J.}~\bibnamefont
  {Aalbers}} \emph {et~al.} (\bibinfo {collaboration} {LZ}),\ }\bibfield
  {title} {\emph {\bibinfo {title} {{First Dark Matter Search Results from the
  LUX-ZEPLIN (LZ) Experiment}}},\ }\href
  {https://doi.org/10.1103/PhysRevLett.131.041002} {\bibfield  {journal}
  {\bibinfo  {journal} {Phys. Rev. Lett.}\ }\textbf {\bibinfo {volume} {131}},\
  \bibinfo {pages} {041002} (\bibinfo {year} {2023})},\ \Eprint
  {https://arxiv.org/abs/2207.03764} {arXiv:2207.03764 [hep-ex]} \BibitemShut
  {NoStop}%
\bibitem [{\citenamefont {Biek{\"o}tter}\ and\ \citenamefont
  {Pierre}(2022)}]{Biekotter:2022ckj}%
  \BibitemOpen
  \bibfield  {author} {\bibinfo {author} {\bibfnamefont {T.}~\bibnamefont
  {Biek{\"o}tter}}\ and\ \bibinfo {author} {\bibfnamefont {M.}~\bibnamefont
  {Pierre}},\ }\bibfield  {title} {\emph {\bibinfo {title} {{Higgs-boson
  visible and invisible constraints on hidden sectors}}},\ }\href
  {https://doi.org/10.1140/epjc/s10052-022-10990-x} {\bibfield  {journal}
  {\bibinfo  {journal} {Eur. Phys. J. C}\ }\textbf {\bibinfo {volume} {82}},\
  \bibinfo {pages} {1026} (\bibinfo {year} {2022})},\ \Eprint
  {https://arxiv.org/abs/2208.05505} {arXiv:2208.05505 [hep-ph]} \BibitemShut
  {NoStop}%
\bibitem [{\citenamefont {Arcadi}\ \emph {et~al.}(2024)\citenamefont {Arcadi},
  \citenamefont {Costa}, \citenamefont {Goudelis},\ and\ \citenamefont
  {Lebedev}}]{Arcadi:2024wwg}%
  \BibitemOpen
  \bibfield  {author} {\bibinfo {author} {\bibfnamefont {G.}~\bibnamefont
  {Arcadi}}, \bibinfo {author} {\bibfnamefont {F.}~\bibnamefont {Costa}},
  \bibinfo {author} {\bibfnamefont {A.}~\bibnamefont {Goudelis}},\ and\
  \bibinfo {author} {\bibfnamefont {O.}~\bibnamefont {Lebedev}},\ }\bibfield
  {title} {\emph {\bibinfo {title} {{Higgs portal dark matter freeze-in at
  stronger coupling: observational benchmarks}}},\ }\href
  {https://doi.org/10.1007/JHEP07(2024)044} {\bibfield  {journal} {\bibinfo
  {journal} {JHEP}\ }\textbf {\bibinfo {volume} {07}},\ \bibinfo {pages}
  {044}},\ \Eprint {https://arxiv.org/abs/2405.03760} {arXiv:2405.03760
  [hep-ph]} \BibitemShut {NoStop}%
\bibitem [{\citenamefont {Escudero~Abenza}\ and\ \citenamefont
  {Hambye}(2025)}]{EscuderoAbenza:2025cfj}%
  \BibitemOpen
  \bibfield  {author} {\bibinfo {author} {\bibfnamefont {M.}~\bibnamefont
  {Escudero~Abenza}}\ and\ \bibinfo {author} {\bibfnamefont {T.}~\bibnamefont
  {Hambye}},\ }\bibfield  {title} {\emph {\bibinfo {title} {{The simplest dark
  matter model at the edge of perturbativity}}},\ }\href
  {https://doi.org/10.1016/j.physletb.2025.139696} {\bibfield  {journal}
  {\bibinfo  {journal} {Phys. Lett. B}\ }\textbf {\bibinfo {volume} {868}},\
  \bibinfo {pages} {139696} (\bibinfo {year} {2025})},\ \Eprint
  {https://arxiv.org/abs/2505.02408} {arXiv:2505.02408 [hep-ph]} \BibitemShut
  {NoStop}%
\bibitem [{\citenamefont {Farakos}\ \emph {et~al.}(1994)\citenamefont
  {Farakos}, \citenamefont {Kajantie}, \citenamefont {Rummukainen},\ and\
  \citenamefont {Shaposhnikov}}]{Farakos:1994kx}%
  \BibitemOpen
  \bibfield  {author} {\bibinfo {author} {\bibfnamefont {K.}~\bibnamefont
  {Farakos}}, \bibinfo {author} {\bibfnamefont {K.}~\bibnamefont {Kajantie}},
  \bibinfo {author} {\bibfnamefont {K.}~\bibnamefont {Rummukainen}},\ and\
  \bibinfo {author} {\bibfnamefont {M.~E.}\ \bibnamefont {Shaposhnikov}},\
  }\bibfield  {title} {\emph {\bibinfo {title} {{3-D physics and the
  electroweak phase transition: Perturbation theory}}},\ }\href
  {https://doi.org/10.1016/0550-3213(94)90173-2} {\bibfield  {journal}
  {\bibinfo  {journal} {Nucl. Phys. B}\ }\textbf {\bibinfo {volume} {425}},\
  \bibinfo {pages} {67} (\bibinfo {year} {1994})},\ \Eprint
  {https://arxiv.org/abs/hep-ph/9404201} {arXiv:hep-ph/9404201} \BibitemShut
  {NoStop}%
\bibitem [{\citenamefont {Kajantie}\ \emph
  {et~al.}(1996{\natexlab{b}})\citenamefont {Kajantie}, \citenamefont {Laine},
  \citenamefont {Rummukainen},\ and\ \citenamefont
  {Shaposhnikov}}]{Kajantie:1995dw}%
  \BibitemOpen
  \bibfield  {author} {\bibinfo {author} {\bibfnamefont {K.}~\bibnamefont
  {Kajantie}}, \bibinfo {author} {\bibfnamefont {M.}~\bibnamefont {Laine}},
  \bibinfo {author} {\bibfnamefont {K.}~\bibnamefont {Rummukainen}},\ and\
  \bibinfo {author} {\bibfnamefont {M.~E.}\ \bibnamefont {Shaposhnikov}},\
  }\bibfield  {title} {\emph {\bibinfo {title} {{Generic rules for high
  temperature dimensional reduction and their application to the standard
  model}}},\ }\href {https://doi.org/10.1016/0550-3213(95)00549-8} {\bibfield
  {journal} {\bibinfo  {journal} {Nucl. Phys. B}\ }\textbf {\bibinfo {volume}
  {458}},\ \bibinfo {pages} {90} (\bibinfo {year} {1996}{\natexlab{b}})},\
  \Eprint {https://arxiv.org/abs/hep-ph/9508379} {arXiv:hep-ph/9508379}
  \BibitemShut {NoStop}%
\bibitem [{\citenamefont {Schicho}\ \emph {et~al.}(2022)\citenamefont
  {Schicho}, \citenamefont {Tenkanen},\ and\ \citenamefont
  {White}}]{Schicho:2022wty}%
  \BibitemOpen
  \bibfield  {author} {\bibinfo {author} {\bibfnamefont {P.}~\bibnamefont
  {Schicho}}, \bibinfo {author} {\bibfnamefont {T.~V.~I.}\ \bibnamefont
  {Tenkanen}},\ and\ \bibinfo {author} {\bibfnamefont {G.}~\bibnamefont
  {White}},\ }\bibfield  {title} {\emph {\bibinfo {title} {{Combining thermal
  resummation and gauge invariance for electroweak phase transition}}},\ }\href
  {https://doi.org/10.1007/JHEP11(2022)047} {\bibfield  {journal} {\bibinfo
  {journal} {JHEP}\ }\textbf {\bibinfo {volume} {11}},\ \bibinfo {pages}
  {047}},\ \Eprint {https://arxiv.org/abs/2203.04284} {arXiv:2203.04284
  [hep-ph]} \BibitemShut {NoStop}%
\bibitem [{\citenamefont {Croon}\ \emph {et~al.}(2021)\citenamefont {Croon},
  \citenamefont {Gould}, \citenamefont {Schicho}, \citenamefont {Tenkanen},\
  and\ \citenamefont {White}}]{Croon:2020cgk}%
  \BibitemOpen
  \bibfield  {author} {\bibinfo {author} {\bibfnamefont {D.}~\bibnamefont
  {Croon}}, \bibinfo {author} {\bibfnamefont {O.}~\bibnamefont {Gould}},
  \bibinfo {author} {\bibfnamefont {P.}~\bibnamefont {Schicho}}, \bibinfo
  {author} {\bibfnamefont {T.~V.~I.}\ \bibnamefont {Tenkanen}},\ and\ \bibinfo
  {author} {\bibfnamefont {G.}~\bibnamefont {White}},\ }\bibfield  {title}
  {\emph {\bibinfo {title} {{Theoretical uncertainties for cosmological
  first-order phase transitions}}},\ }\href
  {https://doi.org/10.1007/JHEP04(2021)055} {\bibfield  {journal} {\bibinfo
  {journal} {JHEP}\ }\textbf {\bibinfo {volume} {04}},\ \bibinfo {pages}
  {055}},\ \Eprint {https://arxiv.org/abs/2009.10080} {arXiv:2009.10080
  [hep-ph]} \BibitemShut {NoStop}%
\bibitem [{\citenamefont {Gould}\ and\ \citenamefont
  {Tenkanen}(2021)}]{Gould:2021oba}%
  \BibitemOpen
  \bibfield  {author} {\bibinfo {author} {\bibfnamefont {O.}~\bibnamefont
  {Gould}}\ and\ \bibinfo {author} {\bibfnamefont {T.~V.~I.}\ \bibnamefont
  {Tenkanen}},\ }\bibfield  {title} {\emph {\bibinfo {title} {{On the
  perturbative expansion at high temperature and implications for cosmological
  phase transitions}}},\ }\href {https://doi.org/10.1007/JHEP06(2021)069}
  {\bibfield  {journal} {\bibinfo  {journal} {JHEP}\ }\textbf {\bibinfo
  {volume} {06}},\ \bibinfo {pages} {069}},\ \Eprint
  {https://arxiv.org/abs/2104.04399} {arXiv:2104.04399 [hep-ph]} \BibitemShut
  {NoStop}%
\bibitem [{\citenamefont {Linde}(1981)}]{Linde:1980tt}%
  \BibitemOpen
  \bibfield  {author} {\bibinfo {author} {\bibfnamefont {A.~D.}\ \bibnamefont
  {Linde}},\ }\bibfield  {title} {\emph {\bibinfo {title} {{Fate of the False
  Vacuum at Finite Temperature: Theory and Applications}}},\ }\href
  {https://doi.org/10.1016/0370-2693(81)90281-1} {\bibfield  {journal}
  {\bibinfo  {journal} {Phys. Lett. B}\ }\textbf {\bibinfo {volume} {100}},\
  \bibinfo {pages} {37} (\bibinfo {year} {1981})}\BibitemShut {NoStop}%
\bibitem [{\citenamefont {Callan}\ and\ \citenamefont
  {Coleman}(1977)}]{Callan:1977pt}%
  \BibitemOpen
  \bibfield  {author} {\bibinfo {author} {\bibfnamefont {C.~G.}\ \bibnamefont
  {Callan}, \bibfnamefont {Jr.}}\ and\ \bibinfo {author} {\bibfnamefont
  {S.~R.}\ \bibnamefont {Coleman}},\ }\bibfield  {title} {\emph {\bibinfo
  {title} {{The Fate of the False Vacuum. 2. First Quantum Corrections}}},\
  }\href {https://doi.org/10.1103/PhysRevD.16.1762} {\bibfield  {journal}
  {\bibinfo  {journal} {Phys. Rev. D}\ }\textbf {\bibinfo {volume} {16}},\
  \bibinfo {pages} {1762} (\bibinfo {year} {1977})}\BibitemShut {NoStop}%
\bibitem [{\citenamefont {Quiros}(1999)}]{Quiros:1999jp}%
  \BibitemOpen
  \bibfield  {author} {\bibinfo {author} {\bibfnamefont {M.}~\bibnamefont
  {Quiros}},\ }in\ \href@noop {} {\emph {\bibinfo {booktitle} {{ICTP Summer
  School in High-Energy Physics and Cosmology}}}}\ (\bibinfo {year} {1999})\
  pp.\ \bibinfo {pages} {187--259},\ \Eprint
  {https://arxiv.org/abs/hep-ph/9901312} {arXiv:hep-ph/9901312} \BibitemShut
  {NoStop}%
\bibitem [{\citenamefont {Athron}\ \emph {et~al.}(2024)\citenamefont {Athron},
  \citenamefont {Bal{\'a}zs}, \citenamefont {Fowlie}, \citenamefont {Morris},\
  and\ \citenamefont {Wu}}]{Athron:2023xlk}%
  \BibitemOpen
  \bibfield  {author} {\bibinfo {author} {\bibfnamefont {P.}~\bibnamefont
  {Athron}}, \bibinfo {author} {\bibfnamefont {C.}~\bibnamefont {Bal{\'a}zs}},
  \bibinfo {author} {\bibfnamefont {A.}~\bibnamefont {Fowlie}}, \bibinfo
  {author} {\bibfnamefont {L.}~\bibnamefont {Morris}},\ and\ \bibinfo {author}
  {\bibfnamefont {L.}~\bibnamefont {Wu}},\ }\bibfield  {title} {\emph {\bibinfo
  {title} {{Cosmological phase transitions: From perturbative particle physics
  to gravitational waves}}},\ }\href
  {https://doi.org/10.1016/j.ppnp.2023.104094} {\bibfield  {journal} {\bibinfo
  {journal} {Prog. Part. Nucl. Phys.}\ }\textbf {\bibinfo {volume} {135}},\
  \bibinfo {pages} {104094} (\bibinfo {year} {2024})},\ \Eprint
  {https://arxiv.org/abs/2305.02357} {arXiv:2305.02357 [hep-ph]} \BibitemShut
  {NoStop}%
\bibitem [{\citenamefont {Auclair}\ \emph {et~al.}(2023)\citenamefont {Auclair}
  \emph {et~al.}}]{LISACosmologyWorkingGroup:2022jok}%
  \BibitemOpen
  \bibfield  {author} {\bibinfo {author} {\bibfnamefont {P.}~\bibnamefont
  {Auclair}} \emph {et~al.} (\bibinfo {collaboration} {LISA Cosmology Working
  Group}),\ }\bibfield  {title} {\emph {\bibinfo {title} {{Cosmology with the
  Laser Interferometer Space Antenna}}},\ }\href
  {https://doi.org/10.1007/s41114-023-00045-2} {\bibfield  {journal} {\bibinfo
  {journal} {Living Rev. Rel.}\ }\textbf {\bibinfo {volume} {26}},\ \bibinfo
  {pages} {5} (\bibinfo {year} {2023})},\ \Eprint
  {https://arxiv.org/abs/2204.05434} {arXiv:2204.05434 [astro-ph.CO]}
  \BibitemShut {NoStop}%
\bibitem [{\citenamefont {De~Curtis}\ \emph {et~al.}(2024)\citenamefont
  {De~Curtis}, \citenamefont {Delle~Rose}, \citenamefont {Guiggiani},
  \citenamefont {Gil~Muyor},\ and\ \citenamefont {Panico}}]{DeCurtis:2024hvh}%
  \BibitemOpen
  \bibfield  {author} {\bibinfo {author} {\bibfnamefont {S.}~\bibnamefont
  {De~Curtis}}, \bibinfo {author} {\bibfnamefont {L.}~\bibnamefont
  {Delle~Rose}}, \bibinfo {author} {\bibfnamefont {A.}~\bibnamefont
  {Guiggiani}}, \bibinfo {author} {\bibfnamefont {{\'A}.}~\bibnamefont
  {Gil~Muyor}},\ and\ \bibinfo {author} {\bibfnamefont {G.}~\bibnamefont
  {Panico}},\ }\bibfield  {title} {\emph {\bibinfo {title} {{Non-linearities in
  cosmological bubble wall dynamics}}},\ }\href
  {https://doi.org/10.1007/JHEP05(2024)009} {\bibfield  {journal} {\bibinfo
  {journal} {JHEP}\ }\textbf {\bibinfo {volume} {05}},\ \bibinfo {pages}
  {009}},\ \Eprint {https://arxiv.org/abs/2401.13522} {arXiv:2401.13522
  [hep-ph]} \BibitemShut {NoStop}%
\bibitem [{\citenamefont {Branchina}\ \emph {et~al.}(2025)\citenamefont
  {Branchina}, \citenamefont {Conaci}, \citenamefont {Delle~Rose},\ and\
  \citenamefont {De~Curtis}}]{Branchina:2025jou}%
  \BibitemOpen
  \bibfield  {author} {\bibinfo {author} {\bibfnamefont {C.}~\bibnamefont
  {Branchina}}, \bibinfo {author} {\bibfnamefont {A.}~\bibnamefont {Conaci}},
  \bibinfo {author} {\bibfnamefont {L.}~\bibnamefont {Delle~Rose}},\ and\
  \bibinfo {author} {\bibfnamefont {S.}~\bibnamefont {De~Curtis}},\ }\bibfield
  {title} {\emph {\bibinfo {title} {{Electroweak phase transition and bubble
  wall velocity in local thermal equilibrium}}},\ }\href
  {https://doi.org/10.1103/lkht-1nv1} {\bibfield  {journal} {\bibinfo
  {journal} {Phys. Rev. D}\ }\textbf {\bibinfo {volume} {112}},\ \bibinfo
  {pages} {095008} (\bibinfo {year} {2025})},\ \Eprint
  {https://arxiv.org/abs/2504.21213} {arXiv:2504.21213 [hep-ph]} \BibitemShut
  {NoStop}%
\bibitem [{\citenamefont {van~de Vis}\ \emph {et~al.}(2026)\citenamefont
  {van~de Vis}, \citenamefont {Schicho}, \citenamefont {Niemi}, \citenamefont
  {Laurent}, \citenamefont {Hirvonen},\ and\ \citenamefont
  {Gould}}]{vandeVis:2025plm}%
  \BibitemOpen
  \bibfield  {author} {\bibinfo {author} {\bibfnamefont {J.}~\bibnamefont
  {van~de Vis}}, \bibinfo {author} {\bibfnamefont {P.}~\bibnamefont {Schicho}},
  \bibinfo {author} {\bibfnamefont {L.}~\bibnamefont {Niemi}}, \bibinfo
  {author} {\bibfnamefont {B.}~\bibnamefont {Laurent}}, \bibinfo {author}
  {\bibfnamefont {J.}~\bibnamefont {Hirvonen}},\ and\ \bibinfo {author}
  {\bibfnamefont {O.}~\bibnamefont {Gould}},\ }\bibfield  {title} {\emph
  {\bibinfo {title} {{WallGo investigates: Theoretical uncertainties in the
  bubble wall velocity}}},\ }\href {https://doi.org/10.1007/JHEP04(2026)041}
  {\bibfield  {journal} {\bibinfo  {journal} {JHEP}\ }\textbf {\bibinfo
  {volume} {04}},\ \bibinfo {pages} {041}},\ \Eprint
  {https://arxiv.org/abs/2510.27691} {arXiv:2510.27691 [hep-ph]} \BibitemShut
  {NoStop}%
\bibitem [{\citenamefont {Tang}\ \emph {et~al.}(2026)\citenamefont {Tang},
  \citenamefont {Huang},\ and\ \citenamefont {Idegawa}}]{Tang:2026dhj}%
  \BibitemOpen
  \bibfield  {author} {\bibinfo {author} {\bibfnamefont {S.}~\bibnamefont
  {Tang}}, \bibinfo {author} {\bibfnamefont {F.~P.}\ \bibnamefont {Huang}},\
  and\ \bibinfo {author} {\bibfnamefont {C.}~\bibnamefont {Idegawa}},\
  }\bibfield  {title} {\emph {\bibinfo {title} {{Precise bubble wall velocity
  in a specific phase transition pattern in the CxSM and beyond}}},\
  }\href@noop {} {\  (\bibinfo {year} {2026})},\ \Eprint
  {https://arxiv.org/abs/2609.11747} {arXiv:2609.11747 [hep-ph]} \BibitemShut
  {NoStop}%
\bibitem [{\citenamefont {Dorsch}\ \emph {et~al.}(2026)\citenamefont {Dorsch},
  \citenamefont {Lewicki},\ and\ \citenamefont {Pinto}}]{Dorsch:2026lmo}%
  \BibitemOpen
  \bibfield  {author} {\bibinfo {author} {\bibfnamefont {G.~C.}\ \bibnamefont
  {Dorsch}}, \bibinfo {author} {\bibfnamefont {M.}~\bibnamefont {Lewicki}},\
  and\ \bibinfo {author} {\bibfnamefont {D.~A.}\ \bibnamefont {Pinto}},\
  }\bibfield  {title} {\emph {\bibinfo {title} {{Benchmarking wall velocities
  in cosmological phase transitions: Fluid Ansatz and WallGo}}},\ }\href@noop
  {} {\  (\bibinfo {year} {2026})},\ \Eprint {https://arxiv.org/abs/2607.18207}
  {arXiv:2607.18207 [astro-ph.CO]} \BibitemShut {NoStop}%
\bibitem [{\citenamefont {Wainwright}(2012)}]{Wainwright:2011kj}%
  \BibitemOpen
  \bibfield  {author} {\bibinfo {author} {\bibfnamefont {C.~L.}\ \bibnamefont
  {Wainwright}},\ }\bibfield  {title} {\emph {\bibinfo {title}
  {{CosmoTransitions: Computing Cosmological Phase Transition Temperatures and
  Bubble Profiles with Multiple Fields}}},\ }\href
  {https://doi.org/10.1016/j.cpc.2012.04.004} {\bibfield  {journal} {\bibinfo
  {journal} {Comput. Phys. Commun.}\ }\textbf {\bibinfo {volume} {183}},\
  \bibinfo {pages} {2006} (\bibinfo {year} {2012})},\ \Eprint
  {https://arxiv.org/abs/1109.4189} {arXiv:1109.4189 [hep-ph]} \BibitemShut
  {NoStop}%
\bibitem [{\citenamefont {Biek{\"o}tter}(2026{\natexlab{b}})}]{fortransition}%
  \BibitemOpen
  \bibfield  {author} {\bibinfo {author} {\bibfnamefont {T.}~\bibnamefont
  {Biek{\"o}tter}},\ }\href@noop {} {\bibinfo {title} {{\textit{fortransition
  -- A Fortran port of the Python package CosmoTransitions}}}},\ \bibinfo
  {howpublished} {\url{https://github.com/thomasbiekoetter/fortransitions}}
  (\bibinfo {year} {2026}{\natexlab{b}})\BibitemShut {NoStop}%
\bibitem [{\citenamefont {Hindmarsh}\ \emph {et~al.}(2014)\citenamefont
  {Hindmarsh}, \citenamefont {Huber}, \citenamefont {Rummukainen},\ and\
  \citenamefont {Weir}}]{Hindmarsh:2013xza}%
  \BibitemOpen
  \bibfield  {author} {\bibinfo {author} {\bibfnamefont {M.}~\bibnamefont
  {Hindmarsh}}, \bibinfo {author} {\bibfnamefont {S.~J.}\ \bibnamefont
  {Huber}}, \bibinfo {author} {\bibfnamefont {K.}~\bibnamefont {Rummukainen}},\
  and\ \bibinfo {author} {\bibfnamefont {D.~J.}\ \bibnamefont {Weir}},\
  }\bibfield  {title} {\emph {\bibinfo {title} {{Gravitational waves from the
  sound of a first order phase transition}}},\ }\href
  {https://doi.org/10.1103/PhysRevLett.112.041301} {\bibfield  {journal}
  {\bibinfo  {journal} {Phys. Rev. Lett.}\ }\textbf {\bibinfo {volume} {112}},\
  \bibinfo {pages} {041301} (\bibinfo {year} {2014})},\ \Eprint
  {https://arxiv.org/abs/1304.2433} {arXiv:1304.2433 [hep-ph]} \BibitemShut
  {NoStop}%
\bibitem [{\citenamefont {Hindmarsh}\ \emph {et~al.}(2015)\citenamefont
  {Hindmarsh}, \citenamefont {Huber}, \citenamefont {Rummukainen},\ and\
  \citenamefont {Weir}}]{Hindmarsh:2015qta}%
  \BibitemOpen
  \bibfield  {author} {\bibinfo {author} {\bibfnamefont {M.}~\bibnamefont
  {Hindmarsh}}, \bibinfo {author} {\bibfnamefont {S.~J.}\ \bibnamefont
  {Huber}}, \bibinfo {author} {\bibfnamefont {K.}~\bibnamefont {Rummukainen}},\
  and\ \bibinfo {author} {\bibfnamefont {D.~J.}\ \bibnamefont {Weir}},\
  }\bibfield  {title} {\emph {\bibinfo {title} {{Numerical simulations of
  acoustically generated gravitational waves at a first order phase
  transition}}},\ }\href {https://doi.org/10.1103/PhysRevD.92.123009}
  {\bibfield  {journal} {\bibinfo  {journal} {Phys. Rev. D}\ }\textbf {\bibinfo
  {volume} {92}},\ \bibinfo {pages} {123009} (\bibinfo {year} {2015})},\
  \Eprint {https://arxiv.org/abs/1504.03291} {arXiv:1504.03291 [astro-ph.CO]}
  \BibitemShut {NoStop}%
\bibitem [{\citenamefont {Hindmarsh}(2018)}]{Hindmarsh:2016lnk}%
  \BibitemOpen
  \bibfield  {author} {\bibinfo {author} {\bibfnamefont {M.}~\bibnamefont
  {Hindmarsh}},\ }\bibfield  {title} {\emph {\bibinfo {title} {{Sound shell
  model for acoustic gravitational wave production at a first-order phase
  transition in the early Universe}}},\ }\href
  {https://doi.org/10.1103/PhysRevLett.120.071301} {\bibfield  {journal}
  {\bibinfo  {journal} {Phys. Rev. Lett.}\ }\textbf {\bibinfo {volume} {120}},\
  \bibinfo {pages} {071301} (\bibinfo {year} {2018})},\ \Eprint
  {https://arxiv.org/abs/1608.04735} {arXiv:1608.04735 [astro-ph.CO]}
  \BibitemShut {NoStop}%
\bibitem [{\citenamefont {Hindmarsh}\ \emph {et~al.}(2017)\citenamefont
  {Hindmarsh}, \citenamefont {Huber}, \citenamefont {Rummukainen},\ and\
  \citenamefont {Weir}}]{Hindmarsh:2017gnf}%
  \BibitemOpen
  \bibfield  {author} {\bibinfo {author} {\bibfnamefont {M.}~\bibnamefont
  {Hindmarsh}}, \bibinfo {author} {\bibfnamefont {S.~J.}\ \bibnamefont
  {Huber}}, \bibinfo {author} {\bibfnamefont {K.}~\bibnamefont {Rummukainen}},\
  and\ \bibinfo {author} {\bibfnamefont {D.~J.}\ \bibnamefont {Weir}},\
  }\bibfield  {title} {\emph {\bibinfo {title} {{Shape of the acoustic
  gravitational wave power spectrum from a first order phase transition}}},\
  }\href {https://doi.org/10.1103/PhysRevD.96.103520} {\bibfield  {journal}
  {\bibinfo  {journal} {Phys. Rev. D}\ }\textbf {\bibinfo {volume} {96}},\
  \bibinfo {pages} {103520} (\bibinfo {year} {2017})},\ \bibinfo {note}
  {[Erratum: Phys.Rev.D 101, 089902 (2020)]},\ \Eprint
  {https://arxiv.org/abs/1704.05871} {arXiv:1704.05871 [astro-ph.CO]}
  \BibitemShut {NoStop}%
\bibitem [{\citenamefont {Caprini}\ and\ \citenamefont
  {Durrer}(2006)}]{Caprini:2006jb}%
  \BibitemOpen
  \bibfield  {author} {\bibinfo {author} {\bibfnamefont {C.}~\bibnamefont
  {Caprini}}\ and\ \bibinfo {author} {\bibfnamefont {R.}~\bibnamefont
  {Durrer}},\ }\bibfield  {title} {\emph {\bibinfo {title} {{Gravitational
  waves from stochastic relativistic sources: Primordial turbulence and
  magnetic fields}}},\ }\href {https://doi.org/10.1103/PhysRevD.74.063521}
  {\bibfield  {journal} {\bibinfo  {journal} {Phys. Rev. D}\ }\textbf {\bibinfo
  {volume} {74}},\ \bibinfo {pages} {063521} (\bibinfo {year} {2006})},\
  \Eprint {https://arxiv.org/abs/astro-ph/0603476} {arXiv:astro-ph/0603476}
  \BibitemShut {NoStop}%
\bibitem [{\citenamefont {Gogoberidze}\ \emph {et~al.}(2007)\citenamefont
  {Gogoberidze}, \citenamefont {Kahniashvili},\ and\ \citenamefont
  {Kosowsky}}]{Gogoberidze:2007an}%
  \BibitemOpen
  \bibfield  {author} {\bibinfo {author} {\bibfnamefont {G.}~\bibnamefont
  {Gogoberidze}}, \bibinfo {author} {\bibfnamefont {T.}~\bibnamefont
  {Kahniashvili}},\ and\ \bibinfo {author} {\bibfnamefont {A.}~\bibnamefont
  {Kosowsky}},\ }\bibfield  {title} {\emph {\bibinfo {title} {{The Spectrum of
  Gravitational Radiation from Primordial Turbulence}}},\ }\href
  {https://doi.org/10.1103/PhysRevD.76.083002} {\bibfield  {journal} {\bibinfo
  {journal} {Phys. Rev. D}\ }\textbf {\bibinfo {volume} {76}},\ \bibinfo
  {pages} {083002} (\bibinfo {year} {2007})},\ \Eprint
  {https://arxiv.org/abs/0705.1733} {arXiv:0705.1733 [astro-ph]} \BibitemShut
  {NoStop}%
\bibitem [{\citenamefont {Caprini}\ \emph {et~al.}(2009)\citenamefont
  {Caprini}, \citenamefont {Durrer},\ and\ \citenamefont
  {Servant}}]{Caprini:2009yp}%
  \BibitemOpen
  \bibfield  {author} {\bibinfo {author} {\bibfnamefont {C.}~\bibnamefont
  {Caprini}}, \bibinfo {author} {\bibfnamefont {R.}~\bibnamefont {Durrer}},\
  and\ \bibinfo {author} {\bibfnamefont {G.}~\bibnamefont {Servant}},\
  }\bibfield  {title} {\emph {\bibinfo {title} {{The stochastic gravitational
  wave background from turbulence and magnetic fields generated by a
  first-order phase transition}}},\ }\href
  {https://doi.org/10.1088/1475-7516/2009/12/024} {\bibfield  {journal}
  {\bibinfo  {journal} {JCAP}\ }\textbf {\bibinfo {volume} {12}},\ \bibinfo
  {pages} {024}},\ \Eprint {https://arxiv.org/abs/0909.0622} {arXiv:0909.0622
  [astro-ph.CO]} \BibitemShut {NoStop}%
\bibitem [{\citenamefont {Roper~Pol}\ \emph {et~al.}(2020)\citenamefont
  {Roper~Pol}, \citenamefont {Mandal}, \citenamefont {Brandenburg},
  \citenamefont {Kahniashvili},\ and\ \citenamefont
  {Kosowsky}}]{RoperPol:2019wvy}%
  \BibitemOpen
  \bibfield  {author} {\bibinfo {author} {\bibfnamefont {A.}~\bibnamefont
  {Roper~Pol}}, \bibinfo {author} {\bibfnamefont {S.}~\bibnamefont {Mandal}},
  \bibinfo {author} {\bibfnamefont {A.}~\bibnamefont {Brandenburg}}, \bibinfo
  {author} {\bibfnamefont {T.}~\bibnamefont {Kahniashvili}},\ and\ \bibinfo
  {author} {\bibfnamefont {A.}~\bibnamefont {Kosowsky}},\ }\bibfield  {title}
  {\emph {\bibinfo {title} {{Numerical simulations of gravitational waves from
  early-universe turbulence}}},\ }\href
  {https://doi.org/10.1103/PhysRevD.102.083512} {\bibfield  {journal} {\bibinfo
   {journal} {Phys. Rev. D}\ }\textbf {\bibinfo {volume} {102}},\ \bibinfo
  {pages} {083512} (\bibinfo {year} {2020})},\ \Eprint
  {https://arxiv.org/abs/1903.08585} {arXiv:1903.08585 [astro-ph.CO]}
  \BibitemShut {NoStop}%
\bibitem [{\citenamefont {Auclair}\ \emph {et~al.}(2022)\citenamefont
  {Auclair}, \citenamefont {Caprini}, \citenamefont {Cutting}, \citenamefont
  {Hindmarsh}, \citenamefont {Rummukainen}, \citenamefont {Steer},\ and\
  \citenamefont {Weir}}]{Auclair:2022jod}%
  \BibitemOpen
  \bibfield  {author} {\bibinfo {author} {\bibfnamefont {P.}~\bibnamefont
  {Auclair}}, \bibinfo {author} {\bibfnamefont {C.}~\bibnamefont {Caprini}},
  \bibinfo {author} {\bibfnamefont {D.}~\bibnamefont {Cutting}}, \bibinfo
  {author} {\bibfnamefont {M.}~\bibnamefont {Hindmarsh}}, \bibinfo {author}
  {\bibfnamefont {K.}~\bibnamefont {Rummukainen}}, \bibinfo {author}
  {\bibfnamefont {D.~A.}\ \bibnamefont {Steer}},\ and\ \bibinfo {author}
  {\bibfnamefont {D.~J.}\ \bibnamefont {Weir}},\ }\bibfield  {title} {\emph
  {\bibinfo {title} {{Generation of gravitational waves from freely decaying
  turbulence}}},\ }\href {https://doi.org/10.1088/1475-7516/2022/09/029}
  {\bibfield  {journal} {\bibinfo  {journal} {JCAP}\ }\textbf {\bibinfo
  {volume} {09}},\ \bibinfo {pages} {029}},\ \Eprint
  {https://arxiv.org/abs/2205.02588} {arXiv:2205.02588 [astro-ph.CO]}
  \BibitemShut {NoStop}%
\bibitem [{\citenamefont {Dahl}\ \emph {et~al.}(2022)\citenamefont {Dahl},
  \citenamefont {Hindmarsh}, \citenamefont {Rummukainen},\ and\ \citenamefont
  {Weir}}]{Dahl:2021wyk}%
  \BibitemOpen
  \bibfield  {author} {\bibinfo {author} {\bibfnamefont {J.}~\bibnamefont
  {Dahl}}, \bibinfo {author} {\bibfnamefont {M.}~\bibnamefont {Hindmarsh}},
  \bibinfo {author} {\bibfnamefont {K.}~\bibnamefont {Rummukainen}},\ and\
  \bibinfo {author} {\bibfnamefont {D.~J.}\ \bibnamefont {Weir}},\ }\bibfield
  {title} {\emph {\bibinfo {title} {{Decay of acoustic turbulence in two
  dimensions and implications for cosmological gravitational waves}}},\ }\href
  {https://doi.org/10.1103/PhysRevD.106.063511} {\bibfield  {journal} {\bibinfo
   {journal} {Phys. Rev. D}\ }\textbf {\bibinfo {volume} {106}},\ \bibinfo
  {pages} {063511} (\bibinfo {year} {2022})},\ \Eprint
  {https://arxiv.org/abs/2112.12013} {arXiv:2112.12013 [gr-qc]} \BibitemShut
  {NoStop}%
\bibitem [{\citenamefont {Dahl}\ \emph {et~al.}(2024)\citenamefont {Dahl},
  \citenamefont {Hindmarsh}, \citenamefont {Rummukainen},\ and\ \citenamefont
  {Weir}}]{Dahl:2024eup}%
  \BibitemOpen
  \bibfield  {author} {\bibinfo {author} {\bibfnamefont {J.}~\bibnamefont
  {Dahl}}, \bibinfo {author} {\bibfnamefont {M.}~\bibnamefont {Hindmarsh}},
  \bibinfo {author} {\bibfnamefont {K.}~\bibnamefont {Rummukainen}},\ and\
  \bibinfo {author} {\bibfnamefont {D.~J.}\ \bibnamefont {Weir}},\ }\bibfield
  {title} {\emph {\bibinfo {title} {{Primordial acoustic turbulence:
  Three-dimensional simulations and gravitational wave predictions}}},\ }\href
  {https://doi.org/10.1103/PhysRevD.110.103512} {\bibfield  {journal} {\bibinfo
   {journal} {Phys. Rev. D}\ }\textbf {\bibinfo {volume} {110}},\ \bibinfo
  {pages} {103512} (\bibinfo {year} {2024})},\ \Eprint
  {https://arxiv.org/abs/2407.05826} {arXiv:2407.05826 [gr-qc]} \BibitemShut
  {NoStop}%
\bibitem [{\citenamefont {Papamakarios}\ \emph {et~al.}(2018)\citenamefont
  {Papamakarios}, \citenamefont {Sterratt},\ and\ \citenamefont
  {Murray}}]{Papamakarios:2018zoy}%
  \BibitemOpen
  \bibfield  {author} {\bibinfo {author} {\bibfnamefont {G.}~\bibnamefont
  {Papamakarios}}, \bibinfo {author} {\bibfnamefont {D.~C.}\ \bibnamefont
  {Sterratt}},\ and\ \bibinfo {author} {\bibfnamefont {I.}~\bibnamefont
  {Murray}},\ }\bibfield  {title} {\emph {\bibinfo {title} {{Sequential Neural
  Likelihood: Fast Likelihood-free Inference with Autoregressive Flows}}},\
  }\href@noop {} {\  (\bibinfo {year} {2018})},\ \Eprint
  {https://arxiv.org/abs/1805.07226} {arXiv:1805.07226 [stat.ML]} \BibitemShut
  {NoStop}%
\bibitem [{\citenamefont {Alsing}\ \emph {et~al.}(2019)\citenamefont {Alsing},
  \citenamefont {Charnock}, \citenamefont {Feeney},\ and\ \citenamefont
  {Wandelt}}]{Alsing:2019xrx}%
  \BibitemOpen
  \bibfield  {author} {\bibinfo {author} {\bibfnamefont {J.}~\bibnamefont
  {Alsing}}, \bibinfo {author} {\bibfnamefont {T.}~\bibnamefont {Charnock}},
  \bibinfo {author} {\bibfnamefont {S.}~\bibnamefont {Feeney}},\ and\ \bibinfo
  {author} {\bibfnamefont {B.}~\bibnamefont {Wandelt}},\ }\bibfield  {title}
  {\emph {\bibinfo {title} {{Fast likelihood-free cosmology with neural density
  estimators and active learning}}},\ }\href
  {https://doi.org/10.1093/mnras/stz1960} {\bibfield  {journal} {\bibinfo
  {journal} {Mon. Not. Roy. Astron. Soc.}\ }\textbf {\bibinfo {volume} {488}},\
  \bibinfo {pages} {4440} (\bibinfo {year} {2019})},\ \Eprint
  {https://arxiv.org/abs/1903.00007} {arXiv:1903.00007 [astro-ph.CO]}
  \BibitemShut {NoStop}%
\bibitem [{\citenamefont {Whittle}(1953)}]{Whittle:1953}%
  \BibitemOpen
  \bibfield  {author} {\bibinfo {author} {\bibfnamefont {P.}~\bibnamefont
  {Whittle}},\ }\bibfield  {title} {\emph {\bibinfo {title} {{The analysis of
  multiple stationary time series}}},\ }\href@noop {} {\bibfield  {journal}
  {\bibinfo  {journal} {J. Roy. Statist. Soc. B}\ }\textbf {\bibinfo {volume}
  {15}},\ \bibinfo {pages} {125} (\bibinfo {year} {1953})}\BibitemShut
  {NoStop}%
\bibitem [{\citenamefont {Franciolini}\ \emph {et~al.}(2025)\citenamefont
  {Franciolini}, \citenamefont {Pieroni}, \citenamefont {Ricciardone},\ and\
  \citenamefont {Romano}}]{Franciolini:2025leq}%
  \BibitemOpen
  \bibfield  {author} {\bibinfo {author} {\bibfnamefont {G.}~\bibnamefont
  {Franciolini}}, \bibinfo {author} {\bibfnamefont {M.}~\bibnamefont
  {Pieroni}}, \bibinfo {author} {\bibfnamefont {A.}~\bibnamefont
  {Ricciardone}},\ and\ \bibinfo {author} {\bibfnamefont {J.~D.}\ \bibnamefont
  {Romano}},\ }\bibfield  {title} {\emph {\bibinfo {title} {{Likelihoods for
  stochastic gravitational wave background data analysis}}},\ }\href
  {https://doi.org/10.1103/sjtb-gnz5} {\bibfield  {journal} {\bibinfo
  {journal} {Phys. Rev. D}\ }\textbf {\bibinfo {volume} {112}},\ \bibinfo
  {pages} {103516} (\bibinfo {year} {2025})},\ \Eprint
  {https://arxiv.org/abs/2505.24695} {arXiv:2505.24695 [gr-qc]} \BibitemShut
  {NoStop}%
\bibitem [{\citenamefont {Flauger}\ \emph {et~al.}(2021)\citenamefont
  {Flauger}, \citenamefont {Karnesis}, \citenamefont {Nardini}, \citenamefont
  {Pieroni}, \citenamefont {Ricciardone},\ and\ \citenamefont
  {Torrado}}]{Flauger:2020qyi}%
  \BibitemOpen
  \bibfield  {author} {\bibinfo {author} {\bibfnamefont {R.}~\bibnamefont
  {Flauger}}, \bibinfo {author} {\bibfnamefont {N.}~\bibnamefont {Karnesis}},
  \bibinfo {author} {\bibfnamefont {G.}~\bibnamefont {Nardini}}, \bibinfo
  {author} {\bibfnamefont {M.}~\bibnamefont {Pieroni}}, \bibinfo {author}
  {\bibfnamefont {A.}~\bibnamefont {Ricciardone}},\ and\ \bibinfo {author}
  {\bibfnamefont {J.}~\bibnamefont {Torrado}},\ }\bibfield  {title} {\emph
  {\bibinfo {title} {{Improved reconstruction of a stochastic gravitational
  wave background with LISA}}},\ }\href@noop {} {\bibfield  {journal} {\bibinfo
   {journal} {JCAP}\ }\textbf {\bibinfo {volume} {01}},\ \bibinfo {pages}
  {059}},\ \Eprint {https://arxiv.org/abs/2009.11845} {arXiv:2009.11845
  [astro-ph.CO]} \BibitemShut {NoStop}%
\bibitem [{\citenamefont {Satterthwaite}(1946)}]{Satterthwaite1946AnAD}%
  \BibitemOpen
  \bibfield  {author} {\bibinfo {author} {\bibfnamefont {F.~E.}\ \bibnamefont
  {Satterthwaite}},\ }\bibfield  {title} {\emph {\bibinfo {title} {{An
  approximate distribution of estimates of variance components}}},\ }\href@noop
  {} {\bibfield  {journal} {\bibinfo  {journal} {Biometrics Bulletin}\ }\textbf
  {\bibinfo {volume} {2}},\ \bibinfo {pages} {110} (\bibinfo {year}
  {1946})}\BibitemShut {NoStop}%
\bibitem [{\citenamefont {Greenberg}\ \emph {et~al.}(2019)\citenamefont
  {Greenberg}, \citenamefont {Nonnenmacher},\ and\ \citenamefont
  {Macke}}]{Greenberg:2019apt}%
  \BibitemOpen
  \bibfield  {author} {\bibinfo {author} {\bibfnamefont {D.~S.}\ \bibnamefont
  {Greenberg}}, \bibinfo {author} {\bibfnamefont {M.}~\bibnamefont
  {Nonnenmacher}},\ and\ \bibinfo {author} {\bibfnamefont {J.~H.}\ \bibnamefont
  {Macke}},\ }in\ \href@noop {} {\emph {\bibinfo {booktitle} {Proc. 36th Int.
  Conf. on Machine Learning (ICML)}}}\ (\bibinfo {year} {2019})\ \Eprint
  {https://arxiv.org/abs/1905.07488} {arXiv:1905.07488 [cs.LG]} \BibitemShut
  {NoStop}%
\bibitem [{\citenamefont {Durkan}\ \emph {et~al.}(2019)\citenamefont {Durkan},
  \citenamefont {Bekasov}, \citenamefont {Murray},\ and\ \citenamefont
  {Papamakarios}}]{Durkan:2019nsf}%
  \BibitemOpen
  \bibfield  {author} {\bibinfo {author} {\bibfnamefont {C.}~\bibnamefont
  {Durkan}}, \bibinfo {author} {\bibfnamefont {A.}~\bibnamefont {Bekasov}},
  \bibinfo {author} {\bibfnamefont {I.}~\bibnamefont {Murray}},\ and\ \bibinfo
  {author} {\bibfnamefont {G.}~\bibnamefont {Papamakarios}},\ }in\ \href@noop
  {} {\emph {\bibinfo {booktitle} {Advances in Neural Information Processing
  Systems 32 (NeurIPS)}}}\ (\bibinfo {year} {2019})\ \Eprint
  {https://arxiv.org/abs/1906.04032} {arXiv:1906.04032 [stat.ML]} \BibitemShut
  {NoStop}%
\bibitem [{\citenamefont {Paige}\ and\ \citenamefont
  {Wood}(2018)}]{Brooks:2016}%
  \BibitemOpen
  \bibfield  {author} {\bibinfo {author} {\bibfnamefont {B.}~\bibnamefont
  {Paige}}\ and\ \bibinfo {author} {\bibfnamefont {F.}~\bibnamefont {Wood}},\
  }\href {https://arxiv.org/abs/1602.06701} {\bibinfo {title} {Inference
  networks for sequential monte carlo in graphical models}} (\bibinfo {year}
  {2018}),\ \Eprint {https://arxiv.org/abs/1602.06701} {arXiv:1602.06701
  [stat.ML]} \BibitemShut {NoStop}%
\bibitem [{\citenamefont {Le}\ \emph {et~al.}(2017)\citenamefont {Le},
  \citenamefont {Baydin},\ and\ \citenamefont {Wood}}]{Le:2016}%
  \BibitemOpen
  \bibfield  {author} {\bibinfo {author} {\bibfnamefont {T.~A.}\ \bibnamefont
  {Le}}, \bibinfo {author} {\bibfnamefont {A.~G.}\ \bibnamefont {Baydin}},\
  and\ \bibinfo {author} {\bibfnamefont {F.}~\bibnamefont {Wood}},\ }\href
  {https://arxiv.org/abs/1610.09900} {\bibinfo {title} {Inference compilation
  and universal probabilistic programming}} (\bibinfo {year} {2017}),\ \Eprint
  {https://arxiv.org/abs/1610.09900} {arXiv:1610.09900 [cs.AI]} \BibitemShut
  {NoStop}%
\bibitem [{\citenamefont {Chala}\ \emph {et~al.}(2024)\citenamefont {Chala},
  \citenamefont {Criado}, \citenamefont {Gil},\ and\ \citenamefont
  {Miras}}]{Chala:2024xll}%
  \BibitemOpen
  \bibfield  {author} {\bibinfo {author} {\bibfnamefont {M.}~\bibnamefont
  {Chala}}, \bibinfo {author} {\bibfnamefont {J.~C.}\ \bibnamefont {Criado}},
  \bibinfo {author} {\bibfnamefont {L.}~\bibnamefont {Gil}},\ and\ \bibinfo
  {author} {\bibfnamefont {J.~L.}\ \bibnamefont {Miras}},\ }\bibfield  {title}
  {\emph {\bibinfo {title} {{Higher-order-operator corrections to
  phase-transition parameters in dimensional reduction}}},\ }\href
  {https://doi.org/10.1007/JHEP10(2024)025} {\bibfield  {journal} {\bibinfo
  {journal} {JHEP}\ }\textbf {\bibinfo {volume} {10}},\ \bibinfo {pages}
  {025}},\ \Eprint {https://arxiv.org/abs/2406.02667} {arXiv:2406.02667
  [hep-ph]} \BibitemShut {NoStop}%
\bibitem [{\citenamefont {Gould}\ and\ \citenamefont
  {Saffin}(2025)}]{Gould:2024jjt}%
  \BibitemOpen
  \bibfield  {author} {\bibinfo {author} {\bibfnamefont {O.}~\bibnamefont
  {Gould}}\ and\ \bibinfo {author} {\bibfnamefont {P.~M.}\ \bibnamefont
  {Saffin}},\ }\bibfield  {title} {\emph {\bibinfo {title} {{Perturbative
  gravitational wave predictions for the real-scalar extended Standard
  Model}}},\ }\href {https://doi.org/10.1007/JHEP03(2025)105} {\bibfield
  {journal} {\bibinfo  {journal} {JHEP}\ }\textbf {\bibinfo {volume} {03}},\
  \bibinfo {pages} {105}},\ \Eprint {https://arxiv.org/abs/2411.08951}
  {arXiv:2411.08951 [hep-ph]} \BibitemShut {NoStop}%
\bibitem [{\citenamefont {Bernardo}\ \emph {et~al.}(2025)\citenamefont
  {Bernardo}, \citenamefont {Klose}, \citenamefont {Schicho},\ and\
  \citenamefont {Tenkanen}}]{Bernardo:2025vkz}%
  \BibitemOpen
  \bibfield  {author} {\bibinfo {author} {\bibfnamefont {F.}~\bibnamefont
  {Bernardo}}, \bibinfo {author} {\bibfnamefont {P.}~\bibnamefont {Klose}},
  \bibinfo {author} {\bibfnamefont {P.}~\bibnamefont {Schicho}},\ and\ \bibinfo
  {author} {\bibfnamefont {T.~V.~I.}\ \bibnamefont {Tenkanen}},\ }\bibfield
  {title} {\emph {\bibinfo {title} {{Higher-dimensional operators at finite
  temperature affect gravitational-wave predictions}}},\ }\href
  {https://doi.org/10.1007/JHEP08(2025)109} {\bibfield  {journal} {\bibinfo
  {journal} {JHEP}\ }\textbf {\bibinfo {volume} {08}},\ \bibinfo {pages}
  {109}},\ \Eprint {https://arxiv.org/abs/2503.18904} {arXiv:2503.18904
  [hep-ph]} \BibitemShut {NoStop}%
\bibitem [{\citenamefont {Chala}\ and\ \citenamefont
  {Guedes}(2025)}]{Chala:2025aiz}%
  \BibitemOpen
  \bibfield  {author} {\bibinfo {author} {\bibfnamefont {M.}~\bibnamefont
  {Chala}}\ and\ \bibinfo {author} {\bibfnamefont {G.}~\bibnamefont {Guedes}},\
  }\bibfield  {title} {\emph {\bibinfo {title} {{The high-temperature limit of
  the SM(EFT)}}},\ }\href {https://doi.org/10.1007/JHEP07(2025)085} {\bibfield
  {journal} {\bibinfo  {journal} {JHEP}\ }\textbf {\bibinfo {volume} {07}},\
  \bibinfo {pages} {085}},\ \Eprint {https://arxiv.org/abs/2503.20016}
  {arXiv:2503.20016 [hep-ph]} \BibitemShut {NoStop}%
\bibitem [{\citenamefont {Chala}\ \emph {et~al.}(2025)\citenamefont {Chala},
  \citenamefont {Gil},\ and\ \citenamefont {Ren}}]{Chala:2025oul}%
  \BibitemOpen
  \bibfield  {author} {\bibinfo {author} {\bibfnamefont {M.}~\bibnamefont
  {Chala}}, \bibinfo {author} {\bibfnamefont {L.}~\bibnamefont {Gil}},\ and\
  \bibinfo {author} {\bibfnamefont {Z.}~\bibnamefont {Ren}},\ }\bibfield
  {title} {\emph {\bibinfo {title} {{Phase transitions in dimensional reduction
  up to three loops}}},\ }\href {https://doi.org/10.1088/1674-1137/adf322}
  {\bibfield  {journal} {\bibinfo  {journal} {Chin. Phys. C}\ }\textbf
  {\bibinfo {volume} {49}},\ \bibinfo {pages} {123105} (\bibinfo {year}
  {2025})},\ \Eprint {https://arxiv.org/abs/2505.14335} {arXiv:2505.14335
  [hep-ph]} \BibitemShut {NoStop}%
\bibitem [{\citenamefont {Espinosa}\ \emph {et~al.}(2010)\citenamefont
  {Espinosa}, \citenamefont {Konstandin}, \citenamefont {No},\ and\
  \citenamefont {Servant}}]{Espinosa:2010hh}%
  \BibitemOpen
  \bibfield  {author} {\bibinfo {author} {\bibfnamefont {J.~R.}\ \bibnamefont
  {Espinosa}}, \bibinfo {author} {\bibfnamefont {T.}~\bibnamefont
  {Konstandin}}, \bibinfo {author} {\bibfnamefont {J.~M.}\ \bibnamefont {No}},\
  and\ \bibinfo {author} {\bibfnamefont {G.}~\bibnamefont {Servant}},\
  }\bibfield  {title} {\emph {\bibinfo {title} {{Energy Budget of Cosmological
  First-order Phase Transitions}}},\ }\href
  {https://doi.org/10.1088/1475-7516/2010/06/028} {\bibfield  {journal}
  {\bibinfo  {journal} {JCAP}\ }\textbf {\bibinfo {volume} {06}},\ \bibinfo
  {pages} {028}},\ \Eprint {https://arxiv.org/abs/1004.4187} {arXiv:1004.4187
  [hep-ph]} \BibitemShut {NoStop}%
\bibitem [{\citenamefont {Gowling}\ and\ \citenamefont
  {Hindmarsh}(2021)}]{Gowling:2021gcy}%
  \BibitemOpen
  \bibfield  {author} {\bibinfo {author} {\bibfnamefont {C.}~\bibnamefont
  {Gowling}}\ and\ \bibinfo {author} {\bibfnamefont {M.}~\bibnamefont
  {Hindmarsh}},\ }\bibfield  {title} {\emph {\bibinfo {title} {{Observational
  prospects for phase transitions at LISA: Fisher matrix analysis}}},\ }\href
  {https://doi.org/10.1088/1475-7516/2021/10/039} {\bibfield  {journal}
  {\bibinfo  {journal} {JCAP}\ }\textbf {\bibinfo {volume} {10}},\ \bibinfo
  {pages} {039}},\ \Eprint {https://arxiv.org/abs/2106.05984} {arXiv:2106.05984
  [astro-ph.CO]} \BibitemShut {NoStop}%
\bibitem [{\citenamefont {Giese}\ \emph {et~al.}(2021)\citenamefont {Giese},
  \citenamefont {Konstandin},\ and\ \citenamefont {van~de
  Vis}}]{Giese:2021dnw}%
  \BibitemOpen
  \bibfield  {author} {\bibinfo {author} {\bibfnamefont {F.}~\bibnamefont
  {Giese}}, \bibinfo {author} {\bibfnamefont {T.}~\bibnamefont {Konstandin}},\
  and\ \bibinfo {author} {\bibfnamefont {J.}~\bibnamefont {van~de Vis}},\
  }\bibfield  {title} {\emph {\bibinfo {title} {{Finding sound shells in LISA
  mock data using likelihood sampling}}},\ }\href
  {https://doi.org/10.1088/1475-7516/2021/11/002} {\bibfield  {journal}
  {\bibinfo  {journal} {JCAP}\ }\textbf {\bibinfo {volume} {11}},\ \bibinfo
  {pages} {002}},\ \Eprint {https://arxiv.org/abs/2107.06275} {arXiv:2107.06275
  [astro-ph.CO]} \BibitemShut {NoStop}%
\bibitem [{\citenamefont {Swendsen}\ and\ \citenamefont
  {Wang}(1986)}]{Swendsen:1986vqb}%
  \BibitemOpen
  \bibfield  {author} {\bibinfo {author} {\bibfnamefont {R.~H.}\ \bibnamefont
  {Swendsen}}\ and\ \bibinfo {author} {\bibfnamefont {J.-S.}\ \bibnamefont
  {Wang}},\ }\bibfield  {title} {\emph {\bibinfo {title} {{Replica Monte Carlo
  Simulation of Spin-Glasses}}},\ }\href
  {https://doi.org/10.1103/PhysRevLett.57.2607} {\bibfield  {journal} {\bibinfo
   {journal} {Phys. Rev. Lett.}\ }\textbf {\bibinfo {volume} {57}},\ \bibinfo
  {pages} {2607} (\bibinfo {year} {1986})}\BibitemShut {NoStop}%
\bibitem [{\citenamefont {Earl}\ and\ \citenamefont
  {Deem}(2005)}]{Earl:2005wvh}%
  \BibitemOpen
  \bibfield  {author} {\bibinfo {author} {\bibfnamefont {D.~J.}\ \bibnamefont
  {Earl}}\ and\ \bibinfo {author} {\bibfnamefont {M.~W.}\ \bibnamefont
  {Deem}},\ }\bibfield  {title} {\emph {\bibinfo {title} {{Parallel tempering:
  Theory, applications, and new perspectives}}},\ }\href
  {https://doi.org/10.1039/b509983h} {\bibfield  {journal} {\bibinfo  {journal}
  {Phys. Chem. Chem. Phys.}\ }\textbf {\bibinfo {volume} {7}},\ \bibinfo
  {pages} {3910} (\bibinfo {year} {2005})},\ \Eprint
  {https://arxiv.org/abs/physics/0508111} {arXiv:physics/0508111} \BibitemShut
  {NoStop}%
\bibitem [{\citenamefont {Vousden}\ \emph {et~al.}(2016)\citenamefont
  {Vousden}, \citenamefont {Farr},\ and\ \citenamefont
  {Mandel}}]{Vousden:2016eeu}%
  \BibitemOpen
  \bibfield  {author} {\bibinfo {author} {\bibfnamefont {W.~D.}\ \bibnamefont
  {Vousden}}, \bibinfo {author} {\bibfnamefont {W.~M.}\ \bibnamefont {Farr}},\
  and\ \bibinfo {author} {\bibfnamefont {I.}~\bibnamefont {Mandel}},\
  }\bibfield  {title} {\emph {\bibinfo {title} {{Dynamic temperature selection
  for parallel tempering in Markov chain Monte Carlo simulations}}},\ }\href
  {https://doi.org/10.1093/mnras/stv2422} {\bibfield  {journal} {\bibinfo
  {journal} {Mon. Not. Roy. Astron. Soc.}\ }\textbf {\bibinfo {volume} {455}},\
  \bibinfo {pages} {1919} (\bibinfo {year} {2016})},\ \Eprint
  {https://arxiv.org/abs/1501.05823} {arXiv:1501.05823} \BibitemShut {NoStop}%
\bibitem [{\citenamefont {Skilling}(2006)}]{Skilling:2006gxv}%
  \BibitemOpen
  \bibfield  {author} {\bibinfo {author} {\bibfnamefont {J.}~\bibnamefont
  {Skilling}},\ }\bibfield  {title} {\emph {\bibinfo {title} {{Nested sampling
  for general Bayesian computation}}},\ }\href
  {https://doi.org/10.1214/06-BA127} {\bibfield  {journal} {\bibinfo  {journal}
  {Bayesian Analysis}\ }\textbf {\bibinfo {volume} {1}},\ \bibinfo {pages}
  {833} (\bibinfo {year} {2006})}\BibitemShut {NoStop}%
\bibitem [{\citenamefont {Tinto}\ and\ \citenamefont
  {Dhurandhar}(2021)}]{Tinto:2020fcc}%
  \BibitemOpen
  \bibfield  {author} {\bibinfo {author} {\bibfnamefont {M.}~\bibnamefont
  {Tinto}}\ and\ \bibinfo {author} {\bibfnamefont {S.~V.}\ \bibnamefont
  {Dhurandhar}},\ }\bibfield  {title} {\emph {\bibinfo {title} {{Time-delay
  interferometry}}},\ }\href {https://doi.org/10.1007/s41114-020-00029-6}
  {\bibfield  {journal} {\bibinfo  {journal} {Living Rev. Rel.}\ }\textbf
  {\bibinfo {volume} {24}},\ \bibinfo {pages} {1} (\bibinfo {year}
  {2021})}\BibitemShut {NoStop}%
\bibitem [{\citenamefont {Hermans}\ \emph {et~al.}(2021)\citenamefont
  {Hermans}, \citenamefont {Delaunoy}, \citenamefont {Rozet}, \citenamefont
  {Wehenkel}, \citenamefont {Begy},\ and\ \citenamefont
  {Louppe}}]{Hermans:2021rqv}%
  \BibitemOpen
  \bibfield  {author} {\bibinfo {author} {\bibfnamefont {J.}~\bibnamefont
  {Hermans}}, \bibinfo {author} {\bibfnamefont {A.}~\bibnamefont {Delaunoy}},
  \bibinfo {author} {\bibfnamefont {F.}~\bibnamefont {Rozet}}, \bibinfo
  {author} {\bibfnamefont {A.}~\bibnamefont {Wehenkel}}, \bibinfo {author}
  {\bibfnamefont {V.}~\bibnamefont {Begy}},\ and\ \bibinfo {author}
  {\bibfnamefont {G.}~\bibnamefont {Louppe}},\ }\bibfield  {title} {\emph
  {\bibinfo {title} {{A Trust Crisis In Simulation-Based Inference? Your
  Posterior Approximations Can Be Unfaithful}}},\ }\href@noop {} {\  (\bibinfo
  {year} {2021})},\ \Eprint {https://arxiv.org/abs/2110.06581}
  {arXiv:2110.06581 [stat.ML]} \BibitemShut {NoStop}%
\bibitem [{\citenamefont {Kingma}\ and\ \citenamefont
  {Ba}(2014)}]{Kingma:2014vow}%
  \BibitemOpen
  \bibfield  {author} {\bibinfo {author} {\bibfnamefont {D.~P.}\ \bibnamefont
  {Kingma}}\ and\ \bibinfo {author} {\bibfnamefont {J.}~\bibnamefont {Ba}},\
  }in\ \href@noop {} {\emph {\bibinfo {booktitle} {{International Conference on
  Learning Representations}}}}\ (\bibinfo {year} {2014})\ \Eprint
  {https://arxiv.org/abs/1412.6980} {arXiv:1412.6980 [cs.LG]} \BibitemShut
  {NoStop}%
\bibitem [{\citenamefont {Polyak}\ and\ \citenamefont
  {Juditsky}(1992)}]{Polyak:1992}%
  \BibitemOpen
  \bibfield  {author} {\bibinfo {author} {\bibfnamefont {B.}~\bibnamefont
  {Polyak}}\ and\ \bibinfo {author} {\bibfnamefont {A.}~\bibnamefont
  {Juditsky}},\ }\bibfield  {title} {\emph {\bibinfo {title} {Acceleration of
  stochastic approximation by averaging}},\ }\href
  {https://doi.org/10.1137/0330046} {\bibfield  {journal} {\bibinfo  {journal}
  {SIAM Journal on Control and Optimization}\ }\textbf {\bibinfo {volume}
  {30}},\ \bibinfo {pages} {838} (\bibinfo {year} {1992})}\BibitemShut
  {NoStop}%
\bibitem [{\citenamefont {Morales-Brotons}\ \emph {et~al.}(2024)\citenamefont
  {Morales-Brotons}, \citenamefont {Vogels},\ and\ \citenamefont
  {Hendrikx}}]{moralesbrotons2024exponentialmovingaverageweights}%
  \BibitemOpen
  \bibfield  {author} {\bibinfo {author} {\bibfnamefont {D.}~\bibnamefont
  {Morales-Brotons}}, \bibinfo {author} {\bibfnamefont {T.}~\bibnamefont
  {Vogels}},\ and\ \bibinfo {author} {\bibfnamefont {H.}~\bibnamefont
  {Hendrikx}},\ }\href {https://arxiv.org/abs/2411.18704} {\bibinfo {title}
  {Exponential moving average of weights in deep learning: Dynamics and
  benefits}} (\bibinfo {year} {2024}),\ \Eprint
  {https://arxiv.org/abs/2411.18704} {arXiv:2411.18704 [cs.LG]} \BibitemShut
  {NoStop}%
\bibitem [{\citenamefont {Wagner-Carena}\ \emph {et~al.}(2021)\citenamefont
  {Wagner-Carena}, \citenamefont {Park}, \citenamefont {Birrer}, \citenamefont
  {Marshall}, \citenamefont {Roodman},\ and\ \citenamefont
  {Wechsler}}]{WagnerCarena:2020tbd}%
  \BibitemOpen
  \bibfield  {author} {\bibinfo {author} {\bibfnamefont {S.}~\bibnamefont
  {Wagner-Carena}}, \bibinfo {author} {\bibfnamefont {J.~W.}\ \bibnamefont
  {Park}}, \bibinfo {author} {\bibfnamefont {S.}~\bibnamefont {Birrer}},
  \bibinfo {author} {\bibfnamefont {P.~J.}\ \bibnamefont {Marshall}}, \bibinfo
  {author} {\bibfnamefont {A.}~\bibnamefont {Roodman}},\ and\ \bibinfo {author}
  {\bibfnamefont {R.~H.}\ \bibnamefont {Wechsler}},\ }\bibfield  {title} {\emph
  {\bibinfo {title} {{Hierarchical Inference With Bayesian Neural Networks: An
  Application to Strong Gravitational Lensing}}},\ }\href
  {https://doi.org/10.3847/1538-4357/abdf59} {\bibfield  {journal} {\bibinfo
  {journal} {Astrophys. J.}\ }\textbf {\bibinfo {volume} {909}},\ \bibinfo
  {pages} {187} (\bibinfo {year} {2021})},\ \Eprint
  {https://arxiv.org/abs/2010.13787} {arXiv:2010.13787 [astro-ph.CO]}
  \BibitemShut {NoStop}%
\bibitem [{\citenamefont {Tejero-Cantero}\ \emph {et~al.}(2020)\citenamefont
  {Tejero-Cantero}, \citenamefont {Boelts}, \citenamefont {Deistler},
  \citenamefont {Lueckmann}, \citenamefont {Durkan}, \citenamefont
  {Gon\c{c}alves}, \citenamefont {Greenberg},\ and\ \citenamefont
  {Macke}}]{TejeroCantero:2020apm}%
  \BibitemOpen
  \bibfield  {author} {\bibinfo {author} {\bibfnamefont {A.}~\bibnamefont
  {Tejero-Cantero}}, \bibinfo {author} {\bibfnamefont {J.}~\bibnamefont
  {Boelts}}, \bibinfo {author} {\bibfnamefont {M.}~\bibnamefont {Deistler}},
  \bibinfo {author} {\bibfnamefont {J.-M.}\ \bibnamefont {Lueckmann}}, \bibinfo
  {author} {\bibfnamefont {C.}~\bibnamefont {Durkan}}, \bibinfo {author}
  {\bibfnamefont {P.~J.}\ \bibnamefont {Gon\c{c}alves}}, \bibinfo {author}
  {\bibfnamefont {D.~S.}\ \bibnamefont {Greenberg}},\ and\ \bibinfo {author}
  {\bibfnamefont {J.~H.}\ \bibnamefont {Macke}},\ }\bibfield  {title} {\emph
  {\bibinfo {title} {{sbi: A toolkit for simulation-based inference}}},\
  }\href@noop {} {\bibfield  {journal} {\bibinfo  {journal} {J. Open Source
  Softw.}\ }\textbf {\bibinfo {volume} {5}},\ \bibinfo {pages} {2505} (\bibinfo
  {year} {2020})},\ \Eprint {https://arxiv.org/abs/2007.09114}
  {arXiv:2007.09114 [stat.ML]} \BibitemShut {NoStop}%
\bibitem [{\citenamefont {Lewis}(2025)}]{Lewis:2019xzd}%
  \BibitemOpen
  \bibfield  {author} {\bibinfo {author} {\bibfnamefont {A.}~\bibnamefont
  {Lewis}},\ }\bibfield  {title} {\emph {\bibinfo {title} {{GetDist: a Python
  package for analysing Monte Carlo samples}}},\ }\href
  {https://doi.org/10.1088/1475-7516/2025/08/025} {\bibfield  {journal}
  {\bibinfo  {journal} {JCAP}\ }\textbf {\bibinfo {volume} {08}},\ \bibinfo
  {pages} {025}},\ \Eprint {https://arxiv.org/abs/1910.13970} {arXiv:1910.13970
  [astro-ph.IM]} \BibitemShut {NoStop}%
\end{thebibliography}%
  \bibliographystyle{apsrev4-2}
\end{document}